\documentclass[journal, 12pt, onecolumn, draftclsnofoot]{IEEEtran}
\usepackage{authblk}
\usepackage{comment}
\usepackage{verbatim}
\usepackage{float}
\usepackage{graphicx}
\usepackage{xcolor}
\usepackage[algoruled]{algorithm2e}
\usepackage{multirow}
\usepackage{graphicx}
\usepackage{colortbl}
\usepackage{tikz}
\usepackage{marvosym}
\usepackage{tikzsymbols}
\usetikzlibrary{automata,arrows,positioning,calc}
\usetikzlibrary{patterns}
\usetikzlibrary{matrix}
\usetikzlibrary{tikzmark}
\usetikzlibrary{fit}
\usetikzlibrary{shadows.blur}
\usetikzlibrary{shapes.symbols}
\usetikzlibrary{shapes.geometric}
\usepackage{bm}
\usepackage{blkarray}

\usetikzlibrary{positioning}
\usepackage{cancel}
\usepackage{amssymb, amsmath}
\usepackage{amsthm}
\usepackage{thmtools}
\usepackage{epstopdf}
\usepackage{stfloats}
\newtheorem{lemma}{Lemma}
\newtheorem{theorem}{Theorem}
\newtheorem{corollary}{Corollary}
\newtheorem{remark}{Remark}

\declaretheorem[style=definition,numbered=yes]{definition}

\newcommand{\gc}{\mbox{\tiny BCGM}}
\newcommand{\usi}{\mbox{\tiny USI}}
\newcommand{\icmap}{\mbox{\tiny MR}}
\newcommand{\de}{\mbox{\rm DE}}
\newcommand{\si}{\mbox{\rm SI}}
\newcommand{\mds}{\mbox{\rm MDS}}
\newcommand{\bI}{\mathbf{I}}
\newcommand{\bzero}{\mathbf{0}}

\def\Mod#1#2{\left\langle #1 \right\rangle_{#2}}

\begin{document}

\title{\Large Blind Interference Alignment for MapReduce: Exploiting Side-information with Reconfigurable Antennas}
\author{\normalsize Yuxiang Lu and Syed A. Jafar\\
{\small Center for Pervasive Communications and Computing (CPCC)}\\
{\small University of California Irvine, Irvine, CA 92697}\\
{\small \it Email: \{yuxiang.lu, syed\}@uci.edu}
\thanks{The results of this work were presented in part at the 2023 57th Asilomar Conference on Signals, Systems and Computers (Asilomar) [DOI: 10.1109/IEEECONF59524.2023.10477052].}
}
\date{}                                           
\maketitle
\begin{abstract}
In order to explore how blind interference alignment (BIA) schemes may take advantage of side-information in computation tasks, we study the degrees of freedom (DoF) of a $K$ user wireless network setting that arises in full-duplex wireless MapReduce  applications. In this setting the receivers are assumed to have reconfigurable antennas and channel knowledge, while the transmitters have neither, i.e., the transmitters lack  channel knowledge and are only equipped with conventional antennas. The central ingredient of the problem formulation is the message structure arising out of the Shuffle phase of MapReduce, whereby each transmitter has a subset of messages that need to be delivered to various receivers, and each receiver has a subset of messages available to it in advance as side-information. We approach this problem by decomposing it into distinctive stages that help identify key ingredients of the overall solution. The novel elements that emerge from the first stage, called broadcast with groupcast messages, include an outer maximum distance separable (MDS) code structure at the transmitter, and an algorithm for iteratively determining groupcast-optimal reconfigurable antenna switching patterns at the receiver to achieve intra-message  (among the symbols of the same message) alignment. The next stage, called unicast with side-information, reveals optimal inter-message (among symbols of different messages) alignment patterns to exploit side-information, and by a relabeling of messages, connects to the desired MapReduce setting.
\end{abstract}

\section{Introduction}
Accelerating trends towards distributed computation \cite{Dean_Ghemawat_MapReduce,sengupta2017fog,ha2019coded,Yu_Epcodes,Yu_Lagrange,Jia_Jafar_CDBC} create new paradigms for re-assessing the capacity of communication networks. One distinguishing aspect of communication networks in the context of distributed computation is the abundance of \emph{side-information}, 
arising naturally as a file is typically processed at multiple computation nodes \cite{Dean_Ghemawat_MapReduce,Li_Chen_Wang_MapReduce,Bi_Wigger_MapReduce,Bi_Cooperative_X,li2017fundamental,li2017scalable,ha2019wireless,yang2019data,xu2021new,han2021over,wang2022dynamic} at intermediate stages of computation.
For wireless networks, that are typically interference-limited, side-information can be especially useful if it enables new robust interference management schemes, e.g.,  interference alignment (IA) constructions that avoid the need for precise channel state information at the transmitters (CSIT). 

High precision CSIT requirements have been the bane of many otherwise promising interference alignment schemes. For example, consider a $K\times K$ wireless network, i.e., a wireless network with $K$ transmitters and $K$ receivers, each equipped with only a single transmit/receive antenna. If there are $K$ independent messages to be delivered, one each from every transmitter to its corresponding receiver, then this is called the $K$ user interference channel. It is known that the $K$ user interference channel has  $K/2$ degrees of freedom under perfect CSIT, that can be achieved by interference alignment schemes \cite{Cadambe_Jafar_int}. However, in sharp contrast, if the CSIT is limited to finite precision then it is also known that the DoF value collapses to $1$, which is achievable with trivial orthogonal access schemes \cite{Arash_Jafar}.

The principle of blind interference alignment (BIA) stands out in this regard due to its minimal CSIT requirements. Introduced originally in \cite{Jafar_BIA} to take advantage of naturally occurring channel coherence patterns, BIA was shown in \cite{Gou_Wang_Jafar_BIA} to be much more powerful when used in conjunction with reconfigurable antennas, especially at the receivers. This is because blindly switching between different antenna modes at the receivers in pre-determined patterns creates predictable, and therefore exploitable, channel coherence patterns. Even though CSIT is lacking, BIA schemes can be surprisingly powerful. For example, consider again the $K\times K$ wireless network but suppose now that there is an independent message from every transmitter to every receiver. This setting is called an $X$ network \cite{Jafar_FnT}. In the $X$ network, if the receivers are equipped with reconfigurable antennas then, even though the transmitters are blind, BIA schemes achieve $K^2/(2K-1)$ DoF \cite{Cadambe_Jafar_X}. Not only is this more than the $K/2$ DoF of the $K$ user interference channel, this is also the maximum DoF value that is possible with \emph{perfect} CSIT in an $X$ network. Thus, from a DoF standpoint, reconfigurable antennas at the receivers are sufficient to unlock the full advantages of interference alignment, with no need for expensive CSIT. Indeed, the advantages of BIA schemes have been explored in a variety of settings that include the MISO BC and $X$-channel \cite{Gou_Wang_Jafar_BIA, Wang_Gou_Jafar_MIMO}, two-cell Z interference MIMO channel \cite{chen2017blind}, cellular networks \cite{BIA_Cellular}, topological interference management framework  \cite{yang2017topologicalBIA}, non-orthogonal multiple access (NOMA) \cite{BIA_NOMA}, and various interference networks \cite{BIA_IC,BIA_MISO_IC_Lu,BIA_MISO_IC_Yang,BIA_SISO_IC_Yang,BIA_SISO_IC__Johnny_Letter,BIA_SISO_IC_Johnny_IT}. A survey by Menon and Selvaprabhu is available in \cite{menon2022BIA_Survey}.

While BIA schemes overcome a critical hurdle by  minimizing the need for CSIT, they introduce new challenges of their own. Among these are the need for reconfigurable antennas which remains an active research
topic, the need for $X$ message sets (such that each transmitter has independent messages for multiple receivers and each receiver desires independent messages from multiple transmitters), the need for long/perfect coherence intervals, and the need to translate the DoF advantage to finite SNRs.  From an information theoretic standpoint, progress on the last two issues, i.e., imperfect coherence and finite SNR performance, with a few notable exceptions, is hindered by their limited analytical tractability, and may perhaps benefit in the future from the powerful machinery of deep neural networks to explore these high dimensional optimizations. 

In terms of the need for $X$ message sets, as communication networks are increasingly used in service of computation tasks, the changes favor BIA schemes as well. When groups of nodes engage in a distributed computation task,  partially processed data needs to be constantly exchanged between them. A prime example of this is the MapReduce setting \cite{Dean_Ghemawat_MapReduce,Li_Chen_Wang_MapReduce,Bi_Wigger_MapReduce,Bi_Cooperative_X,li2017fundamental,li2017scalable,ha2019wireless,yang2019data,xu2021new,han2021over,wang2022dynamic}. The resulting information flows resemble $X$ networks more than interference networks\cite{Bi_Cooperative_X,bi2024dof}, which suits the strengths of BIA schemes. Interestingly, computation networks such as MapReduce also introduce an aspect that is thus far relatively unexplored in the BIA context --- the presence of side-information. This is the question that motivates our work in this paper. How can BIA schemes --- based on blind transmitters and reconfigurable receive antennas --- optimally take advantage of the message and side-information structures of a MapReduce setting? In search for sharp fundamental limits, we will focus primarily on the DoF question, i.e., high SNR performance with idealized coherence models. The concerns of imperfect coherence and finite SNR performance are extremely important, but require a different approach that is beyond the scope of this work.

\subsection{A Motivating Example}
\begin{figure*}
\includegraphics{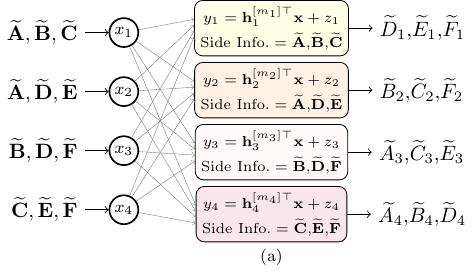}
\hspace{0.8cm}
\includegraphics{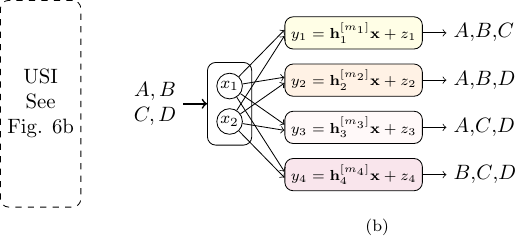}

\caption{(a) A MapReduce network with $\widetilde{\mathbf{A}} = (\widetilde{A}_3,\widetilde{A}_4), \widetilde{\mathbf{B}} = (\widetilde{B}_2,\widetilde{B}_4), \widetilde{\mathbf{C}} = (\widetilde{C}_2,\widetilde{C}_3), \widetilde{\mathbf{D}} = (\widetilde{D}_1,\widetilde{D}_4), \widetilde{\mathbf{E}} = (\widetilde{E}_1,\widetilde{E}_3), \widetilde{\mathbf{F}} = (\widetilde{F}_1,\widetilde{F}_2)$. (b) A corresponding vector BC setting with groupcast messages (BCGM). In both settings, receivers possess reconfigurable antennas and channel knowledge. In both settings no CSIT is assumed. Receivers in the MapReduce network also have side-information, while no side-information is available to the receivers in the BCGM setting.}
\label{fig:mapRedmot}
\end{figure*}

Similar to \cite{Li_Chen_Wang_MapReduce, Bi_Wigger_MapReduce, ha2019wireless}, let us consider MapReduce over a full-duplex wireless network with $K$ users. In particular, we focus on the communication problem that emerges in the Shuffle phase of MapReduce. There are $N_r=\binom{K}{r}$ independent files (super-messages), each available as common information to a distinct subset of $r$ users.  Each file is ``partitioned'' into $K-r$ independent subfiles (messages)\footnote{Actually in MapReduce, each file is ``computed'' into $K$ independent intermediate values (IVAs, what we call messages), through map functions, and these IVAs  are to be delivered, one each, to the $K$ users. However, since $r$ of the IVAs are already available to the $r$ target users who have the original file, only $K-r$ of the IVAs (messages) need to be delivered to the remaining $K-r$ users during the shuffle phase. We focus on these $K-r$ messages.  The  correspondence of our reduced model to the complete wireless MapReduce framework is explicitly illustrated via an example in Appendix \ref{app:mapRed}.} that are to be delivered, one each, to the remaining $K-r$ users.  A simple example is shown in Fig. \ref{fig:mapRedmot} with $K=4, r=2$, so that we have $4$ users and $\binom{4}{2}=6$ independent super-messages (files), labeled as  $\widetilde{\mathbf{A}}, \widetilde{\mathbf{B}}, \widetilde{\mathbf{C}}, \widetilde{\mathbf{D}}, \widetilde{\mathbf{E}}, \widetilde{\mathbf{F}}$ in the figure, each comprised of $K-r=4-2=2$ messages, to be delivered, one each, to the remaining two users. For example, super-message $\widetilde{\mathbf{A}} = (\widetilde{A}_3,\widetilde{A}_4)$ is already available to Users $1,2$ and the  message $\widetilde{A}_3$ is desired by User $3$ while $\widetilde{A}_4$ is desired by User $4$. The physical channel  model is similar to a $K$ user interference channel, but since each user is both a transmitter and a receiver under full-duplex operation,  the same initial set of messages is assumed to be initially available to Transmitter $k$ and Receiver $k$ (since they are co-located). Channel state information at the receivers is assumed to be perfect. No CSIT is assumed at the transmitters. While the transmitters and receivers are co-located, there are several reasons for this assumed disparity of CSI. Even with full-duplex operation it is not uncommon to use separate antennas  exclusively for transmitting and receiving, so the channel from Transmitter $k$ to Receiver $k'$ is not necessarily identical to the channel from Transmitter $k'$ to Receiver $k$. Thus a user may have CSIR but not CSIT. More importantly, however, we wish to avoid fragile schemes that rely strongly on CSIT, and one way to do so is by imposing the assumption of no CSIT on the channel model, which forces the DoF study towards robust solutions, in particular BIA schemes. To facilitate BIA, we assume that the receivers are equipped with reconfigurable antennas.  

\subsection{Overview of Contribution}
The absence of CSIT, the presence of reconfigurable antennas, and the availability of side-information are the main distinguishing features of our work compared to prior works in \cite{Li_Chen_Wang_MapReduce,Bi_Wigger_MapReduce,ha2019wireless, Gou_Wang_Jafar_BIA}.  In \cite{Li_Chen_Wang_MapReduce,Bi_Wigger_MapReduce} perfect CSIT is assumed, and no reconfigurable antennas are involved. In \cite{ha2019wireless}, imperfect CSIT is assumed and no reconfigurable antennas are involved. Prior works on BIA \cite{Gou_Wang_Jafar_BIA,menon2022BIA_Survey} do not explore the availability of side-information. Thus, the main challenge that we face is that \emph{reconfigurable antennas} and message \emph{side-information} have not previously been studied together, so a new approach is needed to determine how to  \emph{optimally combine} these two resources.

The approach we take in this work is to decompose the overall MapReduce optimization problem into distinctive stages, corresponding to smaller, more tractable problems that help us identify the essential ingredients of the  solution. 
\begin{itemize}
\item {\bf Stage 1: BCGM (broadcast with groupcast messages): Groupcast-optimal Intra-message Alignment through Reconfigurable Antennas.}\\ Instead of directly attacking the MapReduce problem (e.g.,  Fig. \ref{fig:mapRedmot}a), we start instead with a simpler problem (e.g., Fig. \ref{fig:mapRedmot}b), which is a broadcast channel (allowing full cooperation among transmitters) with a different message structure (groupcast,  each message has multiple desired destinations and each destination has multiple desired messages), and no side-information. Since  side-information is not available, the focus here is on the optimal use of reconfigurable antennas. Similar to the original BIA \cite{Gou_Wang_Jafar_BIA} coding scheme, the form of interference alignment needed here is {\bf intra}-message alignment, i.e., alignment occurs within symbols of the same message. However,  groupcast-optimal intra-message alignment  introduces novel elements at both the transmit and receive sides. At the transmitter side, the novelty is an MDS code structure as part of the precoding. At the receiver side, the novelty is a new algorithm for \emph{groupcast-optimal} antenna switching. Recall that for BIA schemes, the  alignment of messages is determined not only by the precoding at the transmitter but also by the antenna switching pattern at the receiver. Partially overlapping desired and undesired message sets among receivers (groupcast) impose non-trivial constraints on the switching patterns (Lemma \ref{lem:dimension_received} in this work) and require a new iterative algorithm (see Algorithm \ref{alg:pk}).

\item {\bf Stage 2: USI (unicast with side-information): Optimal Inter-Message Alignment through Side-Information.}\\
The BCGM setting by itself is quite far from MapReduce. For example, the relationship of the BCGM problem in Fig. \ref{fig:mapRedmot}b to its corresponding original MapReduce problem in Fig. \ref{fig:mapRedmot}a is not  obvious --- they have different message sets, and while BCGM does not involve any message side-information, MapReduce does. This connection is established through an intermediate stage, a unicast with side-information setting (USI), represented as a dashed box in Fig. \ref{fig:mapRedmot}, leaving the  cumbersome combinatorial details of this intermediate stage for later (see Fig. \ref{fig:mapRedICuni}b). The main novel aspect that emerges from this stage is the need for {\bf inter}-message alignment (i.e., alignment of symbols from different messages) to exploit side-information. Note that a \emph{digital} form of inter-message alignment (coded multicasting, by adding bits from different messages to create a composite message) that takes advantage of side-information, appears in the wireless MapReduce problem explored in \cite{ha2019wireless} (without reconfigurable antennas) that utilizes the MapReduce scheme of \cite{li2017fundamental}. Interestingly, this is not the DoF optimal inter-message alignment scheme (see Remark \ref{rmk:inter_alignment_only}). This is because a wireless setting allows more possibilities for inter-message alignment, not only by digital means, but also by analog means \cite{Katti_XOR}. This is important because unlike digital alignments that are limited to messages that originate at the same transmitter, analog alignments can be achieved even across independent messages that are distributed among different transmitters, by signal superposition that happens over the air. The feasible space of inter-message alignments further scales with the size of the network. The challenges of inter-message alignment in the USI stage include the understanding of \emph{which} messages should be \emph{optimally} aligned together, how inter-message alignment combines with intra-message alignment, and finally how the setting translates into the original MapReduce problem by a relabeling of messages.
\end{itemize}

This paper is organized as follows. We start with the BCGM setting in Section \ref{sec:form}. The BCGM problem formulation is presented in Section \ref{sec:pfBCGM}, the DoF result for BCGM appears as Theorem \ref{thm:BCGM} in Section \ref{sec:main}, illustrative examples are provided in Sections \ref{sec:ex1} and \ref{sec:ex2}, and a general coding scheme that performs intra-message alignment for BCGM is presented in Section \ref{sec:achievabilityBCGM} as proof of achievability for Theorem \ref{thm:BCGM}. Section \ref{sec:BCGMapplication} makes the connection from BCGM to the MapReduce setting via an intermediate USI setting. The MapReduce problem is formalized in Section \ref{subsec:MapReduce_Form}, and USI is introduced in Section \ref{subsec:USI}. The DoF of the USI setting are stated as Theorem \ref{thm:USI} in Section \ref{sec:DoFUSI}, with an example illustrating the idea of inter-message alignment that exploits the side information. The achievability of Theorem \ref{thm:USI} is proved in Section \ref{sec:achthmusi1}, with an example provided in Section \ref{sec:ex4}. Finally, the DoF result for MapReduce is obtained  as Corollary \ref{cor:MR} in Section \ref{sec:DoFmapreduce}. The converse proofs are provided in Appendix \ref{app:converse}, with the converse of Theorem \ref{thm:BCGM} appearing in Section \ref{app:converse_BCGM} and the converse of Theorem \ref{thm:USI} appearing in Section \ref{app:converse_USI}. Section \ref{sec:conclusion} concludes the paper.

{\it Notation:} For  integers $a, b$ where $a \leq b$, define $[a:b] = \{a, a+1, \cdots, b\}$. Also, let $[b] = [1:b]$ for $b \geq 1$. For any set $\mathcal{A} = \{a_1, a_2, \cdots, a_{|\mathcal{A}|}\} \subset \mathbb{Z}$, we have $a_i < a_{i+1}, \forall i \in [|\mathcal{A}|]$, i.e., any set of integers is an ordered set and we use $\mathcal{A}(i)$ to denote the $i^{th}$ element of $\mathcal{A}$, i.e., $\mathcal{A}(i) = a_i$. We say $\mathcal{A} < \mathcal{B}$ if $\mathcal{B} \subset \mathbb{Z}$, $|\mathcal{B}| = |\mathcal{A}|$ and $\exists i \in [|\mathcal{A}|]$ s.t. $\forall j \in [i-1], \mathcal{A}(j) = \mathcal{B}(j)$ while $\mathcal{A}(i) < \mathcal{B}(i)$. For example, $\mathcal{A} = \{1,2,5\} < \mathcal{B} = \{1,3,4\}$ since $\mathcal{A}(1) = \mathcal{B}(1)$ and $\mathcal{A}(2) < \mathcal{B}(2)$. Let $\binom{[K]}{G} = \{\mathcal{S}_1, \mathcal{S}_2, \cdots, \mathcal{S}_{\binom{K}{G}}\}$ be the ordered set that consists of all the $\binom{K}{G}$ subsets with cardinality $G$ of $[K]$, i.e., $\forall i \in [\scalebox{0.80}{$\binom{K}{G}$}], \mathcal{S}_i \subset [K], |\mathcal{S}_i| = G$, and $\mathcal{S}_1 < \mathcal{S}_2 < \cdots < \mathcal{S}_{\scalebox{0.80}{$\binom{K}{G}$}}$. 
Let \scalebox{0.8}{$(\cdot)^{\top}$} be the transpose operation. For a matrix $\mathbf{A}$, let $\mathbf{A}_{\mathcal{A},:}$ and $\mathbf{A}_{:,\mathcal{B}}$ be the sub-matrices of $\mathbf{A}$ that consist of the rows and columns, whose indices are in $\mathcal{A}$ and $\mathcal{B}$ respectively, of $\mathbf{A}$. $\mbox{rk}(\mathbf{A})$ returns the rank of matrix $\mathbf{A}$, and $\mbox{col-span}(\mathbf{A})$ denotes the subspace spanned by the columns of $\mathbf{A}$. For matrices $\mathbf{A}, \mathbf{B}$ with same number of rows, let $\mathbf{A} \cup \mathbf{B}$ denote the matrix generated by putting columns of $\mathbf{B}$ right to $\mathbf{A}$. For a vector $\mathbf{v}$, let $\mathbf{v}_{\mathcal{A}}$ denote the sub-vector of $\mathbf{v}$ with entries of $\mathbf{v}$ whose indices are in $\mathcal{A}$. Let $\bzero_{r\times c}$ (resp. $\mathbf{1}_{r\times c}$) be the $r \times c$ all zero (resp. one) matrix and $r, c$ will be clear according to the context if not explicitly specified. We use $\mds_{r \times c} \subset \mathbb{C}^{r\times c}, r < c $ to denote the set of all $r \times c$ matrices whose submatrix of any $r$ columns is invertible. Let $\bI_{M}$ denote the $M\times M$ identity matrix and let $\mathbf{e}_M^{i}$ be the $i^{th}$ row of $\bI_M$. Throughout the paper, $\otimes$ stands for the Kronecker product. For any integer $a$, we define $\Mod{a}{M} \triangleq \tilde{a} \in [1:M]$ s.t. $(a - \tilde{a})~\mbox{mod}~M = 0$. This definition is similar to that of modulo operation, just that when the input is multiples of $M$, the output is $M$ instead of $0$. For a vector $\mathbf{a}$, $\Mod{\mathbf{a}}{M}$ means applying the function element-wise to all entries of $\mathbf{a}$.

\section{Broadcast with Groupcast Messages (BCGM): Intra-message Alignment}\label{sec:form}
\subsection{Problem Formulation}\label{sec:pfBCGM}
A BCGM setting is parameterized by a tuple $(K,G,M)$. There is a transmitter (Tx) that is equipped with $M$ conventional antennas, and $K$ receivers (Rx's) each of which is equipped with a single \emph{reconfigurable} antenna that is capable of switching among $M$ independent modes. Over the $t^{th}$ channel use, $t\in\mathbb{N}$, the scalar signal received by Rx-$k$ is expressed as,
\begin{align}
y_k(t)&=\mathbf{h}_k^{[m_k(t)]}(t) \mathbf{x}(t)+z_k(t), &&\forall k\in[K],
\end{align}
where $\mathbf{h}_k^{[m_k(t)]}(t)\in\mathbb{C}^{1\times M}$ is the $1\times M$ channel vector corresponding to receive-antenna mode $m_k(t) \in [M]$ chosen by Rx-$k$ at time $t$, $\mathbf{x}(t)=[x_1(t)~~\cdots~~x_{M}(t)]^\top$ is the $M\times 1$ vector of  symbols sent from the $M$ transmit-antennas, and $z_k(t)$ is the zero-mean unit variance circularly symmetric complex AWGN at Rx-$k$ at time $t$. The transmit power for each antenna is limited to $P$, i.e., $E[|x_{m}(n)|^2]\leq P$ for all $m\in[M]$.  An i.i.d. block fading model is assumed, with coherence time $\mathsf{T}_c$.  The channel vector $\mathbf{h}_k^{[m]}(t)$ associated with the $m^{th}$  mode of Rx-$k$, $m\in[M], k\in[K]$, is drawn i.i.d. uniform according to $\mathcal{CN}(0,1)$ (Rayleigh fading)\footnote{The results generalize to any continuous distribution.} at the beginning of each coherence block, and remains constant for the $\mathsf{T}_c$ consecutive channel-uses  corresponding to that block, after which it changes to another i.i.d. realization for the next block. Mathematically,
\begin{align}
    \mathbf{h}_k^{[m]}(t)=\mathbf{h}_k^{[m]}\left(\left\lceil \frac{t}{\mathsf{T}_c}\right\rceil\mathsf{T}_c\right),&&\forall m\in[M], \forall t\in\mathbb{N}.
\end{align}
Note that an Rx can switch among its $M$ modes in any chosen switching pattern, so even within a coherence block, the channel seen by an Rx will vary according to the selected switching pattern. We will assume that $\mathsf{T}_c\gg K$.  Perfect channel state information at the receivers (CSIR) is assumed. No channel state information at the transmitter (CSIT) is assumed beyond the channel distribution. Next we specify the messages to be transmitted, and define the group size $G$.

There are $N_g = \binom{K}{G}$ independent messages $W_1, W_2, \cdots, W_{N_g}$ each of which is desired by a distinct group of $G$ Rx's out of the $K$ Rxs. The set of $N_g$ groups of Rx's is the set of all size-$G$ subsets of $[K]$,
\begin{align}
    \binom{[K]}{G} = \{\mathcal{S}_1, \mathcal{S}_2, \cdots, \mathcal{S}_{N_g}\}.\label{eq:groupRx}
\end{align}
Then $\forall n \in [N_g]$, message $W_n$ will be groupcast to the $G$ Rx's in group $\mathcal{S}_n$. For any $k \in [K]$, let $\mathcal{V}_k$ denote the indices of all the messages desired by Rx-$k$, i.e.,
\begin{align}
    \mathcal{V}_k = \left\{n \mid  n \in [N_g] \mbox{ s.t. } k \in \mathcal{S}_{n} \right\}.\label{eq:idxDeRxk}
\end{align}
The set of messages desired by Rx-$k, k \in [K]$ is defined as 
\begin{align}
    \de_k^{\gc} = \{W_{n} \mid  n \in \mathcal{V}_k\}.
\end{align}
Note that the number of messages desired by any Rx-$k, k \in [K]$, or equivalently, the number of cardinality-$G$ subsets of $[K]$ that contains $k$, is
\begin{align}
    \nu_g = |\mathcal{V}_k| = \binom{K-1}{G-1} = GN_g/K, && \forall k \in [K].
\end{align}

\begin{figure}
\centering
\begin{tikzpicture}
\def\dx{1.75}
\def\dy{0.4}
\foreach  \i in {1,2}{
	\node (T\i) at (0,{-(\i-1.5)*1.5*\dy}) [circle, draw,  inner sep = 0.03cm] {\footnotesize $x_\i$};
}

\node (Tx) at (0,0) [rectangle, rounded corners, draw, minimum width=0.75cm, minimum height=1.4cm] {};

\node (W) [left=0.5cm of Tx, align=center] {$A, B, C$\\$D,E,F$};
\draw [thick, ->] (W)--(Tx);

\foreach  \i/\j/\u/\c in {1/\mathbf{h}_1/A{,}B{,}C/yellow,2/\mathbf{h}_2/A{,}D{,}E/orange,3/\mathbf{h}_3/B{,}D{,}F/pink, 4/\mathbf{h}_4/C{,}E{,}F/purple}{
	\node (R\i) at (1.5*\dx,{-(\i-2.5)*2*\dy}) [rectangle, rounded corners, draw,  inner sep = 0.1cm, align=center, fill=\c!10] {\scriptsize $y_\i=\j^{[m_\i]} \mathbf{x}+z_\i$};
	\node (W{\i}hat) [right=0.5cm of R\i] {$\u$};
\draw [ ->] (R\i)--(W{\i}hat);

}

\foreach \i in {1,2}{
	\foreach \j in {1,2,3,4}{
		\draw [->] (T\i)--($(R\j.west)+(0,{0.1*(1.5-\i)})$);
		}
}

\node at (2.5,-2.5)[align=center]{\footnotesize $\mathcal{S}_1 = \{1,2\}, \mathcal{S}_2 = \{1,3\}, \mathcal{S}_3 = \{1,4\}$\\ \footnotesize $\mathcal{S}_4 = \{2,3\}, \mathcal{S}_5 = \{2,4\}, \mathcal{S}_6 = \{3,4\}$\\ \footnotesize $W_1 = A, W_2 = B, W_3 = C, W_4 = D, W_5 = E, W_6 = F$};
\end{tikzpicture}
\caption{$(K=4,G=2,M=2)$ BCGM Example.}
\label{fig:BCGM_K4G2M2}
\end{figure}

Fig. \ref{fig:mapRedmot}b shows a simple $(K=4,G=3,M=2)$ setting with $N_g = \binom{4}{3} = 4$ messages, labeled as $A,B,C,D$ for convenience. Each message is desired by $G=3$ Rx's, and each Rx desires $\nu_g=3$ messages.

For any message $W_{n}$, let $|W_{n}(P)|$ denote its alphabet size. For codewords spanning $\mathsf{T}$ channel uses, the rate associated with message $W_{n}$ is $R_{n}^{\gc}(P) \triangleq \frac{\log|W_{n}(P)|}{\mathsf{T}}$. A rate tuple 
\begin{align}
    \mathbf{R}^{\gc} = (R_{1}^{\gc}(P), R_{2}^{\gc}(P), \cdots, R_{N_g}^{\gc}(P))
\end{align}  
is achievable if there exists a sequence (indexed by $\mathsf{T}$) of coding schemes with the specified rates, such that the probability of error $P_e\rightarrow 0$ as $\mathsf{T}\rightarrow\infty$ for the decoding of every message at each of its intended receivers. The capacity region $\mathcal{C}^{\gc}(P)$ is the closure of the set of all achievable rate tuples. A DoF tuple 
\begin{align}
    &\mathbf{d}^{\gc} = (d_{1}^{\gc}, d_{2}^{\gc}, \cdots, d_{N_g}^{\gc}) \notag\\
    \intertext{ is achievable if $\forall n \in [N_g]$,} 
    &\exists \mathbf{R}^{\gc} \in \mathcal{C}^{\gc}(P) \mbox{ s.t. } d_{n}^{\gc} = \lim\limits_{P\rightarrow\infty}\frac{R_{n}^{\gc}(P)}{\log(P)}.
\end{align} 
The DoF region $\mathcal{D}^{\gc}$ is defined as the closure of the set of all achievable DoF tuples. A sum-DoF value $d_{\Sigma}^{\gc}$ is said to be achievable if there exists $\mathbf{d}^{\gc} \in \mathcal{D}^{\gc}$ s.t. $d_{\Sigma}^{\gc} = \sum_{n \in [N_g]} d_{n}^{\gc}$ and the sum-DoF of a BCGM setting $d_{\Sigma}^{\gc,*}$ is the largest achievable sum-DoF.

\subsection{Result: Sum-DoF of the BCGM Setting}\label{sec:main}
The sum-DoF value of the BCGM setting is characterized in the following theorem.
\begin{theorem}\label{thm:BCGM}
The sum-DoF of the $(K,G,M)$ BCGM setting is, 
\begin{align}
d_{\Sigma}^{\gc,*} = \frac{N_gM}{(M-1)\nu_g + N_g}.
\end{align}
\end{theorem}
The achievability of Theorem \ref{thm:BCGM} will be proved in the remainder of this section, starting with a few illustrative examples to introduce the main ideas of the construction. The converse is deferred to Appendix \ref{app:converse_BCGM}.

In the problem formulation, we set the number of transmitter antennas $M_{\mbox{\tiny Tx}}$ equal to the number of modes $M_{\mbox{\tiny Rx}}$ each reconfigurable receiver antenna can switch among ($M_{\mbox{\tiny Tx}} = M_{\mbox{\tiny Rx}} = M$). However, let us note that the  result extends to those settings where $M_{\mbox{\tiny Tx}} \neq M_{\mbox{\tiny Rx}}$, by setting $M = \min\left(M_{\mbox{\tiny Tx}}, M_{\mbox{\tiny Rx}}\right)$ in Theorem \ref{thm:BCGM}. Let us denote by $(K,G,M_{\mbox{\tiny Tx}}, M_{\mbox{\tiny Rx}})$ BCGM, a BCGM setting with $K$ users, groupcast size $G$, number of transmitter antennas $M_{\mbox{\tiny Tx}}$, and number of receiver antenna modes $M_{\mbox{\tiny Rx}}$. Then we have the following corollary.
\begin{corollary}\label{cor:BCGM}
The sum-DoF of the $(K,G,M_{\mbox{\tiny Tx}}, M_{\mbox{\tiny Rx}})$ BCGM setting is, 
\begin{align}
    d_{\Sigma}^{\gc,*} = \frac{N_g\min\left(M_{\mbox{\tiny Tx}}, M_{\mbox{\tiny Rx}}\right)}{\left(\min\left(M_{\mbox{\tiny Tx}}, M_{\mbox{\tiny Rx}}\right)-1\right)\nu_g + N_g}.
\end{align}
\end{corollary}

The achievability is quite straightforward. We let the transmitter only use $\min\left(M_{\mbox{\tiny Tx}}, M_{\mbox{\tiny Rx}}\right)$ transmit antennas, we let the receivers switch among the same number of modes, and utilize the scheme for $(K,G,M=\min\left(M_{\mbox{\tiny Tx}}, M_{\mbox{\tiny Rx}}\right))$ BCGM setting. For the converse argument, please see Remark \ref{rmk:BCGM_Converse_MTx_MRx}. In a nutshell, the converse in Appendix \ref{app:converse_BCGM} is only based on the number of receiver antennas' modes so it is also a valid upper bound to the sum-DoF by replacing $M$ with $M_{\mbox{\tiny Rx}}$ when $M_{\mbox{\tiny Tx}} \geq M_{\mbox{\tiny Rx}}$. The converse for the case $M_{\mbox{\tiny Tx}} < M_{\mbox{\tiny Rx}}$ follows from a straightforward argument that {$M_{\mbox{\tiny Tx}}$ ``copies'' of any receiver can invert the channel matrix}.

\begin{remark}
The DoF value in Theorem \ref{thm:BCGM} matches that in \cite[Theorem 2]{Hachem_Niesen_Diggavi_Multicast_X}. Despite this similarity of expressions, it is difficult to make a direct connection between the two results because the settings are quite different. Indeed, neither the converse nor the achievability in either setting follows from the other. For example, whereas  \cite[Theorem 2]{Hachem_Niesen_Diggavi_Multicast_X} relies strongly on perfect CSIT for asymptotic interference alignment, while assuming no reconfigurable antennas, our setting in Theorem \ref{thm:BCGM} assumes no CSIT, and relies on reconfigurable antennas for non-asymptotic interference alignment. Nevertheless, the matching expressions are noteworthy because they continue a trend of similar observations in \cite{Gou_Wang_Jafar_BIA}, where also it is noted that the DoF of a MISO BC with no CSIT and reconfigurable antennas match that of a corresponding $X$ network with perfect CSIT and no reconfigurable antennas.
\end{remark}
\subsection{Example 1: BIA Scheme for the $(K=4,G=3,M=2)$ Setting shown in Fig. \ref{fig:mapRedmot}b}\label{sec:ex1}
For this simple toy example, the BIA precoding scheme operates over $7$ channel uses corresponding to the $7$ columns shown in Fig. \ref{fig:BCGM_K4G3M2_precoding}. The first $2$ rows show the  signals transmitted from the $2$ transmit antennas, and the last $4$ rows show the  signals received by each of the $4$ users. Noise is omitted for simplicity. Blue color indicates that the reconfigurable antenna at the Rx is in mode-$1$, while red indicates mode-$2$.
\begin{figure*}[htbp]
\begin{align*}
\arraycolsep=1.4pt\def\arraystretch{1.75}
\scalebox{0.9}{$
\begin{array}{r|c|c|c|c|c|c|c|l}
\multicolumn{1}{c}{}&\multicolumn{1}{c}{t=1}&\multicolumn{1}{c}{t=2}&\multicolumn{1}{c}{t=3}&\multicolumn{1}{c}{t=4}&\multicolumn{1}{c}{t=5}&\multicolumn{1}{c}{t=6}&\multicolumn{1}{c}{t=7}\\\cline{2-8}
\mbox{Tx Antenna $1$: }x_1&\bm{\lambda}_1 [A^1~B^1~C^1~D^1]^\top &\bm{\lambda}_2 [A^1~B^1~C^1~D^1]^\top&\bm{\lambda}_3 [A^1~B^1~C^1~D^1]^\top&A^1&B^1&C^1&D^1\\\cline{2-8}
\mbox{Tx Antenna $2$: }x_2&\bm{\lambda}_1[A^2~B^2~C^2~D^2]^\top&\bm{\lambda}_2[A^2~B^2~C^2~D^2]^\top&\bm{\lambda}_3[A^2~B^2~C^2~D^2]^\top&A^2&B^2&C^2&D^2\\\cline{2-8}
\mbox{Rx-$1$: }y_1&\cellcolor{blue!10}\bm{\lambda}_1[a^{1,[1]}~\cdots~d^{1,[1]}]^\top&\cellcolor{blue!10}\bm{\lambda}_2[a^{1,[1]}~\cdots~d^{1,[1]}]^\top&\cellcolor{blue!10}\bm{\lambda}_3[a^{1,[1]}~\cdots~d^{1,[1]}]^\top&\cellcolor{red!10}a^{1,[2]}&\cellcolor{red!10}b^{1,[2]}&\cellcolor{red!10}c^{1,[2]}&\cellcolor{blue!10}d^{1,[1]}&\rightarrow A,B,C\\\cline{2-8}
\mbox{Rx-$2$: }y_2&\cellcolor{blue!10}\bm{\lambda}_1[a^{2,[1]}~\cdots~d^{2,[1]}]^\top&\cellcolor{blue!10}\bm{\lambda}_2[a^{2,[1]}~\cdots~d^{2,[1]}]^\top&\cellcolor{blue!10}\bm{\lambda}_3[a^{2,[1]}~\cdots~d^{2,[1]}]^\top&\cellcolor{red!10}a^{2,[2]}&\cellcolor{red!10}b^{2,[2]}&\cellcolor{blue!10}c^{2,[1]}&\cellcolor{red!10}d^{2,[2]}&\rightarrow A,B,D\\\cline{2-8}
\mbox{Rx-$3$: }y_3&\cellcolor{blue!10}\bm{\lambda}_1[a^{3,[1]}~\cdots~d^{3,[1]}]^\top&\cellcolor{blue!10}\bm{\lambda}_2[a^{3,[1]}~\cdots~d^{3,[1]}]^\top&\cellcolor{blue!10}\bm{\lambda}_3[a^{3,[1]}~\cdots~d^{3,[1]}]^\top&\cellcolor{red!10}a^{3,[2]}&\cellcolor{blue!10}b^{3,[1]}&\cellcolor{red!10}c^{3,[2]}&\cellcolor{red!10}d^{3,[2]}&\rightarrow A,C,D\\\cline{2-8}
\mbox{Rx-$4$: }y_4&\cellcolor{blue!10}\bm{\lambda}_1[a^{4,[1]}~\cdots~d^{4,[1]}]^\top&\cellcolor{blue!10}\bm{\lambda}_2[a^{4,[1]}~\cdots~d^{4,[1]}]^\top&\cellcolor{blue!10}\bm{\lambda}_3[a^{4,[1]}~\cdots~d^{4,[1]}]^\top&\cellcolor{blue!10}a^{4,[1]}&\cellcolor{red!10}b^{4,[2]}&\cellcolor{red!10}c^{4,[2]}&\cellcolor{red!10}d^{4,[2]}&\rightarrow B,C,D\\\cline{2-8}
\end{array}$}
\end{align*}
\caption{The Precoding Scheme for $(4,3,2)$ BCGM setting.}
\label{fig:BCGM_K4G3M2_precoding}
\end{figure*}
The vectors $\bm{\lambda_1},\bm{\lambda_2},\bm{\lambda_3}$ are generic $1\times 4$ vectors, say the rows of any $3\times 4$ matrix, such that all of its $3\times 3$ submatrices are full rank (MDS). For ease of exposition let us choose a Vandermonde structure,
\begin{align}
    \bm{\Lambda} \triangleq 
    \begin{bmatrix}
        \bm{\lambda_1}\\
        \bm{\lambda_2}\\
        \bm{\lambda_3}
    \end{bmatrix}
    =
    \begin{bmatrix}
        1 & 1 & 1 & 1\\
        1 & 2 & 3 & 4\\
        1 & 4 & 9 & 16
    \end{bmatrix}.
\end{align}
Each message is comprised of two symbols, e.g., $A=(A^1,A^2)$. Consider time slot $t=2$. As shown in Fig. \ref{fig:BCGM_K4G3M2_precoding}, Tx Antenna $1$ sends $x_1$ which is a generic linear combination of the symbols $A^1, B^1,C^1,D^1$, with the coefficients specified by the vector $\bm{\lambda_2}$, i.e., $x_1(2)=A^1+2B^1+3C^1+4D^1$. Simultaneously, Tx Antenna $2$ sends the same linear combination of the symbols $A^2,B^2,C^2,D^2$, i.e., $x_2(2)=A^2+2B^2+3C^2+4D^2$. For Rx-$i$, let 
\begin{align}
    a^{i,[m_i]}\triangleq\mathbf{h}_i^{[m_i]}(A^1, A^2)^\top
\end{align}
denote the dot product of the $A=(A^1,A^2)$ vector and the $1\times 2$ channel vector $\mathbf{h}_i^{[m_i]}$ that is seen by Rx-$i$, who has chosen mode $m_i$. Similarly, define $b^{i,[m_i]}, c^{i,[m_i]}, d^{i,[m_i]}$ as the dot products of the same channel vector with $B,C,D$ vectors respectively. Then  in time slot $t=2$, where according to Fig. \ref{fig:BCGM_K4G3M2_precoding} all receivers have chosen mode $1$ (blue), we note that the signal received by Rx-$i$ is  $y_i(2)=\bm{\lambda_2}(a^{i,[1]}, b^{i,[1]}, c^{i,[1]}, d^{i,[1]})^\top=a^{i,[1]}+2b^{i,[1]}+3c^{i,[1]}+4d^{i,[1]}$. Recall that noise is omitted for this intuitive explanation. Consider Rx-$1$, who wants messages $A,B,C$. It will receive its interfering symbol $d^{1,[1]}$ separately again in time slot $t=7$ according to the scheme specified in Fig. \ref{fig:BCGM_K4G3M2_precoding}, so it can cancel the interference from all first $3$ time slots, leaving it with $3$ linear equations in $a^{1,[1]}, b^{1,[1]},c^{1,[1]}$. Since $\bm{\lambda_i}$ are fixed parameters of the coding scheme known to everyone, Rx-$1$ is able to solve $a^{1,[1]}, b^{1,[1]},c^{1,[1]}$. Also note that since all receivers have full  knowledge of their own received channels, it follows that from the two generic linear combinations $a^{1,[1]}$ (solved), $a^{1,[2]}$ (received at time slot $4$), the desired message symbols $A^1, A^2$ can be recovered by Rx-$1$ almost surely (within bounded noise distortion that is inconsequential for DoF). Recovery of all desired symbols by other receivers can be similarly verified in Fig. \ref{fig:BCGM_K4G3M2_precoding}. The BIA principle is evident in the last $4$ columns of  Fig. \ref{fig:BCGM_K4G3M2_precoding}. During these time slots, each receiver switches to mode-$2$ when  desired symbols are being transmitted, so that it receives a  linear combination of the desired symbols that is different from what was received in mode-$1$, i.e., new information, and switches back to mode-$1$ when undesired symbols (interference) is being transmitted, so that it can recover the previously seen interference separately and cancel it from other time slots. The restriction of interference symbols to the same antenna mode constitutes interference alignment, giving the scheme its DoF advantage. The need of the ${\bm \lambda_i}$ parameters is a distinctive and important aspect of the scheme.

\subsection{Example 2: BIA Scheme for a $(K=3,G=2,M=3)$ BCGM}\label{sec:ex2}
While the preceding example is appealing for its simplicity, the general case poses a few additional challenges to receiver antennas' switching pattern design that are not encountered in the previous example. To facilitate the understanding of the general scheme, let us introduce one more example, chosen to be as simple as possible, while allowing us to illustrate the remaining key considerations, before we proceed to the general proof of achievability.

\subsubsection{Conventional BIA solution for $(K=3,G=1,M=3)$}
It will be useful to start with the conventional BIA solution from \cite[Section III.C.3)]{Gou_Wang_Jafar_BIA} corresponding to $(K=3,G=1,M=3)$ as a baseline upon which we can build the solution to the groupcast setting. This setting is shown below.
\begin{figure}[H]
\centering
\begin{tikzpicture}
\def\dx{1.75}
\def\dy{0.4}
\foreach  \i in {1,2,3}{
	\node (T\i) at (0,{-(\i-2)*1.5*\dy}) [circle, draw,  inner sep = 0.03cm] {\footnotesize $x_\i$};
}

\node (Tx) at (0,0) [rectangle, rounded corners, draw, minimum width=0.75cm, minimum height=2.1cm] {};

\node (W) [left=0.5cm of Tx, align=center] {$A, B, C$};
\draw [thick, ->] (W)--(Tx);

\foreach  \i/\j/\u/\c in {1/\mathbf{h}_1/A/yellow,2/\mathbf{h}_2/B/orange,3/\mathbf{h}_3/C/pink}{
	\node (R\i) at (1.5*\dx,{-(\i-2)*2*\dy}) [rectangle, rounded corners, draw,  inner sep = 0.1cm, align=center, fill=\c!10] {\scriptsize $y_\i=\j^{[m_\i]} \mathbf{x}+z_\i$};
	\node (W{\i}hat) [right=0.5cm of R\i] {$\u$};
\draw [ ->] (R\i)--(W{\i}hat);

}

\foreach \i in {1,2,3}{
	\foreach \j in {1,2,3}{
		\draw [->] (T\i)--($(R\j.west)+(0,{0.1*(1.5-\i)})$);
		}
}
\end{tikzpicture}
\caption{$(K=3,G=1, M=3)$ Example.}
\label{fig:subsetcast_N3_nu1_M3}
\end{figure}
Following the construction in \cite{Gou_Wang_Jafar_BIA}, let each message be comprised of $L= M\times(M-1)^{N_g - 1} = 3\times 4 = 12$ streams of symbols. Corresponding to message $A$, let `$\mathbf{a}$' denote its $12$ symbol block as follows,
\begin{align}
    &A \rightarrow \mathbf{a}= \notag\\
    &[\underbrace{A^1~~A^2~~A^3}_{\mathbf{a}^{1\top}} \mid \underbrace{A^4~~A^5~~A^6}_{\mathbf{a}^{2\top}} \mid \underbrace{A^7~~A^8~~A^9}_{\mathbf{a}^{3\top}} \mid
    \underbrace{A^{10}~~A^{11}~~A^{12}}_{\mathbf{a}^{4\top}}]^\top
\end{align} 
i.e., the $12$ symbols are divided into $4$ blocks, each with length $M=3$. Message $B, C$ are encoded into $\mathbf{b}, \mathbf{c}$ and diveded into $4$ blocks similarly. Also, for any $k \in [3], l \in [4], m \in [3]$, let us define
\begin{align}
    a^{k,l,[m]} = \mathbf{h}_k^{[m]} \mathbf{a}^{l}, && b^{k,l,[m]} = \mathbf{h}_k^{[m]} \mathbf{b}^{l}, && c^{k,l,[m]} = \mathbf{h}_k^{[m]} \mathbf{c}^{l}
\end{align}
as the mode-$m$ linear combination of the $l^{th}$ block of $A,B,C$ at Rx-$k$ respectively.

The scheme in \cite{Gou_Wang_Jafar_BIA} requires $20$ time slots for $12$ streams of symbols of all messages to be transmitted to all the destinations. Over the $20$ time slots, the transmitter sends

\begin{align}
&\scalebox{0.8}{$\mathbf{X} = 
\begin{bmatrix}
    \mathbf{x}(1) & \mathbf{x}(2) & \cdots & \mathbf{x}(20)
\end{bmatrix}^\top=$}\notag\\
&\scalebox{0.8}{$
    \underbrace{\begin{bmatrix}
        \color{blue}\bI & \color{blue}\bzero & \color{blue}\bzero & \color{blue}\bzero\\
        \color{blue}\bI & \color{blue}\bzero & \color{blue}\bzero & \color{blue}\bzero\\
        \color{blue}\bzero & \color{blue}\bI & \color{blue}\bzero & \color{blue}\bzero\\
        \color{blue}\bzero & \color{blue}\bI & \color{blue}\bzero & \color{blue}\bzero\\
        \color{blue}\bzero & \color{blue}\bzero & \color{blue}\bI & \color{blue}\bzero\\
        \color{blue}\bzero & \color{blue}\bzero & \color{blue}\bI & \color{blue}\bzero\\
        \color{blue}\bzero & \color{blue}\bzero & \color{blue}\bzero & \color{blue}\bI\\
        \color{blue}\bzero & \color{blue}\bzero & \color{blue}\bzero & \color{blue}\bI\\
        \bI & \bzero & \bzero & \bzero\\
        \bzero & \bI & \bzero & \bzero\\
        \bzero & \bzero & \bI & \bzero\\
        \bzero & \bzero & \bzero & \bI\\
        \bzero & \bzero & \bzero & \bzero\\
        \bzero & \bzero & \bzero & \bzero\\
        \bzero & \bzero & \bzero & \bzero\\
        \bzero & \bzero & \bzero & \bzero\\
        \bzero & \bzero & \bzero & \bzero\\
        \bzero & \bzero & \bzero & \bzero\\
        \bzero & \bzero & \bzero & \bzero\\
        \bzero & \bzero & \bzero & \bzero
    \end{bmatrix}}_{\mathbb{V}_A}
    \hspace{-0.1cm}
    \underbrace{
    \begin{bmatrix}
        A^1\\
        A^2\\
        A^3\\
        \hline
        A^4\\
        A^5\\
        A^6\\
        \hline
        A^7\\
        A^8\\
        A^9\\
        \hline
        A^{10}\\
        A^{11}\\
        A^{12}\\
    \end{bmatrix}}_{\mathbf{a}}
    +
    \underbrace{\begin{bmatrix}
        \color{blue}\bI & \color{blue}\bzero & \color{blue}\bzero & \color{blue}\bzero\\
        \color{blue}\bzero & \color{blue}\bI & \color{blue}\bzero & \color{blue}\bzero\\
        \color{blue}\bI & \color{blue}\bzero & \color{blue}\bzero & \color{blue}\bzero\\
        \color{blue}\bzero & \color{blue}\bI & \color{blue}\bzero & \color{blue}\bzero\\
        \color{blue}\bzero & \color{blue}\bzero & \color{blue}\bI & \color{blue}\bzero\\
        \color{blue}\bzero & \color{blue}\bzero & \color{blue}\bzero & \color{blue}\bI\\
        \color{blue}\bzero & \color{blue}\bzero & \color{blue}\bI & \color{blue}\bzero\\
        \color{blue}\bzero & \color{blue}\bzero & \color{blue}\bzero & \color{blue}\bI\\
        \bzero & \bzero & \bzero & \bzero\\
        \bzero & \bzero & \bzero & \bzero\\
        \bzero & \bzero & \bzero & \bzero\\
        \bzero & \bzero & \bzero & \bzero\\
        \bI & \bzero & \bzero & \bzero\\
        \bzero & \bI & \bzero & \bzero\\
        \bzero & \bzero & \bI & \bzero\\
        \bzero & \bzero & \bzero & \bI\\
        \bzero & \bzero & \bzero & \bzero\\
        \bzero & \bzero & \bzero & \bzero\\
        \bzero & \bzero & \bzero & \bzero\\
        \bzero & \bzero & \bzero & \bzero
    \end{bmatrix}}_{\mathbb{V}_B}
    \hspace{-0.1cm}
    \underbrace{
    \begin{bmatrix}
        B^1\\
        B^2\\
        B^3\\
        \hline
        B^4\\
        B^5\\
        B^6\\
        \hline
        B^7\\
        B^8\\
        B^9\\
        \hline
        B^{10}\\
        B^{11}\\
        B^{12}\\
    \end{bmatrix}}_{\mathbf{b}}
    +
    \underbrace{\begin{bmatrix}
        \color{blue}\bI & \color{blue}\bzero & \color{blue}\bzero & \color{blue}\bzero\\
        \color{blue}\bzero & \color{blue}\bI & \color{blue}\bzero & \color{blue}\bzero\\
        \color{blue}\bzero & \color{blue}\bzero & \color{blue}\bI & \color{blue}\bzero\\
        \color{blue}\bzero & \color{blue}\bzero & \color{blue}\bzero & \color{blue}\bI\\
        \color{blue}\bI & \color{blue}\bzero & \color{blue}\bzero & \color{blue}\bzero\\
        \color{blue}\bzero & \color{blue}\bI & \color{blue}\bzero & \color{blue}\bzero\\
        \color{blue}\bzero & \color{blue}\bzero & \color{blue}\bI & \color{blue}\bzero\\
        \color{blue}\bzero & \color{blue}\bzero & \color{blue}\bzero & \color{blue}\bI\\
        \bzero & \bzero & \bzero & \bzero\\
        \bzero & \bzero & \bzero & \bzero\\
        \bzero & \bzero & \bzero & \bzero\\
        \bzero & \bzero & \bzero & \bzero\\
        \bzero & \bzero & \bzero & \bzero\\
        \bzero & \bzero & \bzero & \bzero\\
        \bzero & \bzero & \bzero & \bzero\\
        \bzero & \bzero & \bzero & \bzero\\
        \bI & \bzero & \bzero & \bzero\\
        \bzero & \bI & \bzero & \bzero\\
        \bzero & \bzero & \bI & \bzero\\
        \bzero & \bzero & \bzero & \bI
    \end{bmatrix}}_{\mathbb{V}_C}
    \hspace{-0.1cm}
    \underbrace{
    \begin{bmatrix}
        C^1\\
        C^2\\
        C^3\\
        \hline
        C^4\\
        C^5\\
        C^6\\
        \hline
        C^7\\
        C^8\\
        C^9\\
        \hline
        C^{10}\\
        C^{11}\\
        C^{12}\\
    \end{bmatrix}}_{\mathbf{c}}$}\label{eq:BCUM_X}
\end{align}

Note that in $\mathbb{V}_A, \mathbb{V}_B, \mathbb{V}_C$, the symbols $\bI$ are $M\times M = 3\times 3$ identity matrices and the symbols $\bzero$ are all-zero matrices of the same size. Within the first $8$ time slots (marked with blue), in each channel use, one block each of $A,B,C$  is chosen, added together, and sent over the $3$ antennas of the transmitter. Within the last $12$ time slots, however, a block of only one message is sent at a time. Let us label the top part of $\mathbb{V}_A, \mathbb{V}_B, \mathbb{V}_C$, that is marked with blue, as $\mathbf{T}_A, \mathbf{T}_B, \mathbf{T}_C$ respectively, and the bottom parts as $\mathbf{U}_A, \mathbf{U}_B, \mathbf{U}_C$ respectively. Thus, during the first $8$ time slots, the transmitter broadcasts 
\begin{align}
    \mathbf{T}_A \mathbf{a} + \mathbf{T}_B \mathbf{b} + \mathbf{T}_C \mathbf{c}.
\end{align}
Let the swtiching pattern of Rx-$1$ be $\mathbf{m}_1 = [m_1(1)~~ m_1(2)\cdots m_1(20)]$. After the first $8$ time slots,  Rx-$k$ obtains, 
\begin{align}
    &\begin{bmatrix}
        y_1(1)\\
        y_1(2)\\
        y_1(3)\\
        y_1(4)\\
        y_1(5)\\
        y_1(6)\\
        y_1(7)\\
        y_1(8)
    \end{bmatrix}
    =
    \begin{bmatrix}
        \mathbf{h}_1^{[m_1(1)]}\mathbf{a}^1\\
        \mathbf{h}_1^{[m_1(2)]}\mathbf{a}^1\\
        \mathbf{h}_1^{[m_1(3)]}\mathbf{a}^2\\
        \mathbf{h}_1^{[m_1(4)]}\mathbf{a}^2\\
        \mathbf{h}_1^{[m_1(5)]}\mathbf{a}^3\\
        \mathbf{h}_1^{[m_1(6)]}\mathbf{a}^3\\
        \mathbf{h}_1^{[m_1(7)]}\mathbf{a}^4\\
        \mathbf{h}_1^{[m_1(8)]}\mathbf{a}^4\\
    \end{bmatrix}
    +
    \begin{bmatrix}
        \mathbf{h}_1^{[m_1(1)]}\mathbf{b}^1\\
        \mathbf{h}_1^{[m_1(2)]}\mathbf{b}^2\\
        \mathbf{h}_1^{[m_1(3)]}\mathbf{b}^1\\
        \mathbf{h}_1^{[m_1(4)]}\mathbf{b}^2\\
        \mathbf{h}_1^{[m_1(5)]}\mathbf{b}^3\\
        \mathbf{h}_1^{[m_1(6)]}\mathbf{b}^4\\
        \mathbf{h}_1^{[m_1(7)]}\mathbf{b}^3\\
        \mathbf{h}_1^{[m_1(8)]}\mathbf{b}^4\\
    \end{bmatrix}
    +
    \begin{bmatrix}
        \mathbf{h}_1^{[m_1(1)]}\mathbf{c}^1\\
        \mathbf{h}_1^{[m_1(2)]}\mathbf{c}^2\\
        \mathbf{h}_1^{[m_1(3)]}\mathbf{c}^3\\
        \mathbf{h}_1^{[m_1(4)]}\mathbf{c}^4\\
        \mathbf{h}_1^{[m_1(5)]}\mathbf{c}^1\\
        \mathbf{h}_1^{[m_1(6)]}\mathbf{c}^2\\
        \mathbf{h}_1^{[m_1(7)]}\mathbf{c}^3\\
        \mathbf{h}_1^{[m_1(8)]}\mathbf{c}^4\\
    \end{bmatrix}\notag\\
    &=
    \begin{bmatrix}
        a^{1,1,[m_1(1)]}\\
        a^{1,1,[m_1(2)]}\\
        a^{1,2,[m_1(3)]}\\
        a^{1,2,[m_1(4)]}\\
        a^{1,3,[m_1(5)]}\\
        a^{1,3,[m_1(6)]}\\
        a^{1,4,[m_1(7)]}\\
        a^{1,4,[m_1(8)]}
    \end{bmatrix}
    +
    \begin{bmatrix}
        b^{1,1,[m_1(1)]}\\
        b^{1,2,[m_1(2)]}\\
        b^{1,1,[m_1(3)]}\\
        b^{1,2,[m_1(4)]}\\
        b^{1,3,[m_1(5)]}\\
        b^{1,4,[m_1(6)]}\\
        b^{1,3,[m_1(7)]}\\
        b^{1,4,[m_1(8)]}
    \end{bmatrix}
    +
    \begin{bmatrix}
        c^{1,1,[m_1(1)]}\\
        c^{1,2,[m_1(2)]}\\
        c^{1,3,[m_1(3)]}\\
        c^{1,4,[m_1(4)]}\\
        c^{1,1,[m_1(5)]}\\
        c^{1,2,[m_1(6)]}\\
        c^{1,3,[m_1(7)]}\\
        c^{1,4,[m_1(8)]}
    \end{bmatrix}\label{eq:subsetcast_N3_M3_y}
\end{align}
In \cite{Gou_Wang_Jafar_BIA}, the switching pattern is specified as $m_1(1), m_1(2), \cdots, m_1(8) = 1,2,1,2,1,2,1,2$, thus \eqref{eq:subsetcast_N3_M3_y} becomes 
\begin{align}
    \begin{bmatrix}
        a^{1,1,[1]}\\
        a^{1,1,[2]}\\
        a^{1,2,[1]}\\
        a^{1,2,[2]}\\
        a^{1,3,[1]}\\
        a^{1,3,[2]}\\
        a^{1,4,[1]}\\
        a^{1,4,[2]}
    \end{bmatrix}
    +
    \begin{bmatrix}
        b^{1,1,[1]}\\
        b^{1,2,[2]}\\
        b^{1,1,[1]}\\
        b^{1,2,[2]}\\
        b^{1,3,[1]}\\
        b^{1,4,[2]}\\
        b^{1,3,[1]}\\
        b^{1,4,[2]}
    \end{bmatrix}
    +
    \begin{bmatrix}
        c^{1,1,[1]}\\
        c^{1,2,[2]}\\
        c^{1,3,[1]}\\
        c^{1,4,[2]}\\
        c^{1,1,[1]}\\
        c^{1,2,[2]}\\
        c^{1,3,[1]}\\
        c^{1,4,[2]}
    \end{bmatrix}.\label{eq:subsetcast_N3_nu1_M3_y}
\end{align}
For the desired message $A$ of Rx-1, we note that $2$ distinct mode linear combinations of all blocks occur in the received signal, while for the undesired messages $B,C$, each block appears only as a one-mode linear combination in the received signal (intra-message alignment). For example, the first block of $A$ is transmitted over time slots $1,2$ and received under different modes $1,2$ respectively. However, the first block of $B$ that transmitted over time slots $1,3$ is received under the same mode $1$. Thus, to recover $A$, Rx-$1$ needs to subtract the interfering linear combinations of blocks of $B,C$ from \eqref{eq:subsetcast_N3_nu1_M3_y} to recover $2$ modes linear combination of each block of $A$, and furthermore download the mode-$3$ linear combination of each block of $A$, i.e., Rx-$1$ needs the following symbols,
\begin{align}
    &a^{1,1,[3]}, a^{1,2,[3]}, a^{1,3,[3]}, a^{1,4,[3]},\notag\\
    &b^{1,1,[1]}, b^{1,2,[2]}, b^{1,3,[1]}, b^{1,4,[2]},\notag\\
    &c^{1,1,[1]}, c^{1,2,[2]}, c^{1,3,[1]}, c^{1,4,[2]}
\end{align}
These symbols are obtained over the last $12$ time slots, where 
\begin{align}
    \mathbf{U}_A \mathbf{a} + \mathbf{U}_B \mathbf{b} + \mathbf{U}_C \mathbf{c}
\end{align}
is broadcast. During the last $12$ time slots, Rx-$1$ switches its antenna to mode-$3$ when blocks of $A$ are broadcast, switches to mode-$1$ when the first and third blocks of $B,C$ are broadcast, and switches to mode-$2$ when the second and fourth blocks of $B,C$ are broadcast. The construction of \cite{Gou_Wang_Jafar_BIA} is thus complete.

\subsubsection{BIA scheme for the BCGM setting with $(K = 3, G = 2, M=3)$}
Now consider the $(K = 3, G = 2, M=3)$ BCGM setting shown in Fig. \ref{fig:subsetcast_N3_nu2_M3} with $N_g = 3, \nu_g = 2$.
\begin{figure}[H]
\centering
\begin{tikzpicture}
\def\dx{1.75}
\def\dy{0.4}
\foreach  \i in {1,2,3}{
	\node (T\i) at (0,{-(\i-2)*1.5*\dy}) [circle, draw,  inner sep = 0.03cm] {\footnotesize $x_\i$};
}

\node (Tx) at (0,0) [rectangle, rounded corners, draw, minimum width=0.75cm, minimum height=2.1cm] {};

\node (W) [left=0.5cm of Tx, align=center] {$A, B, C$};
\draw [thick, ->] (W)--(Tx);

\foreach  \i/\j/\u/\c in {1/\mathbf{h}_1/A{,}B/yellow,2/\mathbf{h}_2/A{,}C/orange,3/\mathbf{h}_3/B{,}C/pink}{
	\node (R\i) at (1.5*\dx,{-(\i-2)*2*\dy}) [rectangle, rounded corners, draw,  inner sep = 0.1cm, align=center, fill=\c!10] {\scriptsize $y_\i=\j^{[m_\i]} \mathbf{x}+z_\i$};
	\node (W{\i}hat) [right=0.5cm of R\i] {$\u$};
\draw [ ->] (R\i)--(W{\i}hat);

}

\foreach \i in {1,2,3}{
	\foreach \j in {1,2,3}{
		\draw [->] (T\i)--($(R\j.west)+(0,{0.1*(1.5-\i)})$);
		}
}
\end{tikzpicture}
\caption{$(K=3,G=2, M=3)$ BCGM Example.}
\label{fig:subsetcast_N3_nu2_M3}
\end{figure}
Following the  construction of \cite{Gou_Wang_Jafar_BIA}, during the first $8$ time slots, we again let the transmitter send $\mathbf{T}_A\mathbf{a} + \mathbf{T}_B\mathbf{b} + \mathbf{T}_C\mathbf{c}$, so that \eqref{eq:subsetcast_N3_M3_y} is received after the first $8$ time slots. However, this time the design of the switching pattern is a bit more challenging. Note that both $A$ and $B$ are Rx-$1$'s desired messages. Intuitively, in \eqref{eq:subsetcast_N3_M3_y}, we hope that only \emph{one} mode linear combination of each block of the undesired message $C$ (interference) occurs (intra-message alignment) similar to the previous example, while \emph{two} distinct mode linear combinations of each block of both $A,B$ also occur. To align the interference, we need,
\begin{align}
    &m_1(1) = m_1(5), m_1(2) = m_1(6),\notag\\
    &m_1(3) = m_1(7), m_1(4) = m_1(8)\label{eq:N3_nu2_M3_alignC}
\end{align}
since block-$1$ of $C$ is sent over time slots $1$ and $5$; block-$2$ of $C$ is sent over time slots $2$ and $6$; $\cdots$, and block-$4$ of $C$ is sent over time slots $4$ and $8$. 

Meanwhile, to make sure that each block of $A$ is received with $2$ distinct modes, we want 
\begin{align}
    &m_1(1) \neq m_1(2), m_1(3) \neq m_1(4),\notag\\
    &m_1(5) \neq m_1(6), m_1(7) \neq m_1(8)\label{eq:N3_nu2_M3_sepA}
\end{align}
since  block-$1$ of $A$ is sent over time slots $1$ and $2$, $\cdots$, and block-$4$ of $A$ is sent over time slots $7$ and $8$. At the same time, to make sure that each block of $B$ is also received with $2$ modes, we want 
\begin{align}
    &m_1(1) \neq m_1(3), m_1(2) \neq m_1(4),\notag\\
    &m_1(5) \neq m_1(7), m_1(6) \neq m_1(8)\label{eq:N3_nu2_M3_sepB}
\end{align}
since  block-$1$ of $B$ is sent over time slots $1$ and  $3$, $\cdots$, and block-$4$ of $B$ is sent over time slots $6$ and $8$. 

One feasible solution that satisfies all the three constraints in \eqref{eq:N3_nu2_M3_alignC} to \eqref{eq:N3_nu2_M3_sepB} is 
\begin{align}
     (m_1(1), m_1(2), \cdots, m_1(8)) = (1,2,2,1,1,2,2,1)
\end{align}
With this switching pattern, Rx-$k$ obtains, 
\begin{align}
    \begin{bmatrix}
        y_1(1)\\
        y_1(2)\\
        y_1(3)\\
        y_1(4)\\
        y_1(5)\\
        y_1(6)\\
        y_1(7)\\
        y_1(8)
    \end{bmatrix}
    =
    \begin{bmatrix}
        a^{1,1,[1]}\\
        a^{1,1,[2]}\\
        a^{1,2,[2]}\\
        a^{1,2,[1]}\\
        a^{1,3,[1]}\\
        a^{1,3,[2]}\\
        a^{1,4,[2]}\\
        a^{1,4,[1]}
    \end{bmatrix}
    +
    \begin{bmatrix}
        b^{1,1,[1]}\\
        b^{1,2,[2]}\\
        b^{1,1,[2]}\\
        b^{1,2,[1]}\\
        b^{1,3,[1]}\\
        b^{1,4,[2]}\\
        b^{1,3,[2]}\\
        b^{1,4,[1]}
    \end{bmatrix}
    +
    \begin{bmatrix}
        c^{1,1,[1]}\\
        c^{1,2,[2]}\\
        c^{1,3,[2]}\\
        c^{1,4,[1]}\\
        c^{1,1,[1]}\\
        c^{1,2,[2]}\\
        c^{1,3,[2]}\\
        c^{1,4,[1]}
    \end{bmatrix}.\label{eq:subsetcast_N3_nu2_M3_y}
\end{align}
Note that Rx-$1$ needs  $c^{1,1,[1]}, c^{1,2,[2]}, c^{1,3,[2]}, c^{1,4,[1]}$ to eliminate the interference from message $C$ in \eqref{eq:subsetcast_N3_nu2_M3_y}. However, even if the interference is eliminated, dimensions of $A$ and $B$ are still aligned. 
To separate $A,B$, we let the transmitter use another $8$ time slots to send,
\begin{align}
    \mathbf{T}_A\mathbf{a} + {\color{red}2} \mathbf{T}_B\mathbf{b} + {\color{red}3} \mathbf{T}_C\mathbf{c},
\end{align}
meanwhile, Rx-$1$ repeats the same switching pattern
\begin{align}
    (m_1(9), m_1(10), \cdots, m_1(16)) = (1,2,2,1,1,2,2,1),
\end{align}
and thus obtains, 
\begin{align}
    \begin{bmatrix}
        y_1(9)\\
        y_1(10)\\
        y_1(11)\\
        y_1(12)\\
        y_1(13)\\
        y_1(14)\\
        y_1(15)\\
        y_1(16)
    \end{bmatrix}
    =
    \begin{bmatrix}
        a^{1,1,[1]}\\
        a^{1,1,[2]}\\
        a^{1,2,[2]}\\
        a^{1,2,[1]}\\
        a^{1,3,[1]}\\
        a^{1,3,[2]}\\
        a^{1,4,[2]}\\
        a^{1,4,[1]}
    \end{bmatrix}
    +{\color{red}2}
    \begin{bmatrix}
        b^{1,1,[1]}\\
        b^{1,2,[2]}\\
        b^{1,1,[2]}\\
        b^{1,2,[1]}\\
        b^{1,3,[1]}\\
        b^{1,4,[2]}\\
        b^{1,3,[2]}\\
        b^{1,4,[1]}
    \end{bmatrix}
    +{\color{red}3}
    \begin{bmatrix}
        c^{1,1,[1]}\\
        c^{1,2,[2]}\\
        c^{1,3,[2]}\\
        c^{1,4,[1]}\\
        c^{1,1,[1]}\\
        c^{1,2,[2]}\\
        c^{1,3,[2]}\\
        c^{1,4,[1]}
    \end{bmatrix}.\label{eq:subsetcast_N3_nu2_M3_y_2}
\end{align}
Again, the same $c^{1,1,[1]}, c^{1,2,[2]}, c^{1,3,[2]}, c^{1,4,[1]}$ are needed to eliminate the interference. They will be downloaded in the future. For now, suppose they are already available to Rx-$1$, thus after eliminating the interference, stacking a row (e.g., the first row) of \eqref{eq:subsetcast_N3_nu2_M3_y} and the same row of \eqref{eq:subsetcast_N3_nu2_M3_y_2} together, Rx-$1$ finds,
\begin{align}
    \begin{bmatrix}
        y_1(1) - c^{1,1,[1]}\\
        y_1(9) - 3c^{1,1,[1]}
    \end{bmatrix}
    =
    \begin{bmatrix}
        {\color{red}1} & {\color{red}1}\\
        {\color{red}1} & {\color{red}2}
    \end{bmatrix}
    \begin{bmatrix}
        a^{1,1,[1]}\\
        b^{1,1,[1]}
    \end{bmatrix}
\end{align}

Due to the invertibility of the $2\times 2$ matrix, $a^{1,1,[1]}, b^{1,1,[1]}$ can be recovered. Similarly, $a^{1,l,[1]}, a^{1,l,[2]}$, $b^{1,l,[1]}, b^{1,l,[2]}$ can be recovered for arbitrary $l \in [4]$. 

Therefore, the remaining task for Rx-$1$ is to obtain the desired symbols $a^{1,l,[3]}, b^{1,l,[3]}$ for all $l \in [4]$ and to obtain the interfering symbols $c^{1,1,[1]}, c^{1,2,[2]}, c^{1,3,[2]}, c^{1,4,[1]}$. This is again done by letting the transmitter broadcast $\mathbf{U}_A \mathbf{a} + \mathbf{U}_B \mathbf{b} + \mathbf{U}_C \mathbf{c}$ and letting Rx-$1$'s switching pattern be $3,3,3,3,3,3,3,3,1,2,2,1$ during the last $12$ time slots.

Overall, in the $8+8+12 = 28$ time slots, the broadcast information is as follows.
\begin{align}
    \mathbf{X} =
    \underbrace{
    \begin{bmatrix}
        {\color{red} 1}\mathbf{T}_A\\
        {\color{red} 1}\mathbf{T}_A\\
        \mathbf{U}_A
    \end{bmatrix}}_{\mathbb{V}_A}
    \mathbf{a}
    +
    \underbrace{
    \begin{bmatrix}
        {\color{red} 1}\mathbf{T}_B\\
        {\color{red} 2}\mathbf{T}_B\\
        \mathbf{U}_B
    \end{bmatrix}}_{\mathbb{V}_B}
    \mathbf{b}
    +
    \underbrace{
    \begin{bmatrix}
        {\color{red} 1}\mathbf{T}_C\\
        {\color{red} 3}\mathbf{T}_C\\
        \mathbf{U}_C
    \end{bmatrix}}_{\mathbb{V}_C}
    \mathbf{c}\label{eq:K3G2M3BCGMprecoding}
\end{align}
It is not difficult to verify that following the same considerations, during the first $8$ time slots, the switching pattern of Rx-$2$ and Rx-$3$ can be chosen as,
\begin{align}
    m_2(1), m_2(2), \cdots, m_2(8) = 1,2,1,2,2,1,2,1\\
    m_3(1), m_3(2), \cdots, m_3(8) = 1,1,2,2,2,2,1,1
\end{align}
The switching pattern of the remaining time slots can be found accordingly. This completes the BIA construction for this BCGM example.

\begin{remark}
    Due to the iterative structure of $\mathbf{T}_A, \mathbf{T}_B, \mathbf{T}_C$,  the switching patterns $(m_k(1), m_k(2), \cdots, m_k(8))$ for $k =1,2,3$ can be found iteratively. For example, for Rx-$2$, the fact that message $A$ is desired determines $m_2(1)\neq m_2(2)$ which can be set as $1,2$, respectively. The fact that $B$ is undesired, and the fact that same blocks of $B$ are sent over $t = 1,2$ and $t = 3,4$ determines $m_2(3) = 1, m_2(4) = 2$ as a repetition of $m_2(1), m_2(2)$. The status of message $C$ as a desired message and the fact that same blocks of $C$ are sent over $t = 1,2,3,4$ and $t = 5,6,7,8$ determines $m_2(5)=2=\Mod{m_2(1) + 1}{2}, m_2(6)=1=\Mod{m_2(2) + 1}{2}, m_2(7)=2=\Mod{m_2(3)+ 1}{2}, m_2(8)=1=\Mod{m_2(4)+ 1}{2}$ by shifting the switching pattern over the first $4$ time slots as a whole. Similarly, for Rx-$3$, the fact that $A$ is undesired determines $m_3(1) = m_3(2) = 1$. The desired status of message $B$ determines $m_3(3) = m_3(4) = 2$ by shifting $m_3(1), m_3(2)$ as a whole. The desired status of $C$ determines $m_3(5), \cdots, m_3(8)$ by shifting $m_3(1), \cdots, m_3(4)$ as a whole.
\end{remark}
This idea of shifting or repeating switching patterns according to whether a message is desired or not is quite important for designing the switching pattern for the general problem.

\subsection{Theorem \ref{thm:BCGM}: Proof of Achievability}\label{sec:achievabilityBCGM}
Will consider precoding  over $\mathsf{T}_{p}$ channel uses where $\mathsf{T}_p \ll \mathsf{T}_c$. Thus for any $m \in [M]$, we write $\mathbf{h}_k^{[m]}(t)$ as $\mathbf{h}_k^{[m]}$ since it is a constant vector during the $\mathsf{T}_{p}$ channel uses. In our description of the scheme in Section \ref{sec:genscheme}, we will ignore the AWGN for simplicity. In Section \ref{subsec:BCGM_property} we will show that with AWGN, the sum-DoF value $\frac{N_g M}{(M-1)\nu_g + N_g}$ is achievable. 

\subsubsection{BIA Precoding Scheme for General $(K, G, M)$ BCGM}\label{sec:genscheme}
Recall that  a $(K,G,M)$ BCGM problem has $N_g = \binom{K}{G}$ messages and each receiver desires $\nu_g = \binom{K-1}{G-1}$ messages. The scheme only depends on $N_g$ and $\nu_g$. The precoding structure at the transmitters' side will be closely related to that of the unicast problem in \cite{Gou_Wang_Jafar_BIA} with $N_g$ messages (or equivalently $N_g$ Rx's), with a newly introduced MDS structure being the primary difference. At the receivers' side, the new switching pattern will be determined iteratively according to Algorithm \ref{alg:pk} that satisfies the condition \eqref{eq:alignSepSuff} in Lemma \ref{lem:dimension_received}. The main building blocks of our precoding scheme are specified as follows.
\begin{enumerate}
    \item $\bm{\Lambda} = \{\lambda_{i,j}\}_{i \in [\nu_g], j \in [N_g]} \in \mds_{\nu_g \times N_g}$ whose first row is $\mathbf{1}_{1\times M}$\footnote{Actually, we only need $\bm{\Lambda}$ to be a $\nu_g \times N_g$ MDS matrix. We add the restriction that the first row should be all $1$ for ease of presenting the scheme.}.
    \item Same as \cite{Gou_Wang_Jafar_BIA}, for any $n \in [N_g]$, message $W_n$ is encoded into $L = M\ell$ independent streams, where $\ell = (M-1)^{N_g - 1}$.
    \begin{align}
        &\bigg[\underbrace{W_n^1 ~~ W_n^2 \cdots W_n^M}_{\triangleq \mathbf{w}_n^{1\top}} ~~ \mid~~\underbrace{W_n^{M+1} ~~ W_n^{M+2} \cdots W_n^{2M}}_{\triangleq \mathbf{w}_n^{2\top}}~~\mid\notag\\
        &\cdots\mid~~\underbrace{W_n^{(\ell-1)M+1}~~ W_n^{(\ell-1)M+2} \cdots W_n^{(\ell-1)M+M}}_{\triangleq \mathbf{w}_n^{\ell\top}}\bigg]^\top\notag\\
        &\triangleq \mathbf{w}_{n} \in \mathbb{C}^{M\ell \times 1}
    \end{align}
The streams of symbols are partitioned into $\ell$ blocks $\mathbf{w}_{n}^{l} \in \mathbb{C}^{M\times 1}, \forall l \in [\ell]$. For any $n \in [N_g], k \in [K], l \in [\ell], m \in [M]$, let $w_{n}^{k,l,[m]}$ be the \emph{mode-$m$ linear combination} of the $\ell^{th}$ block of message $W_n$ at Rx-$k$, defined by Rx-$k$'s channel vector while operating in mode-$m$, i.e., 
    \begin{align}
        w_{n}^{k,l,[m]} = \mathbf{h}_k^{[m]}\mathbf{w}_{n}^{l}.\label{eq:subsetcast_mode_info}
    \end{align}
    \item For any $n \in [N_g]$, define a function $f_n(t)\colon [(M-1)\ell] \rightarrow [\ell]$ such that, 
    \begin{align}
        f_n(t) \triangleq \left\lfloor \frac{t-1}{(M-1)^n} \right\rfloor(M-1)^{n-1} + \Mod{t}{(M-1)^{n-1}},
    \end{align}
    and a matrix $\mathbf{\Pi}_n \in \mathbb{C}^{(M-1)\ell \times \ell}$ where 
    \begin{align}
        \mathbf{\Pi}_n &= 
       \begin{bmatrix}
           \mathbf{e}_{\ell}^{f_n(1)}\\
           \mathbf{e}_{\ell}^{f_n(2)}\\
           \vdots\\
           \mathbf{e}_{\ell}^{f_n\big((M-1)\ell\big)}
       \end{bmatrix}.
    \end{align}
    \item For any $n \in [N_g]$, let $\mathbf{T}_n \in \mathbb{C}^{M(M-1)\ell \times M\ell}$ be the collection of the precoding matrices for message $W_n$ over the first $(M-1)\ell$ time slots, where 
    \begin{align}
        \mathbf{T}_n = \mathbf{\Pi}_n\otimes \bI_M.\label{eq:BIA_X_T}
    \end{align}
    For any $t \in [(M-1)\ell]$ the $M$ antennas of the transmitter broadcast the vector,
    \begin{align}
        \mathbf{x}(t) &= \sum_{n\in [N_g]}\bigg(\mathbf{T}_{n}\bigg)_{[(t-1)M+1:tM], :}\mathbf{w}_n\notag\\
        &=\sum_{n\in[N_g]} (\mathbf{e}_{\ell}^{f_n(t)}\otimes \bI_M) \mathbf{w}_{n}
       =\sum_{n\in[N_g]}\mathbf{w}_{n}^{f_n(t)}\label{eq:subsetcast_xt}
    \end{align}
    This simply means that \textbf{at time slot $t$, only the $M$ symbols of the $\big(f_n(t)\big)^{th}$ block of message $W_n$ are transmitted}, and each one of the $M$ symbols is sent from a distinct antenna. Note that by this definition, $\mathbf{T}_n$ is exactly the same as the first $M(M-1)\ell$ rows of the precoding matrix for message $W_n$ in the $(N_g, M)$ unicast problem studied in \cite{Gou_Wang_Jafar_BIA}.
   \item For any $k \in [K]$, Rx-$k$'s switching pattern $\mathbf{p}_k = (p_k(1), p_k(2), \cdots, p_k\big((M-1)\ell\big))$ for the first $(M-1)\ell$ time slots that is specified by Algorithm \ref{alg:pk}.
\end{enumerate}

\begin{algorithm*}
    \SetKwInOut{Input}{input}\SetKwInOut{Output}{output}
    \caption{Rx-$k$'s switching pattern for $t \in [(M-1)\ell]$}\label{alg:pk}
    \Input{$\mathcal{V}_k$ that denotes Rx-$k$'s desired messages}
    \Output{Rx-$k$'s switching pattern $\mathbf{p}_k$}
    \uIf{$1 \notin \mathcal{V}_k$} {$p_k(1), p_k(2), \cdots, p_k(M-1) \gets \mathbf{1}_{1\times (M-1)}$\;}
    \Else{$p_k(1), p_k(2), \cdots, p_k(M-1) \gets [1~2~3~\cdots~M-1]$\;}
    $n \gets 2$\;
    \While{$n \leq N_g$}{
        $i \gets 1$\;
        \uIf{$n \in \mathcal{V}_k$}{
            \While{$\ i \leq M-2$}{
                $p_k\bigg(i(M-1)^{n-1}+1\bigg), p_k\bigg(i(M-1)^{n-1}+2\bigg), \cdots, p_k\bigg((i+1)(M-1)^{n-1}\bigg)
                \gets \Mod{p_k(1) + i}{M-1},\Mod{p_k(2) + i}{M-1}, \cdots, \Mod{p_k\bigg((M-1)^{n-1}\bigg) + i}{M-1}$\;
                $i \gets i+1$\;
            }
        }
        \Else{
            \While{$i \leq M-2$}{
                $p_k\bigg(i(M-1)^{n-1}+1\bigg), p_k\bigg(i(M-1)^{n-1}+2\bigg), \cdots, p_k\bigg((i+1)(M-1)^{n-1}\bigg)
                \gets p_k(1), p_k(2), \cdots, p_k\bigg((M-1)^{n-1}\bigg)$\;
                $i \gets i+1$\;
            }
        }
        $n \gets n+1$\;
    }
\end{algorithm*}

Let us first analyze the signal produced at Rx-$k$ because of the precoding structure and switching pattern specified thus far, over the first $(M-1)\ell$ time slots. The full precoding structure and switching pattern for the remaining time slots will be specified subsequently.

At time slot $t \in [(M-1)\ell=(M-1)^{N_g}]$, the receiver Rx-$k \in [k]$, operating in mode-$p_k(t)$, receives 
\begin{align}
    y_k(t) &= \mathbf{h}_k^{p_k(t)}\mathbf{x}(t)\notag\\
    &\overset{\eqref{eq:subsetcast_xt}}{=}\sum_{n\in[N_g]}\mathbf{h}_k^{p_k(t)}\mathbf{w}_n^{f_n(t)} \overset{\eqref{eq:subsetcast_mode_info}}{=} \sum_{n \in [N_g]}w_n^{k,f_n(t),[p_k(t)]}\notag\\
    &=\underbrace{\sum_{n \in \mathcal{V}_k} w_n^{k,f_n(t),[p_k(t)]}}_{\mbox{desired}} + \underbrace{\sum_{n \in [N_g] \setminus \mathcal{V}_k}w_n^{k,f_n(t),[p_k(t)]}}_{\mbox{interference}}.
\end{align}
Thus, during the first $(M-1)\ell$ time slots, the transmitter sends
\begin{align}
    \begin{bmatrix}
        \mathbf{x}(1)\\
        \vdots\\
        \mathbf{x}\big((M-1)\ell\big)
    \end{bmatrix}
    =
    \sum_{n \in [N_g]} \mathbf{T}_n \mathbf{w}_n = \sum_{n \in [N_g]}  \mathbf{\Pi}_n \otimes \bI_M \mathbf{w}_n,
\end{align}
and Rx-$k$ receives,
\begin{align}
    &\begin{bmatrix}
        y_k(1)\\
        \vdots\\
        y_k\big((M-1)\ell\big)
    \end{bmatrix}
    =
    \underbrace{
    \sum_{n \in \mathcal{V}_k}
    \begin{bmatrix}
        w_n^{k,f_n(1),[p_k(1)]}\\
        \vdots\\
        w_n^{k,f_n\big((M-1)\ell\big),\left[p_k\big((M-1)\ell\big)\right]}
    \end{bmatrix}}_{\mbox{desired}}\notag\\
    &\quad\quad\quad+
    \underbrace{
    \sum_{n \in [N_g] \setminus \mathcal{V}_k}
    \begin{bmatrix}
        w_n^{k,f_n(1),[p_k(1)]}\\
        \vdots\\
        w_n^{k,f_n\big((M-1)\ell\big),\left[p_k\big((M-1)\ell\big)\right]}
    \end{bmatrix}
    }_{\mbox{interference}}.\label{eq:subsetcast_top_y}
\end{align}

For ease of studying the property of the received signal, fixing any $n \in [N_g]$, let us represent $t \in [(M-1)\ell = (M-1)^{N_g}]$ and $l \in [\ell=(M-1)^{N_g - 1}]$ as 
\begin{align}
    &t = t_n^{h,i,j} = h(M-1)^n + i(M-1)^{n-1} + j\\
    &l = l_n^{h,j} = h(M-1)^{n-1} + j\\
    &\mbox{where}~~h \in [0:(M-1)^{N_g - n} - 1], \notag\\
    &\quad\quad\quad i \in [0:M-2], j \in [(M-1)^{n-1}].
\end{align}
One can verify that,
\begin{align}
    &f_n(t_n^{h,i,j}) = h(M-1)^{n-1} + j = l_n^{h,j}\Longrightarrow\notag\\
    &f_n(t_n^{h,0,j}) = f_n(t_n^{h,1,j}) = \cdots = f_n(t_n^{h,M-2,j}) = l_n^{h,j}.
\end{align}

With these representations, we have the following lemma regarding $\mathbf{p}_k$ and $y_k(1), \cdots, y_k\big((M-1)\ell\big)$.
\begin{lemma}\label{lem:dimension_received}
    For $k \in [K]$, $n \in [N_g]$, the switching pattern $\mathbf{p}_k$ satisfies 
    \begin{align}
    &\forall h \in [0:(M-1)^{N_g - n} - 1], j \in [(M-1)^{n-1}]&\notag\\
    &\mbox{when}~n \in [N_g] \setminus \mathcal{V}_k,\notag\\
    &p_k(t_n^{h,0,j}) = \cdots = p_k(t_n^{h,M-2,j}) \triangleq m_{k,n}^{l = l_n^{h,j}} \in [M-1];\notag\\
    &\mbox{when}~n \in \mathcal{V}_k,\notag\\
    &p_k(t_n^{h,i,j}) = \Mod{\underbrace{p_k(t_n^{h,0,j})}_{\triangleq m_{k,n}^{l = l_n^{h,j}}} + i}{M-1}, \forall i \in [0:M-2].\label{eq:alignSepSuff}
    \end{align}
\end{lemma}
\proof See Appendix \ref{app:dimension_received}. $\hfill\blacksquare$

Thus, in $\left[y_k(1)\cdots y_k\big((M-1)\ell\big)\right]^\top$ specified in \eqref{eq:subsetcast_top_y}, for any $n \in \mathcal{V}_k$, $M-1$ modes of linear combination $w_n^{k,l,[1]}, w_n^{k,l,[2]}, \cdots, w_n^{k,l,[M-1]}$ occur in the desired part for any block $l \in [\ell]$, while for any $n \in [N_g] \setminus \mathcal{V}_k$, only $\mbox{mode}-m_{k,n}^l$ linear combination $w_n^{k,l,[m_{k,n}^l]}$ occurs in the interference part for the block $l \in [\ell]$. Mathematically,
    \begin{align}
        &\left\{w_n^{k,f_n(1),[p_k(1)]},\cdots, w_n^{k,f_n((M-1)\ell),[p_k((M-1)\ell)]}\right\} =\notag\\
        &\begin{cases}
            \left\{w_n^{k,l,[m_{k,n}^l]}\right\}_{l \in [\ell]}, & \forall n \in [N_g]\setminus \mathcal{V}_k\\
            \left\{w_n^{k,l,[1]}, w_n^{k,l,[2]}, \cdots, w_n^{k,l,[M-1]}\right\}_{l \in [\ell]}, & \forall n \in \mathcal{V}_k
        \end{cases}.\label{eq:alignSep}
    \end{align}
To cancel the interference in \eqref{eq:subsetcast_top_y}, for each block ($M$ symbols) of an undesired message, Rx-$k$ only needs \emph{one} mode linear combination (intra-message alignment). Specifically, Rx-$k$ needs,
\begin{align}
    w_n^{k,l,[m_{k,n}^l]}, \forall n \in [N_g] \setminus \mathcal{V}_k, l \in [\ell].\label{eq:interference_subsetcast}
\end{align}
This is accomplished by letting the transmitter send pure $\mathbf{w}_n^l$ through the $M$ antennas at some time slot in the future and having Rx-$k$ switch its antenna to mode-$m_{k,n}^l$. Thus, the scheme should ensure that Rx-$k$ will obtain the interference in the future so it can cancel it from all received symbols. For now, suppose Rx-$k$ already has the interference terms, so it can cancel the interference and obtain, 
\begin{align}
    \begin{bmatrix}
        \tilde{y}_k(1)\\
        \vdots\\
        \tilde{y}_k\big((M-1)\ell\big)
    \end{bmatrix}
    =
    \sum_{n \in \mathcal{V}_k}
    \begin{bmatrix}
        w_n^{k,f_n(1),[p_k(1)]}\\
        \vdots\\
        w_n^{k,f_n\big((M-1)\ell\big),\left[p_k\big((M-1)\ell\big)\right]}
    \end{bmatrix}.
\end{align}
This, however, still does not guarantee the decodability of the desired messages. The transmitter and the receivers will `repeat' the process specified above another $\nu_g - 1$ times, with the difference being that, $\mathbf{T}_n$ will be scaled by some $\lambda$ to precode the message $W_n$. Specifically, for each $v \in [2:\nu_g]$, the transmitter uses $(M-1)\ell$ time slots to send,
\begin{align}
    \begin{bmatrix}
        \mathbf{x}\big((v-1)(M-1)\ell+1\big)\\
        \vdots\\
        \mathbf{x}\big(v(M-1)\ell\big)
    \end{bmatrix}
    &= \sum_{n\in[N_g]} \lambda_{v,n}\mathbf{T}_n\mathbf{w}_n\notag\\
    &= \sum_{n\in[N_g]} \lambda_{v,n}\mathbf{\Pi}\otimes\bI_{M}\mathbf{w}_n.
\end{align}
Again, following the switching pattern $\mathbf{p}_k$, Rx-$k$ obtains, 
\begin{align}
    &\begin{bmatrix}
        y_k\big((v-1)(M-1)\ell+1\big)\\
        \vdots\\
        y_k\big(v(M-1)\ell\big)
    \end{bmatrix}\notag\\
    &=\underbrace{
    \sum_{n \in \mathcal{V}_k}\lambda_{v,n}
    \begin{bmatrix}
        w_n^{k,f_n(1),[p_k(1)]}\\
        \vdots\\
        w_n^{k,f_n\big((M-1)\ell\big),\left[p_k\big((M-1)\ell\big)\right]}
    \end{bmatrix}}_{\mbox{desired}}\notag\\
    &\quad+\underbrace{
    \sum_{n \in [N_g] \setminus \mathcal{V}_k}\lambda_{v,n}
    \begin{bmatrix}
        w_n^{k,f_n(1),[p_k(1)]}\\
        \vdots\\
        w_n^{k,f_n\big((M-1)\ell\big),\left[p_k\big((M-1)\ell\big)\right]}
    \end{bmatrix}
    }_{\mbox{interference}}.\label{eq:subsetcast_nug_y}
\end{align}
The interference can again be eliminated by subtracting the same combinations in \eqref{eq:interference_subsetcast}. Note that no `new' interference is encountered, highlighting the interference alignment aspect of the coding scheme.

After eliminating the interference, for any $v \in [\nu_g]$, Rx-$k$ obtains, 
\begin{align}
    &\begin{bmatrix}
        \tilde{y}_k\big((v-1)(M-1)\ell+1\big)\\
        \vdots\\
        \tilde{y}_k\big(v(M-1)\ell\big)
    \end{bmatrix}\notag\\
    &=
    \sum_{n \in \mathcal{V}_k}\lambda_{v,n}
    \begin{bmatrix}
        w_n^{k,f_n(1),[p_k(1)]}\\
        \vdots\\
        w_n^{k,f_n\big((M-1)\ell\big),\left[p_k\big((M-1)\ell\big)\right]}
    \end{bmatrix}\label{eq:subsetcast_nug_y_desired}
\end{align}

Recall that in the first round of broadcasting linear combinations of all the messages (e.g., the first row of \eqref{eq:K3G2M3BCGMprecoding}), we have $\lambda_{1,1} = \cdots = \lambda_{1,N_g} = 1$, because the first row of $\bm{\Lambda}$ is an all one vector. Stacking the $t^{th}$ rows, where $t \in [(M-1)\ell]$, of \eqref{eq:subsetcast_nug_y_desired} for all $v \in [\nu_g]$ together, Rx-$k$ obtains, 
\begin{align}
&\begin{bmatrix}
    \tilde{y}_k(t)\\
    \tilde{y}_k\big((M-1)\ell + t\big)\\
    \vdots\\
    \tilde{y}_k\big((\nu_g - 1)(M-1)\ell + t\big)
\end{bmatrix}
\hspace{-0.16cm}
=
\hspace{-0.16cm}
\begin{bmatrix}
    \sum_{n \in \mathcal{V}_k}\lambda_{1,n}w_n^{k,f_n(t),[p_k(t)]}\\
    \sum_{n \in \mathcal{V}_k}\lambda_{2,n}w_n^{k,f_n(t),[p_k(t)]}\\
    \vdots\\
    \sum_{n \in \mathcal{V}_k}\lambda_{\nu_g,n}w_n^{k,f_n(t),[p_k(t)]}
\end{bmatrix}\notag\\
&=
\underbrace{
\begin{bmatrix}
    \lambda_{1,n_1} & \lambda_{1,n_2} & \cdots & \lambda_{1,n_{\nu_g}}\\
    \lambda_{2,n_1} & \lambda_{2,n_2} & \cdots & \lambda_{2,n_{\nu_g}}\\
    \vdots & \vdots & \vdots & \vdots\\
    \lambda_{\nu_g,n_1} & \lambda_{\nu_g,n_2} & \cdots & \lambda_{\nu_g,n_{\nu_g}}
\end{bmatrix}}_{\bm{\Lambda}_{:,\mathcal{V}_k}}
\begin{bmatrix}
    w_{n_1}^{k,f_{n_1}(t),[p_k(t)]}\\
    w_{n_2}^{k,f_{n_2}(t),[p_k(t)]}\\
    \vdots\\
    w_{n_{\nu_g}}^{k,f_{n_{\nu_g}}(t),[p_k(t)]}
\end{bmatrix}
\end{align}
where we let the $\nu_g$ elements of $\mathcal{V}_k$, i.e., the indices of the $\nu_g$ messages desired by Rx-$k$, be denoted by,
\begin{align}
    \mathcal{V}_k = \{n_1, n_2, \cdots, n_{\nu_g}\}.
\end{align}
Note that $\bm{\Lambda}_{:,\mathcal{V}_k}$ is invertible due the MDS property of $\bm{\Lambda}$ and the fact that $|\mathcal{V}_k| = \nu_g$. Thus, $\forall n \in \mathcal{V}_k, t \in [(M-1)\ell]$, Rx-$k$ is able to recover $w_{n}^{k,f_{n}(t),[p_k(t)]}$ by inverting $\bm{\Lambda}_{:,\mathcal{V}_k}$. According to \eqref{eq:alignSep}, Rx-$k$ is thus able to recover,
\begin{align}
    \begin{bmatrix}
        w_n^{k,l,[1]}\\
        \vdots\\
        w_n^{k,l,[M-1]}
    \end{bmatrix}
    =
    \begin{bmatrix}
        \mathbf{h}_k^{[1]}\\
        \vdots\\
        \mathbf{h}_k^{[M-1]}
    \end{bmatrix}
    \mathbf{w}_n^{l}, && \forall n \in \mathcal{V}_k, l \in [\ell].
\end{align}
To fully recover the $l^{th}$ block of a desired message $W_n$, one more combination, 
\begin{align}
    w_n^{k,l,[M]} = \mathbf{h}_k^{[M]}\mathbf{w}_n^{l},
\end{align}
needs to be downloaded by Rx-$k$. Also, recall that, the interference symbols in \eqref{eq:interference_subsetcast} have not been downloaded yet. We will deal with them together next.

Let the transmitter now use another $N_g\ell$ time slots to broadcast each block ($M$ symbols) of the $N_g$ messages. Specifically, for any $t \in [N_g\ell]$, 
\begin{align}
    \mathbf{x}\big(\nu_g(M-1)\ell +t\big) = \mathbf{w}_{\left\lfloor\frac{t-1}{\ell}\right\rfloor + 1}^{\Mod{t}{\ell}}.
\end{align}
Rx-$k$ switches its antenna to mode-$M$ if the transmitter is broadcasting a block of a desired message at the current time slot, or switches to mode-$m_{k,n}^l$ if the transmitter is broadcasting the $l^{th}$ block of an undesired message $W_n$. Thus, 
\begin{align}
    &\begin{bmatrix}
        \mathbf{x}\big(\nu_g(M-1)\ell+1\big)\\
        \mathbf{x}\big(\nu_g(M-1)\ell+2\big)\\
        \vdots\\
        \mathbf{x}\big(\nu_g(M-1)\ell+ N_g\ell\big)
    \end{bmatrix}\notag\\
    &=
    \sum_{n \in [N_g]}
    \underbrace{(
    \begin{bmatrix}
        \bzero_{(n-1)\ell\times\ell}\\
        \bI_{\ell}\\
        \bzero_{(N_g - n)\ell\times \ell}
    \end{bmatrix}\otimes \bI_M)}_{\triangleq \mathbf{U}_n}
    \mathbf{w}_n
    =
    \sum_{n \in [N_g]}
    \mathbf{U}_n
    \mathbf{w}_n,\label{eq:BIA_X_U}
\end{align}
and for any $t \in [N_g\ell]$, Rx-$k$ sets the mode to $m_k\big(\nu_g(M-1)\ell+t\big)$ at time slot $\nu_g(M-1)\ell+t$. Specifically, $\forall t \in [N_g\ell]$,
\begin{align}
    &m_k(\nu_g(M-1)\ell+t)\notag\\
    &=\begin{cases}
        m_{k,\left\lfloor \frac{t-1}{\ell} \right\rfloor + 1}^{\Mod{t}{\ell}}  &  \left\lfloor \frac{t-1}{\ell} \right\rfloor + 1 \in [N_g] \setminus \mathcal{V}_k\\
        M & \left\lfloor \frac{t-1}{\ell} \right\rfloor + 1 \in \mathcal{V}_k
    \end{cases}.\label{eq:pattern_last}
\end{align}
This guarantees that the mode-$M$ combination of all blocks of all desired messages and the interference in \eqref{eq:interference_subsetcast} can be recovered. Thus, after $\mathsf{T}_p = \nu_g(M-1)\ell + N_g\ell$ time slots, for all $n \in \mathcal{V}_k, l \in \ell$, Rx-$k$ obtains, 
\begin{align}
    \begin{bmatrix}
        w_n^{k,l,[1]}\\
        \vdots\\
        w_n^{k,l,[M]}
    \end{bmatrix}
    =
    \underbrace{
    \begin{bmatrix}
        \mathbf{h}_k^{[1]}\\
        \vdots\\
        \mathbf{h}_k^{[M]}
    \end{bmatrix}
    }_{\triangleq\mathbf{H}_k}
    \mathbf{w}_n^l.\label{eq:BIA_BCGM_Eq_MIMO}
\end{align}
Since $\mathbf{H}_k$ is invertible \emph{almost surely}, $\{\mathbf{w}_n^l\}_{n \in \mathcal{V}_k, l \in [\ell]}$ can be recovered almost surely, i.e., for all $n \in \mathcal{V}_k$, $W_n$ can be recovered by Rx-$k$. 

Overall, in $\mathsf{T}_p = \nu_g(M-1)\ell + N_g\ell$ time slots, the transmitter sends, 
\begin{align}
    \mathbf{X} = 
    \begin{bmatrix}
        \mathbf{x}(1)\\
        \vdots\\
        \mathbf{x}\big(\nu_g(M-1)\ell + N_g\ell\big)
    \end{bmatrix}
    =
    \sum_{n \in [N_g]}
    \underbrace{
    \begin{bmatrix}
        \mathbf{T}_n\\
        \lambda_{2,n}\mathbf{T}_n\\
        \vdots\\
        \lambda_{\nu_g,n}\mathbf{T}_n\\
        \mathbf{U}_n
    \end{bmatrix}}_{= \mathbb{V}_n}
    \mathbf{w}_n,\label{eq:BIA_X_V}
\end{align}
and the Rx-$k$'s switching pattern over the first $\nu_g(M-1)\ell$ time slots is 
\begin{align}
    m_k(1), m_k(2), \cdots, m_k\big(\nu_g(M-1)\ell\big) = [\underbrace{\mathbf{p}_k\cdots\mathbf{p}_k}_{\ell}],
\end{align} 
while the switching pattern over the last $N_g\ell$ time slots is specified by \eqref{eq:pattern_last}.

From a dimension counting perspective, note that since over $\mathsf{T}_p = \nu_g(M-1)\ell + N_g\ell$ channel uses, $L = M\ell$ symbols of each messages are delivered, the DoF achieved for any message $n \in [N]$ is 
\begin{align}
    d_n = \frac{M\ell}{\mathsf{T}_p} = \frac{M\ell}{\nu_g(M-1)\ell + N_g\ell} = \frac{M}{(M-1)\nu_g + N_g}
\end{align}
so the sum-DoF value achieved is $d_{\Sigma} = \frac{N_g M}{(M-1)\nu_g + N_g}$.

\subsubsection{Key Properties of BIA Precoding Scheme}\label{subsec:BCGM_property}
Let us represent our BIA precoding scheme for the BCGM setting in a slightly more general form and specify its properties to complete the proof that the sum-DoF $d_{\Sigma} = \frac{N_g M}{(M-1)\nu_g + N_g}$ is achievable. The generalized representation will be useful to apply the scheme later in a different context in Section \ref{subsec:USI}.

Specifically, according to \eqref{eq:BIA_X_T}, \eqref{eq:BIA_X_U} and \eqref{eq:BIA_X_V}, the signal sent from the $M$ antennas of the transmitter over $\mathsf{T}_p$ channel uses, can be represented as 
\begin{align}
    &\mathbf{X} = 
    \begin{bmatrix}
        \mathbf{x}(1)&\mathbf{x}(2)&\cdots&\mathbf{x}(\mathsf{T}_p)
    \end{bmatrix}^\top\notag\\
    &=
    \sum_{n\in[N_g]}
    \underbrace{
    \begin{bmatrix}
        \alpha_{n,1}^1 \bI_M & \alpha_{n,1}^2 \bI_M & \cdots & \alpha_{n,1}^{\ell} \bI_M\\
        \alpha_{n,2}^1 \bI_M & \alpha_{n,2}^2 \bI_M & \cdots & \alpha_{n,2}^{\ell} \bI_M\\
        \vdots\\
        \alpha_{n,\mathsf{T}_p}^1 \bI_M & \alpha_{n,\mathsf{T}_p}^2 \bI_M & \cdots & \alpha_{n,\mathsf{T}_p}^{\ell} \bI_M\\
    \end{bmatrix}}_{\mathbb{V}_n}
    \mathbf{w}_n\label{eq:BIA_BCGM_X}
\end{align}
where $\mathbb{V}_n$ is the precoding matrix for message $W_n$, and the $\alpha$'s can be $0,1$ or some $\lambda$ as an entry of the $\nu_g \times N_g$ MDS matrix $\bm{\Lambda}$.

Let $\mathbf{z}_k = [z_k(1)~~z_k(2)~~\cdots~~z_k(\mathsf{T}_p)]^\top$ be the AWGN over the $\mathsf{T}_p$ channel uses at Rx-$k$. The signal received at $\mbox{Rx}-k, k\in [K]$ can then be represented as \eqref{eq:BIA_BCGM_Y_compact}, where in step \eqref{eq:BIA_BCGM_Y} $\mathbf{H}_k$ is as defined in \eqref{eq:BIA_BCGM_Eq_MIMO}, and we use the fact that $\mathbf{e}_{M}^{m} \mathbf{H}_k = \mathbf{h}_k^{[m]}$ for any $m \in [M]$.
\begin{figure*}[ht]
\begin{align}
    &\mathbf{y}_k = \mbox{Block-Diag}(\mathbf{h}_k^{[m_k(1)]},\mathbf{h}_k^{[m_k(2)]},\cdots,\mathbf{h}_k^{[m_k(\mathsf{T}_p)]})\mathbf{X} + \mathbf{z}_k\\
    &=\sum_{n\in[N_g]}\mbox{Block-Diag}(\mathbf{h}_k^{[m_k(1)]},\mathbf{h}_k^{[m_k(2)]},\cdots,\mathbf{h}_k^{[m_k(\mathsf{T}_p)]})\mathbb{V}_n \mathbf{w}_n + \mathbf{z}_k \label{eq:BIA_BCGM_YV}\\
    &=\sum_{n\in[N_g]}
    \begin{bmatrix}
        \alpha_{n,1}^1 \mathbf{h}_k^{[m_k(1)]} & \alpha_{n,1}^2 \mathbf{h}_k^{[m_k(1)]} & \cdots & \alpha_{n,1}^{\ell} \mathbf{h}_k^{[m_k(1)]}\\
        \alpha_{n,2}^1 \mathbf{h}_k^{[m_k(2)]} & \alpha_{n,2}^2 \mathbf{h}_k^{[m_k(2)]} & \cdots & \alpha_{n,2}^{\ell} \mathbf{h}_k^{[m_k(2)]}\\
        \vdots\\
        \alpha_{n,\mathsf{T}_p}^1 \mathbf{h}_k^{[m_k(\mathsf{T}_p)]} & \alpha_{n,\mathsf{T}_p}^2 \mathbf{h}_k^{[m_k(\mathsf{T}_p)]} & \cdots & \alpha_{n,\mathsf{T}_p}^{\ell} \mathbf{h}_k^{[m_k(\mathsf{T}_p)]}\\
    \end{bmatrix}
    \mathbf{w}_n
    +
    \mathbf{z}_k\\
    &=\sum_{n\in[N_g]}
    \underbrace{
    \begin{bmatrix}
        \alpha_{n,1}^1 \mathbf{e}_M^{m_k(1)} & \alpha_{n,1}^2 \mathbf{e}_M^{m_k(1)} & \cdots & \alpha_{n,1}^{\ell} \mathbf{e}_M^{m_k(1)}\\
        \alpha_{n,2}^1 \mathbf{e}_M^{m_k(2)} & \alpha_{n,2}^2 \mathbf{e}_M^{m_k(2)} & \cdots & \alpha_{n,2}^{\ell} \mathbf{e}_M^{m_k(2)}\\
        \vdots\\
        \alpha_{n,\mathsf{T}_p}^1 \mathbf{e}_M^{m_k(\mathsf{T}_p)} & \alpha_{n,\mathsf{T}_p}^2 \mathbf{e}_M^{m_k(\mathsf{T}_p)} & \cdots & \alpha_{n,\mathsf{T}_p}^{\ell} \mathbf{e}_M^{m_k(\mathsf{T}_p)}\\
    \end{bmatrix}}_{\triangleq \mathbb{E}_{k,n}\in \mathbb{C}^{\mathsf{T}_p \times M\ell}}
    \underbrace{
    \begin{bmatrix}
        \mathbf{H}_k & & & \\
        &\mathbf{H}_k & & &\\
        & & \ddots & \\
        & & & \mathbf{H}_k
    \end{bmatrix}}_{\# \mbox{ of } \mathbf{H}_k = \ell, \triangleq \mathbb{H}_k}
    \mathbf{w}_n
    +
    \mathbf{z}_k\label{eq:BIA_BCGM_Y}\\
    &= \sum_{n \in \mathcal{V}_k} \mathbb{E}_{k,n}\underbrace{\mathbb{H}_k\mathbf{w}_n}_{\mbox{desired}} + \sum_{n \in [N_g]\setminus\mathcal{V}_k} \mathbb{E}_{k,n}\underbrace{\mathbb{H}_k\mathbf{w}_n}_{\mbox{interference}} + \mathbf{z}_k\label{eq:BIA_BCGM_Y_compact}
\end{align}
\hrule
\end{figure*}

Note that in our scheme, all the $M\ell \times M\ell$ symbols of the desired message are recovered at Rx-$k$, while every block of $M$ symbols of the undesired message (interference) is aligned into $1$ dimension at Rx-$k$. This implies the following properties of the corresponding matrices in \eqref{eq:BIA_BCGM_Y_compact}.
\begin{align}
    \begin{cases}
        \mbox{rk}(\mathbb{E}_{k,n}) = M\ell, \forall n \in \mathcal{V}_k\\
        \mbox{rk}(\mathbb{E}_{k,n}) = \ell, \forall n \in [N_g] \setminus \mathcal{V}_k\\
        \mbox{rk}(\bigcup_{n \in [N_g]}\mathbb{E}_{k,n}) = \sum_{n \in [N_g]} \mbox{rk}(\mathbb{E}_{k,n}) = \mathsf{T}_p
    \end{cases}.\label{eq:BIA_sep}
\end{align}

Note that according to \eqref{eq:BIA_sep} there exists $\mathbb{D}_{k,n} \in \mathbb{C}^{\mathsf{T}_p \times M\ell}, \forall n \in \mathcal{V}_k$ s.t. 
\begin{align}
    \mathbb{D}_{k,n}^\top \mathbb{E}_{k,n} = \mathbf{I}_{M\ell},~~~ \mathbb{D}_{k,n}^\top \mathbb{E}_{k,\bar{n}} = \mathbf{0}_{M\ell \times M\ell},\notag\\
    \forall n \in \mathcal{V}_k, \bar{n} \in [N_g], \bar{n} \neq n.\label{eq:BIA_dec}
\end{align}

For a specific desired message $W_{n}$ where $n \in \mathcal{V}_k$, after applying $\mathbb{D}_{k,n}^\top$ to $\mathbf{y}_k$, Rx-$k$ obtains, 
\begin{align}
    \mathbb{D}_{k,n}^\top \mathbf{y}_k = \mathbb{H}_{k,n}\mathbf{w}_{n} + \mathbb{D}_{k,n}^\top \mathbf{z}_k
    =
    \begin{bmatrix}
        \mathbf{H}_k \mathbf{w}_n^{1}\\
        \mathbf{H}_k \mathbf{w}_n^{2}\\
        \vdots\\
        \mathbf{H}_k \mathbf{w}_n^{\ell_n}
    \end{bmatrix}
    + \mathbb{D}_{k,n}^\top \mathbf{z}_k
\end{align}
where $\mathbb{D}_{k,n}^\top \mathbf{z}_k$ is AWGN whose variance is determined by $\mathbb{D}_{k,n}$, which is further determined by $\mathbb{E}_{k,n}$'s. That is to say, after $\mathsf{T}_p$ channel uses, for any block of $M$ symbols of $W_{n}$, i.e., $\mathbf{w}_n^l, l \in [\ell_n]$, Rx-$k$ sees a point to point $M\times M$ MIMO channel $\mathbf{H}_k\mathbf{w}_n^l$ plus some AWGN, whose DoF is known to be the rank of $\mathbf{H}_{k}$ which is equal to $M$ almost surely. Thus, for any message $W_n$, the DoF value achieved is $d_n^{\gc} = M\ell_n/\mathsf{T}_p$ and the sum-DoF achieved is $d_{\Sigma}^{\gc}=N_gd_n^{\gc} = N_g M /\left((M-1)\nu_g + N_g\right)$.

\section{From BCGM to MapReduce via USI}\label{sec:BCGMapplication}
In this section, we first formalize a symmetric wireless MapReduce network following the formulation in \cite{Li_Chen_Wang_MapReduce,Bi_Wigger_MapReduce}. In order to determine the optimal \emph{inter-message alignment} that maximally leverages the side information in the wireless MapReduce network, we then relabel the messages to formulate an equivalent unicast with side information (USI) problem. This relabeling at the same time makes the integration of inter-message alignment and the intra-message alignment  apparent.

\subsection{$(K,r,M)$ MapReduce: Formulation}\label{subsec:MapReduce_Form}
In a $(K,r,M)$ MapReduce network, there are $K$ Tx's each of which has only one conventional antenna. There are also $K$ Rx's each equipped with a reconfigurable antenna that is able to switch among $M$ independent modes. Over the $t^{th}$ channel use, the scalar signal received at Rx-$k$ is 
\begin{align}
y_k(t)&=\mathbf{h}_k^{[m_k(t)]}(t) \mathbf{x}(t)+z_k(t), &&\forall k\in[K].
\end{align}
Here $\mathbf{h}_k^{[m_k(t)]}(t)\in\mathbb{C}^{1\times K}$ is the $1\times K$ channel vector corresponding to receive-antenna mode $m_k(t) \in [M]$ chosen by Rx-$k$ at time $t$, $\mathbf{x}(t)=[x_1(t)~~\cdots~~x_{K}(t)]^\top$ is the $K\times 1$ vector of  symbols sent from the $K$ Txs (Tx-$k$ corresponds to $x_k(t)$), and $z_k(t)$ is the zero-mean unit variance circularly symmetric complex AWGN at Rx-$k$ at time $t$. The channel fading model, and the power constraint are the same as those in the BCGM setting. Also as before, let $\mathsf{T}_c$ be the channel coherence time. There is no CSIT while all the Rx's have perfect CSIR.

Let us next specify the messages. There are $N_r = \binom{K}{r}$ independent super-messages indexed as $\widetilde{\bf W}_\mathcal{T}$,  $\forall\mathcal{T}\subset\binom{[K]}{r}$. Each super-message is a set of $K-r$ independent messages, $\widetilde{\bf W}_\mathcal{T} \triangleq \{\widetilde{W}_{\mathcal{T},\bar{t}}, ~\bar{t} \in [K]\setminus \mathcal{T}\}$. Thus, the total number of independent messages in the MapReduce problem is $N_m=N_r(K-r)$. Define,
\begin{align}
\si_k^{\icmap}&\triangleq \{\widetilde{\mathbf{W}}_{\mathcal{T}} \mid \forall \mathcal{T} \in \scalebox{0.8}{$\binom{[K]}{r}$}, k \in \mathcal{T}\}
\end{align}
as the set of messages  available as side-information to Rx-$k$. Tx-$k$ has knowledge of only the messages in $\si_k^{\icmap}$, thus $x_k(t)$ can only depend on these messages. The set of messages desired by Rx-$k$ is specified as,
\begin{align}
\de_k^{\icmap} &= \{\widetilde{W}_{\mathcal{T},k} \mid \forall \mathcal{T} \in \scalebox{0.8}{$\binom{[K]}{r}$}, k\in[K]\setminus\mathcal{T}\}.\label{eq:deicmap}
\end{align}
In plain words, each super-message is known to $r$ Tx-Rx pairs, and its constituent $K-r$ messages are desired by the remaining $K-r$ receivers, respectively.

\begin{figure*}
    %\begin{subfigure}[!htbp]{0.42\textwidth}
      %\centering
    \includegraphics{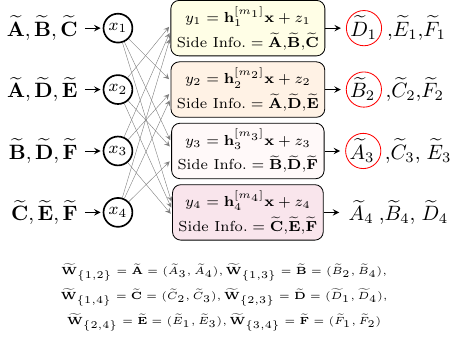}
    \hspace{0.45cm}
    \includegraphics{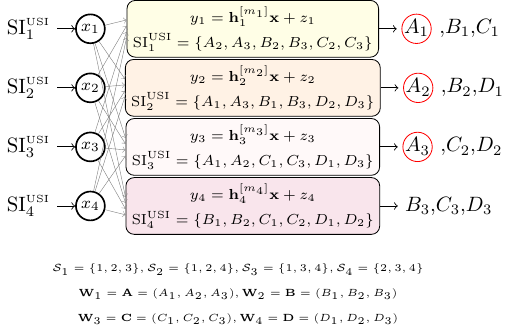}
  \caption{(a) $(K=4, r=2,M)$ MapReduce setting. (b) Equivalent $(K=4, G=3,M)$ USI setting.}\label{fig:mapRedICuni}
\end{figure*}

One example of a $(K=4, r=2, M)$ MapReduce problem is shown in Fig. \ref{fig:mapRedICuni}a. The super-message $\widetilde{\bf W}_{\{1,2\}}$ which is relabeled as $\widetilde{\mathbf{A}}$ for ease of reference, is known to Tx-$1,2$ and Rx-$1,2$, and is comprised of $K-r=2$ messages $\widetilde{W}_{\{1,2\},3} = \widetilde{A}_3, \widetilde{W}_{\{1,2\},4} = \widetilde{A}_4$, such that Rx-$3$ desires $\widetilde{A}_3$ while Rx-$4$ desires $\widetilde{A}_4$.

For any message $W_{\mathcal{T},\bar{t}}, \mathcal{T} \subset \binom{[K]}{r}, \bar{t} \in [K]\setminus \mathcal{T}$, let $|W_{\mathcal{T},\bar{t}}(P)|$ denote its alphabet size ($P$ is the power constraint). For codewords spanning $\mathsf{T}$ channel uses, the rate associated with $W_{\mathcal{T},\bar{t}}$ is $R_{\mathcal{T},\bar{t}}^{\icmap}(P) \triangleq \frac{\log|W_{\mathcal{T},\bar{t}}(P)|}{\mathsf{T}}$. The rate is achievable if there exists a sequence of coding schemes with the specified rate, such that all the messages can be decoded at their intended Rx's with negligible probability of error. Due to the fact that a message in this MapReduce network corresponds to an intermediate value (IVA) in the original MapReduce setting \cite{Li_Chen_Wang_MapReduce,Bi_Wigger_MapReduce}, and all the IVAs are assumed to have the same number of bits of information, we require that in the MapReduce network, all the messages are transmitted at the same rate, i.e.,
\begin{align}
    R_{\mathcal{T},\bar{t}}^{\icmap}(P) = R^{\icmap}(P), \forall \mathcal{T} \subset \binom{[K]}{r}, \bar{t} \in [K]\setminus \mathcal{T}.
\end{align}
A DoF per message value $d^{\icmap}$ is said to be achievable if there exists an achievable rate $R^{\icmap}(P)$ such that
\begin{align}
    d^{\icmap} = \lim_{P \rightarrow \infty} \frac{R^{\icmap}(P)}{P},
\end{align}
and the DoF per message of this network $d^{\icmap,*}$ is the largest achievable DoF per message value. Since the messages are transmitted at the same rate, the sum-DoF of this network is simply 
\begin{align}
    d_{\Sigma}^{\icmap,*} = (K-r)\binom{K}{r} d^{\icmap,*}.
\end{align}
\begin{remark}
    Compared with the $r$-fold cooperation network in \cite[Section II.B]{Bi_Wigger_MapReduce} whose channel model is specified in \cite[Eq. (5)]{Bi_Wigger_MapReduce} where there is no direct link (link from Tx-$k$ to Rx-$k$), our network with direct links is equivalent since all the messages known by Tx-$k$ are also known to Rx-$k$ as side information and thus any signal transmitted from Tx-$k$ can be eliminated by Rx-$k$. 
\end{remark}

\subsection{$(K,G,M)$ Unicast with Side Information (USI): Inter-message Alignment}\label{subsec:USI}
To properly leverage the side information at each Rx to improve the communication efficiency, we relabel each message in a $(K,r,M)$ MapReduce network and regroup them into new super-messages. We call the setting after the relabeling a $(K,G,M)$ \emph{unicast with side information} (USI) setting with $G = r+1$. Let us first specify what we mean by a $(K,G=r+1,M)$ USI setting, and then show the exact way we relabel the messages.

In a $(K,G=r+1,M)$ USI setting, there are $N_g\triangleq \binom{K}{G}$ groups and $N_g$ independent super-messages, namely $\mathbf{W}_n$,  $\forall n\in [N_g]$. For any $n \in [N_g]$, the $G$ Rx's in group $\mathcal{S}_n$ (as defined in \eqref{eq:groupRx}) correspond to the super-message $\mathbf{W}_n$. Similar to the $(K,G,M)$ BCGM setting, for any $k \in [K]$, let $\mathcal{V}_k$ be the indices of all the super-messages corresponding to Rx-$k$ --- same as defined in \eqref{eq:idxDeRxk}. Each super-message consists of $G$ independent \emph{messages}, i.e., $\mathbf{W}_{n} \triangleq \{W_{n,g}, g \in [G]\}$. The $g^{th}$ Rx in the group $\mathcal{S}_n$, i.e., Rx-$\mathcal{S}_n(g)$, desires message $W_{n,g}$. At the same time, the messages $\mathbf{W}_n \setminus \{W_{n,g}\}$, i.e., all the messages within the super-message $\mathbf{W}_n$ other than $W_{n,g}$ that desired by Rx-$\mathcal{S}_n(g)$, are known to Tx-$\mathcal{S}_n(g)$ and also available to Rx-$\mathcal{S}_n(g)$ as \emph{side-information} prior to the transmission. In total, there are $N_u = GN_g$ messages and each Rx desires a total of $\nu_g=GN_g/K$ messages. For any $n \in \mathcal{V}_k$, let the index of the message within the super-message $\mathbf{W}_n$ that is required by Rx-$k$ be $g_n^k \in [G]$, i.e., 
\begin{align}
    g_n^k \triangleq g \in [G] \mbox{ s.t. } \mathcal{S}_n(g) = k, \forall k \in [K], n \in \mathcal{V}_k.
\end{align}

The set of messages desired by Rx-$k$ is denoted by $\de_k^{\usi}$, and the set of messages available to Rx-$k$ as side-information is denoted by $\si_k^{\usi}$, 
\begin{align}
    \de_k^{\usi} &= \left\{W_{n,g} \mid \forall n \in \mathcal{V}_k, g \in [G] \mbox{ s.t. } \mathcal{S}_n(g) = k\right\}\notag\\
    &= \left\{W_{n,g_n^k} \mid \forall n \in \mathcal{V}_k\right\},\label{eq:deicuni}\\
    \si_k^{\usi} &= \left\{W_{n,\overline{g}} \mid \forall n \in \mathcal{V}_k, \overline{g} \in [G] \mbox{ s.t. } \mathcal{S}_n(\overline{g}) \neq k\right\}\notag\\
    &= \left\{W_{n,\bar{g}} \mid \forall n \in \mathcal{V}_k, \bar{g} \in [G], \bar{g} \neq g_n^k\right\}.\label{eq:sibcuni}
\end{align}
Note that Tx-$k$ only has the messages in $\si_k^{\usi}$, thus $x_k(t)$ only depends on $\si_k^{\usi}$. 

An example of $(K=4, G=3, M)$ USI setting is shown in Fig. \ref{fig:mapRedICuni}b, where we have $N_g=4$ groups $\mathcal{S}_1 = \{1,2,3\}, \mathcal{S}_2 = \{1,2,4\}, \mathcal{S}_3 = \{1,3,4\}, \mathcal{S}_4 = \{2,3,4\}$ and $N_g=4$ super-messages $\mathbf{W}_{1}, \mathbf{W}_{2}, \mathbf{W}_{3}, \mathbf{W}_{4}$, renamed as $\mathbf{A},\mathbf{B},\mathbf{C}, \mathbf{D}$ respectively for convenience, correspond to each group. Each super-message is comprised of $G=3$ messages, each Rx desires  $\nu_g=3$ messages. Take the super-message $\mathbf{W}_{2}$ which is relabeled as ${\bf B}$ for example. It is comprised of $3$ messages $W_{2,1} = B_1, W_{2,2} = B_2, W_{2,3} = B_3$ and corresponds to the group $\mathcal{S}_2 = \{1,2,4\}$. The first Rx within the group, i.e., Rx-$1$ desires $W_{2,1} = B_1$ and knows $W_{2,2} = B_2, W_{2,3} = B_3$. Similarly, the second Rx in $\mathcal{S}_2$, i.e., Rx-$2$ desires $B_2$ and knows $B_1, B_3$, Rx-$4$ which is the $3^{rd}$ Rx in group $\mathcal{S}_2$, desires $B_3$ and knows $B_1, B_2$. Note that Tx-$1$ only knows the messages in the set  $\si_1^{\usi} = \{A_2, A_3, B_2, B_3, C_2, C_3\}$. Thus, the transmitted symbols $x_1(t)$ can only depend on messages $\si_1^{\usi}$.

Let us now  specify how to relabel the messages in $(K,r,M)$ MapReduce to form a $(K,G=r+1,M)$ USI. We have the following lemma.
\begin{lemma}\label{lem:MR2USI}
A $(K,r,M)$ MapReduce setting with super-messages $\widetilde{\mathbf{W}}_{\mathcal{T}}$, $\forall \mathcal{T}\subset \binom{[K]}{r}$ is equivalent to a $(K, G=r+1, M)$ USI setting with super-messages $\mathbf{W}_{n}$, $\forall n \in [N_g]$  following the relabeling,
\begin{align}
	\mathbf{W}_{n} &= \bigg(W_{n, 1}, W_{n, 2}, \cdots, W_{n, G}\bigg)\notag\\
	&\triangleq \bigg(\widetilde{W}_{\mathcal{S}_n\setminus\{\mathcal{S}_n(1)\}, \mathcal{S}_n(1)}, \widetilde{W}_{\mathcal{S}_n\setminus\{\mathcal{S}_n(2)\}, \mathcal{S}_n(2)},\notag\\
    &\quad\quad\quad \cdots, \widetilde{W}_{\mathcal{S}_n\setminus\{\mathcal{S}_n(G)\}, \mathcal{S}_n(G)}\bigg).\label{eq:map2uni}
\end{align}   
\end{lemma}
\proof 
Consider any specific $n,g$ where $n \in [N_g]$ and $g \in [G]$. The set $\mathcal{S}_n\setminus\{\mathcal{S}_n(g)\}$ has cardinality $r = G-1$. The message $\widetilde{W}_{\mathcal{S}_n\setminus\{\mathcal{S}_n(g)\}, \mathcal{S}_n(g)}$ is desired by Rx-$\mathcal{S}_n(g)$. Meanwhile, for all $\bar{g} \in [G], \bar{g} \neq g$, $\widetilde{W}_{\mathcal{S}_{n}\setminus\{\mathcal{S}_n(\bar{g})\}, \mathcal{S}_n(\bar{g})}$ is available to Rx-$\mathcal{S}_n(g)$ as side information, because in the MapReduce setting, $\widetilde{W}_{\mathcal{S}_{n}\setminus\{\mathcal{S}_n(\bar{g})\}, \mathcal{S}_n(\bar{g})}$ that is desired by Rx-$\mathcal{S}_n(\bar{g})$, is available as side information to every Rx whose index is in $\mathcal{T} = \mathcal{S}_{n}\setminus\{\mathcal{S}_n(\bar{g})\}$. Note that $\mathcal{S}_n(g) \in \mathcal{S}_{n}\setminus\{\mathcal{S}_n(\bar{g})\}$. Thus, according to the mapping, for any $g \in [G]$, $W_{n, g} = \widetilde{W}_{\mathcal{S}_n\setminus\{\mathcal{S}_n(g)\}, \mathcal{S}_n(g)}$ is desired by Rx-$\mathcal{S}_n(g)$, while $\mathbf{W}_{n}\setminus\{W_{n,g}\} = \eqref{eq:map2uni} \setminus \{\widetilde{W}_{\mathcal{S}_n\setminus\{\mathcal{S}_n(g)\}, \mathcal{S}_n(g)}\}$ are known to Rx-$\mathcal{S}_n(g)$, thus establishing the equivalence between the two settings. 
Also, note that the mapping is one to one, as there are $(K-r)\binom{K}{r} = \frac{K!}{r!(K-r-1)!}$ messages in $(K,r,M)$ MapReduce, which is equal to the number of messages, $G\binom{K}{G} = \frac{K!}{(G-1)!(K-G)!}$, in a $(K,G,M)$ USI setting, with $G = r + 1$.$\hfill\blacksquare$

\begin{remark}
    Since USI is just a MapReduce with relabeled messages, the rates, DoF per message, sum-DoF of a USI setting are defined, and DoF results of USI translate to MapReduce. Thus, we only need to study USI.
\end{remark}

\subsubsection{Example 3: Inter-message Alignment for $(K=4, G=3, M=1)$ USI}\label{sec:inter_alignment_only}
Let us consider the $(K=4, G=3, M=1)$ USI setting whose network topology is shown in Fig. \ref{fig:mapRedICuni}b as a simple example, in order to make the main idea of inter-message alignment intuitively transparent. Since we have $M = 1$, there are no reconfigurable antennas, and therefore no intra-message alignment to be realized by antenna mode switching, which  greatly simplifies the setting for this example. Let us show that the sum-DoF value equal to $G=3$ can be achieved in this example. Firstly, the $3$ messages in super-message $\mathbf{A}$, namely $A_1, A_2, A_3$ which are circled with red in Fig. \ref{fig:mapRedICuni}b, are simultaneously transmitted by Tx-$2$, Tx-$3$, Tx-$1$ respectively. All $4$ receivers obtain some linear combination of $A_1,A_2, A_3$, and by subtracting their respective side-information, Rx-$1,2,3$ are able to retrieve their desired symbols $A_1,A_2,A_3$, respectively. Meanwhile, the $3$ messages from super-message $\mathbf{A}$ are interference to Rx-$4$, where they are aligned into one linear combination due to the superposition property of the channel. This is a simple case of \emph{inter-message alignment}. Super-messages $\mathbf{B}, \mathbf{C}, \mathbf{D}$ can be delivered to their destinations similarly, and the sum-DoF achieved is $G=3$.

The alignment of $A_1, A_2, A_3$ reflects back to the alignment of $\widetilde{D}_1, \widetilde{B}_2, \widetilde{A}_3$ in the original MapReduce setting in Fig. \ref{fig:mapRedICuni}a. The relabeling makes the inter-message alignment based on the side information structure transparent.

\begin{remark}\label{rmk:inter_alignment_only}
    It is not difficult to see that the sum-DoF $G$ is achievable for arbitrary $(K,G,M=1)$ USI setting by every time transmitting the $G$ messages within a super-message simultaneously. Since $(K,r,M)$ MapReduce is equivalent to $(K,G=r+1,M)$ USI, this shows that the sum-DoF $r+1$ is achievable in a wireless MapReduce problem with no CSIT, with only conventional receiver antennas ($M=1$). Notably, the state-of-the-art MapReduce scheme that requires neither CSIT nor reconfigurable antenna in \cite{ha2019wireless} only achieves the sum-DoF $r$. The improvement from $r$ to $r+1$ is based on alignment of messages originating across different transmitters (e.g., there in no single transmitter has all of $A_1, A_2, A_3$), whereas \cite{ha2019wireless} employs inter-message alignment within messages originating at the same transmitter.
\end{remark}

\subsection{DoF of the USI Setting}\label{sec:DoFUSI}
It is now increasingly apparent that a $(K,G,M)$ USI setting is similar to a $(K,G,M)$ BCGM setting. Specifically, we may view every message $W_n$ in a BCGM setting as the aligned version of the super-message $\mathbf{W}_n$ in  USI. Intuitively, in USI, if $\sum_{g \in [G]}W_{n,g}$ is received by the $G$ Rxs in the corresponding group, they will all be able to decode their desired messages by subtracting their side information out. We will show that the BIA precoding scheme for a $(K,G,M)$ BCGM setting yields a precoding scheme for a $(K,G,M)$ USI setting when $M \leq G-1$. Intuitively, in USI, every message will be precoded and broadcast from the first $M$ Txs who have that message, and messages belong to the same super-message will be aligned due to the fact that they will be precoded with the same precoding matrix. Specifically, we have the following theorem.
\begin{theorem}\label{thm:USI}
    For the $(K,G,M)$ USI, the sum-DoF 
    \begin{align}
        &d_{\Sigma}^{\usi,*} = \frac{GN_g M}{(M-1)\nu_g + N_g} && M \leq G-1,\\
        &\frac{GN_g (G-1)}{(G-2)\nu_g + N_g} \leq d_{\Sigma}^{\usi,*} \leq \frac{GN_g M}{(M-1)\nu_g + N_g}  && M \geq G.\label{eq:thm_USI}
    \end{align}
\end{theorem}

Compared with the sum-DoF of BCGM specified in Theorem \ref{thm:BCGM}, there is an extra gain by a factor of $G$. Intuitively, this is due to the inter-message alignment scheme specified in Section \ref{sec:inter_alignment_only} that efficiently utilizes the side information available to each user. 

\begin{figure}
    \centering
    \includegraphics{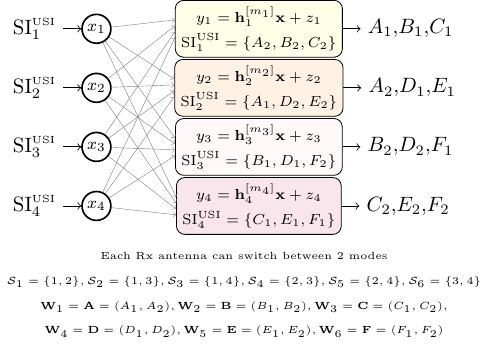}
    \caption{$(K=4, G=2,M=2)$ USI}\label{fig:ICuni_open}
\end{figure}

\begin{remark} How to optimally exploit reconfigurable antennas jointly with side information is still an open problem in the regime $M \geq G$. In this regime, though each receive antenna has $M$ modes, each message is only available to $G-1$ transmit antennas, which is strictly less than $M$. Though we can still combine the inter-message alignment in Section \ref{sec:inter_alignment_only} with the intra-message alignment in the achievable scheme of Corollary \ref{cor:BCGM} where $\min\left(M_{\mbox{\tiny Tx}} = G-1, M_{\mbox{\tiny Rx}} = M\right) = G-1$, to obtain an achievable scheme, its optimality remains open. As a concrete example, a $(K=4, G=2, M=2)$ USI setting that falls into this regime is shown in Fig. \ref{fig:ICuni_open}. In this case, the sum-DoF value that is achieved is simply $G=2$, which only exploits the side information (inter-message alignment). Since each message is only known by just one transmitter, each antenna only utilizes one mode. Thus, the reconfigurability of the receive antennas is not fully utilized.
\end{remark}

Let us now prove the achievability by applying the BIA precoding scheme for $(K,G,M)$ BCGM to $(K,G,M)$ USI. The converse will be proved in Appendix \ref{app:converse_USI}.

\subsection{Proof of Achievability for Theorem \ref{thm:USI}}\label{sec:achthmusi1}
\subsubsection{Achievability for the Case: $M \leq G-1$}\label{subsec:mleq}
Let $\mathsf{T}_p, \ell$, the precoding matrices $\mathbb{V}_n, n \in [N_g]$, the matrices $\mathbb{E}_{k,n}, \mathbb{D}_{k,n}, k \in [K], n \in [N_g]$ and the switching pattern $\mathbf{m}_k, k \in [K]$ be the same as those in the achievable scheme for the BCGM setting. For any $n \in [N_g], g\in [G]$, message $W_{n, g}$ is encoded into $L = M\ell$ independent streams $\mathbf{w}_{n,g} = [W_{n,g}^1 ~~ \cdots ~~ W_{n,g}^{L}]^\top$. Let $\mathcal{M}_{n,g} \subset [K], |\mathcal{M}_{n,g}| = M$ denote the first $M$ Tx's who have the message $W_{n,g}$. Let us introduce the following definition.
\begin{definition}\label{def:XM}
    $\forall n \in [N_g], g \in [G]$, let $\mathbf{X}_{n,g} \in \mathbb{C}^{K\mathsf{T}_p \times L}$ be defined to be the overall signal transmitted by the $K$ Tx's over $\mathsf{T}_p$ channel uses when the Tx's in $\mathcal{M}_{n,g}$ simulate the $M$ transmit antennas in BCGM setting and transmit $\mathbb{V}_n \mathbf{w}_{n,g}$, with $\mathbb{V}_n$ as specified in \eqref{eq:BIA_BCGM_X}, while the remaining $K - M$ Tx's transmit nothing. 
\end{definition}

Let the signal transmitted by the $K$ Tx's over $\mathsf{T}_p$ be 
\begin{align}
    \mathbf{X} = \sum_{n \in [N_g]}\sum_{g \in [G]} \mathbf{X}_{n,g} \in \mathbb{C}^{K\mathsf{T}_p \times L},
\end{align}
then following the switching pattern $\mathbf{m}_k$, Rx-$k, k\in [K]$ received signal can be represented as \eqref{eq:USI_simulatesM3} 
\begin{figure*}[htbp]
\begin{align}
    &\mathbf{y}_k = \mbox{Block-Diag}(\mathbf{h}_k^{[m_k(1)]}, \mathbf{h}_k^{[m_k(2)]}, \cdots, \mathbf{h}_k^{[m_k(\mathsf{T}_p)]})\mathbf{X} + \mathbf{z}_k\\
    &= \sum_{n \in [N_g]}\sum_{g \in [G]}\mbox{Block-Diag}(\mathbf{h}_k^{[m_k(1)]}, \mathbf{h}_k^{[m_k(2)]}, \cdots, \mathbf{h}_k^{[m_k(\mathsf{T}_p)]}) \mathbf{X}_{n,g} + \mathbf{z}_k\\
    &= \sum_{n \in [N_g]}\sum_{g \in [G]}\mbox{Block-Diag}\left(\left(\mathbf{h}_k^{[m_k(1)]}\right)_{\mathcal{M}_{n,g}}, \left(\mathbf{h}_k^{[m_k(2)]}\right)_{\mathcal{M}_{n,g}}, \cdots, \left(\mathbf{h}_k^{[m_k(\mathsf{T}_p)]}\right)_{\mathcal{M}_{n,g}}\right) \mathbb{V}_n \mathbf{w}_{n,g} + \mathbf{z}_k\label{eq:USI_simulatesM1}\\
    &=\sum_{n \in [N_g]}\sum_{g \in [G]} \mathbb{E}_{k,n} \mathbb{H}_{k,n,g} \mathbf{w}_{n,g} + \mathbf{z}_k\label{eq:USI_simulatesM2}\\
    &=\sum_{n \in [\mathcal{V}_k]}\mathbb{E}_{k,n}\underbrace{\sum_{g \in [G]}\mathbb{H}_{k,n,g}\mathbf{w}_{n,g}}_{\mbox{desired $+$ side information}} + \sum_{n \in [N_g]\setminus\mathcal{V}_k} \mathbb{E}_{k,n}\underbrace{\sum_{g \in [G]}\mathbb{H}_{k,n,g}\mathbf{w}_{n,g}}_{\mbox{interference}} + \mathbf{z}_k\label{eq:USI_simulatesM3}
\end{align}
\hrule
\end{figure*}
where $\mathbf{z}_k \in \mathbb{C}^{\mathsf{T}_p \times 1}$ is the AWGN at Rx-$k$, \eqref{eq:USI_simulatesM1} results from the fact that in $\mathbf{X}_{n,g}$, only Tx's in $\mathcal{M}_{n,g}$ transmit $\mathbb{V}_n \mathbf{w}_{n,g}$ while all other Tx's transmit nothing according to Definition \ref{def:XM}, \eqref{eq:USI_simulatesM2} follows from the same reasoning as used from \eqref{eq:BIA_BCGM_YV} to \eqref{eq:BIA_BCGM_Y}, and we have defined, 
\begin{align}
    &\mathbf{H}_{k,n,g} = 
    \begin{bmatrix}
        \left(\mathbf{h}_{k}^{[1]}\right)_{\mathcal{M}_{n,g}}\\
        \left(\mathbf{h}_{k}^{[2]}\right)_{\mathcal{M}_{n,g}}\\
        \vdots\\
        \left(\mathbf{h}_{k}^{[M]}\right)_{\mathcal{M}_{n,g}}\\
    \end{bmatrix} \in \mathbb{C}^{M\times M},\notag\\
    &\mathbb{H}_{k,n,g} = 
    \begin{bmatrix}
        \mathbf{H}_{k,n,g} & & &\\
        & \mathbf{H}_{k,n,g} & &\\
        & & \ddots & \\
        & & & \mathbf{H}_{k,n,g}
    \end{bmatrix} \in \mathbb{C}^{M\ell \times M\ell}.
\end{align}

For any $n \in \mathcal{V}_k$, Rx-$k$ can apply the $\mathbb{D}_{k,n}^\top$ to $\mathbf{y}_k$ in \eqref{eq:USI_simulatesM3} and get 
\begin{align}
    &\mathbb{D}_{k,n}^\top\mathbf{y}_k = \sum_{g \in [G]}\mathbb{H}_{k,n,g} \mathbf{w}_{n,g} + \mathbb{D}_{k,n}^\top \mathbf{z}_k\notag\\
    &= \underbrace{\mathbb{H}_{k,n,g_n^k} \mathbf{w}_{n,g_n^k}}_{\mbox{desired}} + \underbrace{\sum_{g \in [G], g \neq g_n^k}\mathbb{H}_{k,n,g} \mathbf{w}_{n,g}}_{\mbox{side information}} + \underbrace{\mathbb{D}_{k,n}^\top \mathbf{z}_k}_{\mbox{AWGN}}\label{eq:BIA_IC_projection}
\end{align}

Note that for the super-message $\mathbf{W}_n, n \in \mathcal{V}_k$, Rx-$k$ knows $W_{n,g}, g \neq g_n^k$ as side information. It can thus subtract the known symbols out from \eqref{eq:BIA_IC_projection}. Again, for any block of $M$ symbols of $W_{n,g_n^k}$, Rx-$k$ sees a point to point AWGN $M\times M$ MIMO channel, and the DoF per message achieved is $d^{\usi} = M\ell/\mathsf{T}_p$. The sum-DoF achieved is thus $d_{\Sigma}^{\usi} = GN_gd = GN_gM/\left((M-1)\nu_g + N_g\right)$.

\subsubsection{Example 4: Converting the BIA  Scheme from BCGM to USI}\label{sec:ex4}
Let us show how to combine the BIA precoding scheme for the $(K=4, G=3, M=2)$ BCGM setting shown in Fig. \ref{fig:mapRedmot}b with the inter-message alignment mentioned in Section \ref{sec:inter_alignment_only} to get an efficient precoding scheme for the corresponding USI setting shown in Fig. \ref{fig:mapRedICuni}b.  Each message is encoded into $2$ streams of coded symbols as follows, 
\begin{align}
	&A_i \rightarrow [A_i^1~~ A_i^2] && B_i \rightarrow [B_i^1~~ B_i^2]\notag\\
    &C_i \rightarrow [C_i^1~~ C_i^2] && D_i \rightarrow [D_i^1~~ D_i^2].
\end{align}
Transmitters Tx-$1,2,3,4$ then prepare $\bm{W}^1, \bm{W}^2, \bm{W}^3, \bm{W}^4$, respectively, as follows.
\begin{align}
	\bm{W}^1&\triangleq\begin{bmatrix}{\color{red}{A}_2^1}+{\color{blue}{A}_3^1} & {B}_2^1+{B}_3^1 & {C}_2^1+ {C}_3^1 & 0\end{bmatrix}^\top\notag\\
    &\triangleq \begin{bmatrix}w_{1,A} & w_{1,B} & w_{1,C} & w_{1,D}\end{bmatrix}^\top\notag\\
	\bm{W}^2&\triangleq \begin{bmatrix}{\color{violet}{A}_1^1}+{\color{blue}{A}_3^2} & {B}_1^1+{B}_3^2 & 0 & {D}_2^1+{D}_3^1\end{bmatrix}^\top\notag\\
    &\triangleq \begin{bmatrix}w_{2,A} & w_{2,B} & w_{2,C} & w_{2,D}\end{bmatrix}^\top\notag\\
	\bm{W}^3&\triangleq\begin{bmatrix}{\color{violet}{A}_1^2} + {\color{red}{A}_2^2} & 0 & {C}_1^1+ {C}_3^2 & {D}_1^1+{D}_3^2\end{bmatrix}^\top\notag\\
    &\triangleq \begin{bmatrix}w_{3,A} & w_{3,B} & w_{3,C} & w_{3,D}\end{bmatrix}^\top\notag\\
	\bm{W}^4&\triangleq\begin{bmatrix}0 & {B}_1^2+{B}_2^2 & {C}_1^2+{C}_2^2 & {D}_1^2+{D}_2^2\end{bmatrix}^\top\notag\\
    &\triangleq \begin{bmatrix}w_{4,A} & w_{4,B} & w_{4,C} & w_{4,D}\end{bmatrix}^\top\label{eq:K4G3Tinfo}
\end{align}

The precoding structure of the achievable scheme is shown over $7$ channel-uses in Fig. \ref{fig:USI_K4G3M2_precoding} where the noise is ignored for ease of understanding. It is not difficult to see the precoding structure is quite similar to the one in in Fig. \ref{fig:BCGM_K4G3M2_precoding}. Again, let $\bm{\lambda}_1 = [{\color{red}1}~~{\color{red}1}~~{\color{red}1}~~{\color{red}1}]$, $\bm{\lambda}_2 = [{\color{red}1}~~{\color{red}2}~~{\color{red}3}~~{\color{red}4}]$, $\bm{\lambda}_3= [{\color{red}1}~~{\color{red}4}~~{\color{red}9}~~{\color{red}16}]$. The blue blocks mean that the corresponding receivers operate in mode-$1$, while the red blocks mean that the corresponding receivers operate in mode-$2$.
\begin{figure*}[htbp]
\begin{align*}
\arraycolsep=1.4pt\def\arraystretch{1.75}
\scalebox{0.9}{$
\begin{array}{r|c|c|c|c|c|c|c|l}
\multicolumn{1}{c}{}&\multicolumn{1}{c}{t=1}&\multicolumn{1}{c}{t=2}&\multicolumn{1}{c}{t=3}&\multicolumn{1}{c}{t=4}&\multicolumn{1}{c}{t=5}&\multicolumn{1}{c}{t=6}&\multicolumn{1}{c}{t=7}\\\cline{2-8}
x_1&\bm{\lambda}_1 \bm{W}^1&\bm{\lambda}_2 \bm{W}^1&\bm{\lambda}_3 \bm{W}^1&w_{1,A}&w_{1,B}&w_{1,C}&w_{1,D}\\\cline{2-8}
x_2&\bm{\lambda}_1 \bm{W}^2&\bm{\lambda}_2 \bm{W}^2&\bm{\lambda}_3 \bm{W}^2&w_{2,A}&w_{2,B}&w_{2,C}&w_{2,D}\\\cline{2-8}
x_3&\bm{\lambda}_1 \bm{W}^3&\bm{\lambda}_2 \bm{W}^3&\bm{\lambda}_3 \bm{W}^3&w_{3,A}&w_{3,B}&w_{3,C}&w_{3,D}\\\cline{2-8}
x_4&\bm{\lambda}_1 \bm{W}^4&\bm{\lambda}_2 \bm{W}^4&\bm{\lambda}_3 \bm{W}^4&w_{4,A}&w_{4,B}&w_{4,C}&w_{4,D}\\\cline{2-8}
y_1&
\cellcolor{blue!10}\bm{\lambda}_1 [a^{1,[1]}~~\cdots~~d^{1,[1]}]^\top
&\cellcolor{blue!10}\bm{\lambda}_2 [a^{1,[1]}~~\cdots~~d^{1,[1]}]^\top 
&\cellcolor{blue!10}\bm{\lambda}_3 [a^{1,[1]}~~\cdots~~d^{1,[1]}]^\top 
&\cellcolor{red!10}a^{1,[2]}&\cellcolor{red!10}b^{1,[2]}&\cellcolor{red!10}c^{1,[2]}&\cellcolor{blue!10}d^{1,[1]}&\rightarrow (a^{1,[m]},b^{1,[m]},c^{1,[m]})_{m\in[2]}\\\cline{2-8}
y_2&
\cellcolor{blue!10}\bm{\lambda}_1 [a^{2,[1]}~~\cdots~~d^{2,[1]}]^\top
&\cellcolor{blue!10}\bm{\lambda}_2 [a^{2,[1]}~~\cdots~~d^{2,[1]}]^\top
&\cellcolor{blue!10}\bm{\lambda}_3 [a^{2,[1]}~~\cdots~~d^{2,[1]}]^\top
&\cellcolor{red!10}a^{2,[2]}&\cellcolor{red!10}b^{2,[2]}&\cellcolor{blue!10}c^{2,[1]}&\cellcolor{red!10}d^{2,[2]}&\rightarrow (a^{2,[m]},b^{2,[m]},d^{2,[m]})_{m\in[2]}\\\cline{2-8}
y_3&
\cellcolor{blue!10}\bm{\lambda}_1 [a^{3,[1]}~~\cdots~~d^{3,[1]}]^\top
&\cellcolor{blue!10}\bm{\lambda}_2 [a^{3,[1]}~~\cdots~~d^{3,[1]}]^\top
&\cellcolor{blue!10}\bm{\lambda}_3 [a^{3,[1]}~~\cdots~~d^{3,[1]}]^\top
&\cellcolor{red!10}a^{3,[2]}&\cellcolor{blue!10}b^{3,[1]}&\cellcolor{red!10}c^{3,[2]}&\cellcolor{red!10}d^{3,[2]}&\rightarrow (a^{3,[m]},c^{3,[m]},d^{3,[m]})_{m\in[2]}\\\cline{2-8}
y_4&
\cellcolor{blue!10}\bm{\lambda}_1 [a^{4,[1]}~~\cdots~~d^{4,[1]}]^\top
&\cellcolor{blue!10}\bm{\lambda}_2 [a^{4,[1]}~~\cdots~~d^{4,[1]}]^\top
&\cellcolor{blue!10}\bm{\lambda}_3 [a^{4,[1]}~~\cdots~~d^{4,[1]}]^\top
&\cellcolor{blue!10}a^{4,[1]}&\cellcolor{red!10}b^{4,[2]}&\cellcolor{red!10}c^{4,[2]}&\cellcolor{red!10}d^{4,[2]}&\rightarrow (b^{4,[m]},c^{4,[m]},d^{4,[m]})_{m\in[2]}\\\cline{2-8}
\end{array}$}
\end{align*}
\caption{Precoding scheme for the $(4,3,2)$ USI setting in Fig. \ref{fig:mapRedICuni}b.}
\label{fig:USI_K4G3M2_precoding}
\end{figure*}

Let us intuitively explain how this scheme works. First, recall that this is a unicast with side information problem. For example, $A_1, A_2, A_3$ need to be unicast to Rx-$1,2,3$, who have $(A_2, A_3)$, $(A_1, A_3)$, $(A_1, A_2)$ as side information, respectively. 

Then let us establish the connection between the precoding scheme in Fig. \ref{fig:USI_K4G3M2_precoding} and the one in Fig. \ref{fig:BCGM_K4G3M2_precoding} through a thought experiment. Suppose $A_2 = A_3 = B_2 = B_3 = C_2 = C_3 = D_2 = D_3 = 0$, while the remaining  $A_1, B_1, C_1, D_1$ correspond to $A,B,C,D$ in the BCGM setting. Consider the message $A_1$, we have $w_{1,A} = w_{4,A} = 0$ and $w_{2,A} = {\color{violet}A_1^1}, w_{3,A} = {\color{violet}A_1^2}$. Thus, $A_1$ is precoded by  Tx-$2,3$ (see $x_2, x_3$ in Fig. \ref{fig:USI_K4G3M2_precoding}) the same way as $A$ is precoded by the $M=2$ transmitter antennas in Fig. \ref{fig:BCGM_K4G3M2_precoding}. Similarly, $B_1$ is precoded by  Tx-$2,4$ ($w_{1,B} = w_{3,B} = 0$) the same way as $B$ precoded by the two transmitter antennas in Fig. \ref{fig:BCGM_K4G3M2_precoding}. Similarly, $C_1, D_1$ are  precoded by two transmitters the same way as $C, D$ are precoded in the BCGM setting. Thus, we should expect, e.g., $A_1$ is {\bf groupcast} to Rx-$\{1,2,3\}$ just as $A$ in BCGM. Similar reasoning applies to $B_1, C_1, D_1$. We may repeat the thought experiment by only considering $A_2, \cdots, D_2$ while all other messages are $0$. Under this case, $A_2, B_2, C_2, D_2$ are groupcast. Similarly, $A_3, \cdots, D_3$ are groupcast if all other messages are $0$. 

Once the connection between USI and BCGM is clear, the inter-message alignment can be easily shown. Note that since the messages that belongs to the same super-message, say, $A_1, A_2, A_3$, are precoded in the same way by a corresponding set of $M=2$ transmitters, they are \emph{mixed} at the receivers' side. Thus, as interfering messages at Rx-$4$, they are \emph{inter-aligned}. On the other hand, at Rx-$1$, $A_1$ can be decoded by subtracting $A_2, A_3$ which are available as side information, from the mixed $A_1, A_2, A_3$. The decoding of $A_2, A_3$ at Rx-$2,3$, respectively, is similarly achieved.

The precise forms of the symbols, e.g., $w_{1,A}, w_{2,A}, w_{3,A}$ are determined as follows. Each message will be precoded at the (first) $M=2$ transmitters who have it. For example, $A_2$ is precoded by Tx-$1$ and Tx-$3$, and thus $A_2^1, A_2^2$ are included in $w_{1,A}, w_{3,A}$ respectively which are marked red in \eqref{eq:K4G3Tinfo}. Similarly, $A_3^1, A_3^2$ which are marked blue are included in $w_{1,A}, w_{2,A}$ respectively, as Tx-$1$ and Tx-$2$ are the first two that have $A_3$. Following the same rule, $A_1^1$ and $A_1^2$ are involved in $w_{2,A}, w_{3,A}$ respectively.

Let us then analyze the received signal at Rx-$k$. Note that if at a time slot $t, t\in[3]$, Tx-$1,2,3,4$ send $\bm\lambda_t\bm{W}^1, \bm\lambda_t \bm{W}^2, \bm\lambda_t \bm{W}^3, \bm\lambda_t \bm{W}^4$, respectively, then Rx-$k$, operating in antenna mode $m$, receives (ignore AWGN),
\begin{align}
y_k(t) 
&=
\mathbf{h}_k^{[m]}
\begin{bmatrix}
    \bm\lambda_t\bm{W}^1 & \bm\lambda_t\bm{W}^2 & \bm\lambda_t\bm{W}^3 & \bm\lambda_t\bm{W}^4
\end{bmatrix}^\top\\
&=
\mathbf{h}_k^{[m]}
\begin{bmatrix}
    \bm{W}^1 & \bm{W}^2 & \bm{W}^3 & \bm{W}^4
\end{bmatrix}^\top \bm\lambda_t^\top\notag\\
&=
\mathbf{h}_{k}^{[m]}
\begin{bmatrix}
	w_{1,A} & w_{1,B} & w_{1,C} & w_{1,D}\\
	\hline
	w_{2,A} & w_{2,B} & w_{2,C} & w_{2,D}\\
	\hline
	w_{3,A} & w_{3,B} & w_{3,C} & w_{3,D}\\
	\hline
	w_{4,A} & w_{4,B} & w_{4,C} & w_{4,D}\\
\end{bmatrix}
\bm\lambda_t^\top\\
&\triangleq
\begin{bmatrix}
	a^{k,[m]} & b^{k,[m]} & c^{k,m} & d^{k,[m]}
\end{bmatrix}
\bm\lambda_t^\top
\end{align}
where we define 
\begin{align}
	a^{k,[m]} &\triangleq 
 \mathbf{h}_{k}^{[m]}
 \begin{bmatrix}
     w_{1,A} & w_{2,A} & w_{3,A} & w_{4,A}
 \end{bmatrix}^\top\\
 &\overset{\eqref{eq:K4G3Tinfo}}{=}
 \left(\mathbf{h}_k^{[m]}\right)_{\{2,3\}}
 \begin{bmatrix}
    {\color{violet}{A}_1^1}\\{\color{violet}{A}_1^2}
 \end{bmatrix}
 +
 \left(\mathbf{h}_k^{[m]}\right)_{\{1,3\}}
 \begin{bmatrix}
    {\color{red}{A}_2^1}\\{\color{red}{A}_2^2}
 \end{bmatrix}\notag\\
 &\quad\quad+
 \left(\mathbf{h}_k^{[m]}\right)_{\{1,2\}}
 \begin{bmatrix}
    {\color{blue}{A}_1^1}\\{\color{blue}{A}_1^2}
 \end{bmatrix}, \forall m \in [2],\label{eq:USI_K4G3_infoA}
\end{align}
i.e., a linear combination of $w_{1,A}, \cdots, w_{4,A}$ determined by the channel operated in mode $m$ at Rx-$k$. $b^{k,[m]}, c^{k,[m]}, d^{k,[m]}$ are similarly defined.

Also note that, if at time slot $t$, Tx-$1,2,3,4$ send $w_{1,A}, w_{2,A}, w_{3,A}, w_{4,A}$, respectively, then Rx-$k$, operating in antenna mode $m$, receives (ignore AWGN),
\begin{align}
y_k(t) &=
 \mathbf{h}_{k}^{[m]}
 \begin{bmatrix}
     w_{1,A} & w_{2,A} & w_{3,A} & w_{4,A}
 \end{bmatrix}^\top = a^{k,[m]}.
\end{align}
$b^{k,[m]}, c^{k,[m]}, d^{k,[m]}$ are received by Rx-$k$ similarly.

For the decodability of the desired message, let us take Rx-$1$ for example. Since the structure of the precoding scheme in Fig. \ref{fig:USI_K4G3M2_precoding} is the same as the precoding scheme in Fig. \ref{fig:BCGM_K4G3M2_precoding}, we know that Rx-$1$ is able to decode $(a^{1,[m]}, b^{1,[m]}, c^{1,[m]})_{m \in [2]}$. Take message $A_1$ for example. Rx-$1$ can subtract the side information $A_2, A_3$ from $a^{k,[1]}, a^{k,[2]}$ and, according to \eqref{eq:USI_K4G3_infoA}, obtains,
\begin{align}
\begin{bmatrix}
    \tilde{a}^{1,[1]}\\
    \tilde{a}^{1,[2]}
\end{bmatrix}
=
\begin{bmatrix}
    \left(\mathbf{h}_1^{[1]}\right)_{\{2,3\}}\\
    \left(\mathbf{h}_1^{[2]}\right)_{\{2,3\}}\\
\end{bmatrix}
\begin{bmatrix}
    {\color{violet}A_3^1}\\
    {\color{violet}A_3^2}
\end{bmatrix}
\end{align}
The channel matrix above is invertible almost surely because the channel vectors associated with the two modes are generic. Thus $A_1^1, A_1^2$ can be decoded. All desired messages are decoded similarly. The DoF achieved for each message is $2/7$.

\subsubsection{Achievability for the case $M \geq G$}
For $M \geq G$, since each message is only available to $M_{\mbox{\tiny Tx}} = G-1$ transmit antennas, which is strictly less than each receive antenna's modes $M_{\mbox{\tiny Rx}} = M$, we always have the choice to utilize the intra-message alignment structure from Corollary \ref{cor:BCGM} by letting the receivers switch among $M' = \min\left(M_{\mbox{\tiny Tx}}, M_{\mbox{\tiny Rx}}\right) = G-1$ modes, and apply the scheme of Section \ref{subsec:mleq}.

\subsection{DoF of MapReduce}\label{sec:DoFmapreduce}
Combining Theorem \ref{thm:USI} and Lemma \ref{lem:MR2USI} (the equivalence of $(K,r,M)$ MapReduce and $(K,G=r+1,M)$ USI), we have the following corollary that states our DoF result for the symmetric MapReduce problem defined in this paper, where there are $\binom{K}{r}$ super-messages (files), each of which is assigned to $r$ out of $K$ users and contains $K-r$ messages (IVAs to be sent during shuffle phase) to be sent to the remaining users.
\begin{corollary}\label{cor:MR}
    For the $(K,r,M)$ symmetric MapReduce problem defined in Section \ref{subsec:MapReduce_Form}, the sum-DoF 
\begin{align}
d_{\Sigma}^{\icmap,*} = \frac{(r+1)\binom{K}{r+1}M}{(M-1)\binom{K-1}{r}+\binom{K}{r+1}} = \frac{K(r+1)M}{(M-1)(r+1) + K},&\notag\\
\hfill\mbox{for}~M \leq r;&\notag\\
\frac{K(r+1)r}{(r-1)(r+1) + K} \leq d_{\Sigma}^{\icmap,*} \leq \frac{K(r+1)M}{(M-1)(r+1) + K},&\notag\\
\mbox{for}~ M \geq r+1.&\label{eq:MRDoF}
\end{align}
\end{corollary}
Corollary \ref{cor:MR} is acquired by directly setting $G=r+1$ in \eqref{eq:thm_USI}.

As a special case with $M=1$, i.e., allowing only conventional antennas, we  obtain the following corollary for the symmetric MapReduce problem with no CSIT.
\begin{corollary}\label{cor:MR_conventional}
    The sum-DoF of the symmetric $(K,r,M=1)$ MapReduce network defined in Section \ref{subsec:MapReduce_Form} is $r+1$.
\end{corollary}

\section{Conclusion}\label{sec:conclusion}
We found the sum-DoF of a MISO BC with groupcast messages (BCGM) without CSIT, where each receiver is equipped with a reconfigurable antenna. The coding scheme was then applied to the wireless MapReduce network, with the unicast with side information (USI) setting as an intermediate step for identifying the desired inter-message alignments. Considering that BIA schemes harness the power of interference alignment with no CSIT overhead, exploiting the benefits of side-information to further strengthen such schemes is a promising path towards discovering robust interference management schemes for wireless networks when used for distributed computation tasks. New achievable schemes emerge from this study, that require both inter-message and intra-message alignments. The progress made here points the way for the next steps, exploring DoF when the channel coherence time and side-information are further restricted.

\appendices
\section{Proof of Lemma \ref{lem:dimension_received}}\label{app:dimension_received}
Here we  prove that the $\mathbf{p}_k$ generated by the Algorithm \ref{alg:pk} satisfies \eqref{eq:alignSepSuff}. Let us explain the way we generate $\mathbf{p}_k$. Essentially, the first $n$ messages determine the first $(M-1)^n$ entries of $\mathbf{p}_k$. 

First of all, \eqref{eq:alignSepSuff} is true for $n=1, h = 0$. Note that $(M-1)^{1-1} = 1$ and thus $j \in [1:1]$ and
\begin{align}
    1 = t_1^{0,0,1}, 2 = t_1^{0,1,1}, \cdots, M-1 = t_1^{0,M-2,1}.
\end{align}
Combining with the initialization of Algorithm \ref{alg:pk}, \eqref{eq:alignSepSuff} is true.

Let us now prove \eqref{eq:alignSepSuff} is true for arbitrary $n \in [N_g], h = 0, j \in [(M-1)^{n-1}]$. Suppose the first $(M-1)^{n-1}$ entries of $\mathbf{p}_k$ have already been determined by the first $n-1$ messages. Then the $n^{th}$ message enters the outer `while loop' in the algorithm. Suppose $W_n$ is not Rx-$k$'s desired message, i.e., $n \in [N_g] \setminus \mathcal{V}_k$. The first $(M-1)^n$ entries of $\mathbf{p}_k$ are shown in Table \ref{tab:repeat_interference}. Note that in Table \ref{tab:repeat_interference}, the first sub-table that corresponds to the first $(M-1)^{n-1}$ entries of $\mathbf{p}_k$, is  determined by the previous $n-1$ messages. The next $M-2$ sub-tables are just repetitions of the first sub-table which are determined by the $n^{th}$ message. Column-wise, $p_k(t)$ remains unchanged.

\begin{table*}
    \centering
    \caption{The first $(M-1)^n$ entries of $\mathbf{p}_k$ when $n \in [N_g] \setminus \mathcal{V}_k$}
    \resizebox{0.9\textwidth}{!}{
    \begin{tabular}{|c|c|c|c|c|c|}
    \hline
    $t$ & $1 = t_n^{0,0,1}$ & $2 = t_n^{0,0,2}$ & $\cdots$ & $(M-1)^{n-1} = t_n^{0,0,(M-1)^{n-1}}$ & \multirow{2}{*}{determined}\\\cline{1-5}  
    $p_k(t)$ & $p_k(1)$ & $p_k(2)$ & $\cdots$ & $p_k\big((M-1)^{n-1}\big)$ & \\\hline\hline
    $t$ & $(M-1)^{n-1} + 1 = t_n^{0,1,1}$ & $(M-1)^{n-1} + 2 = t_n^{0,1,2}$ & $\cdots$ & $2(M-1)^{n-1} = t_n^{0,1,(M-1)^{n-1}}$ & \multirow{2}{*}{repeat} \\\cline{1-5}  
    $p_k(t)$ & $p_k(1)$ & $p_k(2)$ & $\cdots$ & $p_k\big((M-1)^{n-1}\big)$ & \\\hline\hline
    $\vdots$ & $\vdots$ & $\vdots$ & $\vdots$ & $\vdots$ & $\vdots$\\\hline\hline
    $t$ & $(M-2)(M-1)^{n-1} + 1 = t_n^{0,M-2,1}$ & $(M-2)(M-1)^{n-1} + 2 = t_n^{0,M-2,2}$ & $\cdots$ & $(M-1)^{n} = t_n^{0,M-2,(M-1)^{n-1}}$ & \multirow{2}{*}{repeat} \\\cline{1-5}  
    $p_k(t)$ & $p_k(1)$ & $p_k(2)$ & $\cdots$ & $p_k\big((M-1)^{n-1}\big)$ & \\\hline
    \end{tabular}}
    \label{tab:repeat_interference}
\end{table*}

On the other hand, suppose $W_n$ is Rx-$k$'s desired message, i.e., $n \in \mathcal{V}_k$. The first $(M-1)^n$ entries of $\mathbf{p}_k$ now take the form in Table \ref{tab:shift_desired}. Again, in Table \ref{tab:shift_desired}, the first sub-table that corresponds to the first $(M-1)^{n-1}$ entries of $\mathbf{p}_k$ is determined by the previous $n-1$ messages. The next $M-2$ sub-tables are just the first sub-table \emph{shifted} by some value, which is determined by the $n^{th}$ message. Column-wise, $p_k(t_n^{0,i,j}) = \Mod{p_k(t_n^{0,0,j}) + i}{M-1}$.

\begin{table*}
    \centering
    \caption{The first $(M-1)^n$ entries of $\mathbf{p}_k$ when $n \in \mathcal{V}_k$}
    \resizebox{0.9\textwidth}{!}{
    \begin{tabular}{|c|c|c|c|c|c|}
    \hline
    $t$ & $1 = t_n^{0,0,1}$ & $2 = t_n^{0,0,2}$ & $\cdots$ & $(M-1)^{n-1} = t_n^{0,0,(M-1)^{n-1}}$ & \multirow{2}{*}{determined}\\\cline{1-5}  
    $p_k(t)$ & $p_k(1)$ & $p_k(2)$ & $\cdots$ & $p_k\big((M-1)^{n-1}\big)$ & \\\hline\hline
    $t$ & $(M-1)^{n-1} + 1 = t_n^{0,1,1}$ & $(M-1)^{n-1} + 2 = t_n^{0,1,2}$ & $\cdots$ & $2(M-1)^{n-1} = t_n^{0,1,(M-1)^{n-1}}$ & \multirow{2}{*}{shift} \\\cline{1-5}  
    $p_k(t)$ & $\Mod{p_k(1) + 1}{M-1}$ & $\Mod{p_k(2) + 1}{M-1}$ & $\cdots$ & $\Mod{p_k\big((M-1)^{n-1}\big) + 1}{M-1}$ & \\\hline\hline
    $\vdots$ & $\vdots$ & $\vdots$ & $\vdots$ & $\vdots$ & $\vdots$\\\hline\hline
    $t$ & $(M-2)(M-1)^{n-1} + 1 = t_n^{0,M-2,1}$ & $(M-2)(M-1)^{n-1} + 2 = t_n^{0,M-2,2}$ & $\cdots$ & $(M-1)^{n} = t_n^{0,M-2,(M-1)^{n-1}}$ & \multirow{2}{*}{shift} \\\cline{1-5}  
    $p_k(t)$ & $\Mod{p_k(1)+ (M-2)}{M-1}$ & $\Mod{p_k(2)+ (M-2)}{M-1}$ & $\cdots$ & $\Mod{p_k\big((M-1)^{n-1}\big)+ (M-2)}{M-1}$ & \\\hline
    \end{tabular}}
    \label{tab:shift_desired}
\end{table*}

Let us then prove \eqref{eq:alignSepSuff} is true for arbitrary $n \in [N_g], h \in [0:(M-1)^{N_g - n} - 1], j \in [(M-1)^{n-1}]$. It can be easily verify that, for arbitrary $h \in [0:(M-1)^{N_g - n} - 1]$
\begin{align}
    &p_k\big(h(M-1)^n + 1\big), p_k\big(h(M-1)^n + 2\big),\notag\\
    &\quad\quad\quad\cdots, p_k\big((h+1)(M-1)^n\big)\notag\\
    = &\Mod{p_k(1) + z}{M-1}, \Mod{p_k(2)+z}{M-1},\notag\\
    &\quad\quad\quad\cdots, \Mod{p_k\big((M-1)^n\big) + z}{M-1}
\end{align}
where $z \in [0:M-1]$ can be viewed as a random number. This is because according to the algorithm, $p_k\big(h(M-1)^n + 1\big), \cdots, p_k\big((h+1)(M-1)^n\big)$ is generated by repeating or shifting $p_k(1), \cdots, p_k\big((M-1)^n\big)$ as a whole each time by  messages whose indices are greater than $n$. Note that,
\begin{align}
    &p_k\big(h(M-1)^n + 1\big), p_k\big(h(M-1)^n + 2\big),\notag\\
    &\quad\quad\quad\cdots, p_k\big((h+1)(M-1)^n \big)\notag\\
    = &p_k\big(t_n^{h,0,1}\big), \cdots, p_k\big(t_n^{h,M-2,(M-1)^{n-1}}\big).
\end{align}
Thus, for arbitrary $n \in [N_g], h \in [0:(M-1)^{N_g - n} - 1], j \in [(M-1)^{n-1}]$, and for all $i \in [0:M-2]$ the difference between $p_k(t_n^{h,i,j})$ and $p_k(t_n^{h,0,j})$ is the same as the difference between $p_k(t_n^{0,i,j})$ and $p_k(t_n^{0,0,j})$. \eqref{eq:alignSepSuff} is thus proved. \eqref{eq:alignSep} is a direct consequence of \eqref{eq:alignSepSuff}. Basically, given any $n \in [N_g]$, for any $h \in [0:(M-1)^{N_g - n} - 1], j \in [(M-1)^{n-1}]$ and $l = l_n^{h,j} \in [\ell]$, we have \eqref{eq:BIA_BCGM_Appendix}.
\begin{figure*}[hb]
\hrule
\begin{align}
&\left\{w_n^{k,f_n(t_n^{h,0,j}), [p_k(t_n^{h,0,j})]}, w_n^{k,f_n(t_n^{h,1,j}), [p_k(t_n^{h,1,j})]}, \cdots, w_n^{k,f_n(t_n^{h,M-2,j}), [p_k(t_n^{h,M-2,j})]}\right\} \notag\\
&= \begin{cases}
\left\{w_n^{k,l,[m_{k,n}^{l}]}, w_n^{k,l,[m_{k,n}^{l}]}, \cdots, w_n^{k,l,[m_{k,n}^{l}]}\right\} & n \in [N_g] \setminus \mathcal{V}_k\\
\left\{w_n^{k,l,[m_{k,n}^{l}]}, w_n^{k,l,\left[\Mod{m_{k,n}^{l} + 1}{M-1}\right]}, \cdots, w_n^{k,l,\left[\Mod{m_{k,n}^{l} + (M-2)}{M-1}\right]}\right\} & n \in \mathcal{V}_k
\end{cases}\notag\\
&= \begin{cases}
\left\{w_n^{k,l,[m_{k,n}^{l}]}\right\} & n \in [N_g] \setminus \mathcal{V}_k\\
\left\{w_n^{k,l,[1]}, w_n^{k,l,[2]}, \cdots, w_n^{k,l,[M-1]}\right\} & n \in \mathcal{V}_k
\end{cases}\label{eq:BIA_BCGM_Appendix}
\end{align}
\end{figure*}

\section{Correspondence to a Standard Wireless MapReduce Problem}\label{app:mapRed}
The $K$ Tx-Rx pairs (users) correspond to $K$ computation nodes in a standard MapReduce problem. The $N_r = \binom{K}{r}$ super-messages $\widetilde{\mathbf{W}}_{\mathcal{T}}, \forall \mathcal{T} \in \binom{[K]}{r}$ can be regarded as $N_r$ files $w_1, \cdots, w_{N_r}$. There are $K$ functions $\phi_1\left(w_{[N_r]}\right), \phi_2\left(w_{[N_r]}\right), \cdots, \phi_K\left(w_{[N_r]}\right)$ where $\phi_k$ is assigned to node $k$ for any $k \in [K]$.\footnote{In \cite{Li_Chen_Wang_MapReduce}, there are $Q$ functions to be computed, however we consider a symmetric case in this paper where $Q=K$ that is similar to \cite{Bi_Wigger_MapReduce}.}

In a standard MapReduce problem, there are three phases:

\textbf{Map Phase}: For any $n \in [N_r]$, file $w_n$ will be used to compute $K$ intermediate values (IVAs) $a_{1,n}, a_{2,n}, \cdots, a_{K,n}$ through $K$ map functions $u_{1,n}, u_{2,n}, \cdots, u_{K,n}$, i.e., 
\begin{align}
    a_{k,n} = u_{k,n}(w_n), \forall k \in [K], n \in [N_r].
\end{align}
We consider a symmetric MapReduce problem so that each file $w_n, n \in [N_r]$ is assigned to a distinct group of $r$ nodes, say nodes $\mathcal{T}_n$, where $\left\{\mathcal{T}_1, \mathcal{T}_2, \cdots, \mathcal{T}_{N_r}\right\} = \binom{[K]}{r}$ (super-message $\widetilde{\mathbf{W}}_{\mathcal{T}}$ is known to Tx-Rx pairs $\mathcal{T}$). The $K$ IVAs $\left\{a_{k,n}\right\}_{k \in [K]}$ for file $w_n$ will be computed at all the $r$ nodes $\mathcal{T}_n$ who are assigned $w_n$. Thus, $r$ is the computation load.

\textbf{Shuffle Phase}: In the shuffle phase, for any $n \in [N_r]$, IVA $a_{k,n}$ should be delivered to node $k$ for any $k \in [K]$, through the wireless network. Note that in the map phase, $w_n$ is already assigned to $r$ nodes $\mathcal{T}_n$, thus, the IVAs
\begin{align}
[\mbox{Redundant}]: && \left\{{\color{black!50}a_{t,n}} \mid t \in \mathcal{T}_n\right\}.
\end{align}
are already available (computed) at the destinations and are redundant. Consequently, we only need to transmit the effective IVAs 
\begin{align}
[\mbox{Effective}]: && \left\{a_{\bar{t},n} \mid \bar{t} \in [K]\setminus\mathcal{T}_n\right\}.
\end{align}
The \emph{effective IVAs} computed from \emph{files} $w_n$ which are delivered to nodes $\mathcal{T}_n$ correspond to the \emph{messages} contained in \emph{super-message} $\widetilde{\mathbf{W}}_{\mathcal{T}_n}$, i.e., we have $a_{\bar{t},n} = \widetilde{W}_{\mathcal{T}_n, \bar{t}}, \forall n \in [N_r], \bar{t} \in [K] \setminus \mathcal{T}_n$.

\textbf{Reduce Phase}: In this phase, for any $k \in [K]$, node $k$ computes 
\begin{align}
    \phi_k(w_{[N_r]}) = v_k(a_{k,1}, a_{k,2}, \cdots, a_{k,N_r})
\end{align}
where $v_k$ is a reduce function. Since this paper only considers the communications cost of the wireless MapReduce problem, this phase has no effect in the wireless MapReduce network abstraction.

Below is a concrete example where \cite[Fig. 3]{Li_Chen_Wang_MapReduce} is abstracted to the example in Fig. \ref{fig:mapRedICuni}a.

In \cite[Fig. 3]{Li_Chen_Wang_MapReduce}, there are $N_r=6$ files and $K = 4$ users. Each user knows $3$ files, e.g., node $1$ knows $w_1, w_2, w_3$, and each file is available to $r = 2$ users, e.g., $w_1$ is available to user $1$ and $2$. Each user wants to compute $1$ function of the $6$ files, e.g., user $1$ wants to compute $\phi_1(w_1, w_2, \cdots, w_6) = v_1(a_{1,1}, a_{1,2}, \cdots, a_{1,6})$ where $v_1(\cdot)$ is the reduce function and $a_{1,1}, \cdots, a_{1,6}$ are the six IVAs corresponding to the $6$ files generated by $6$ map functions, that are required by user $1$. Thus, each user must get the IVAs that cannot be computed by itself, from the other users. For example, user $1$ does not have $w_4, w_5, w_6$ so it must collect $a_{1,4}, a_{1,5}, a_{1,6}$ from the other users. And also, for message, say $w_1$, the IVAs ${\color{black!50}a_{1,1}}, {\color{black!50}a_{2,1}}$ are redundant as $w_1$ is available to user $1$ and $2$ at the map phase.

The entire notation mapping is as follows,
\begin{align}
	w_1 \rightarrow ({\color{black!50}a_{1,1}}, {\color{black!50}a_{2,1}}, a_{3,1}, a_{4,1}) &\equiv \widetilde{\bf A} = (\widetilde{A}_3, \widetilde{A}_4)\notag\\
	w_2 \rightarrow ({\color{black!50}a_{1,2}}, a_{2,2}, {\color{black!50}a_{3,2}}, a_{4,2}) &\equiv \widetilde{\bf B} = (\widetilde{B}_2, \widetilde{B}_4)\notag\\
	w_3 \rightarrow ({\color{black!50}a_{1,3}}, a_{2,3}, a_{3,3}, {\color{black!50}a_{4,3}}) &\equiv \widetilde{\bf C} = (\widetilde{C}_2, \widetilde{C}_3)\notag\\
	w_4 \rightarrow (a_{1,4}, {\color{black!50}a_{2,4}}, {\color{black!50}a_{3,4}}, a_{4,4}) &\equiv \widetilde{\bf D} = (\widetilde{D}_1, \widetilde{D}_4)\notag\\
	w_5 \rightarrow (a_{1,5}, {\color{black!50}a_{2,5}}, a_{3,5},{\color{black!50} a_{4,5}}) &\equiv \widetilde{\bf E} = (\widetilde{E}_1, \widetilde{E}_3)\notag\\
	w_6 \rightarrow (a_{1,6}, a_{2,6}, {\color{black!50}a_{3,6}}, {\color{black!50}a_{4,6}}) &\equiv \widetilde{\bf F} = (\widetilde{F}_1, \widetilde{F}_2)\notag
\end{align}
where gray elements are those discarded redundant IVAs and they are  not shown in the ``Broadcast'' part in  \cite[Fig. 3]{Li_Chen_Wang_MapReduce} as well.

\section{Converse Proofs for Theorem \ref{thm:BCGM} and Theorem \ref{thm:USI}}\label{app:converse}
Suppose a scheme takes $\mathsf{T}$ channel uses and the switching pattern of Rx-$k, k\in [K]$, is $\mathbf{m}_k = (m_k(1), m_k(2), \cdots, m_k(\mathsf{T}))$. 
We allow $\mathsf{T}$ to approach infinity so that $\mathsf{T} \geq \mathsf{T}_c$, i.e., channel coherence time. Let $Q = \left\lceil\frac{\mathsf{T}}{\mathsf{T}_c}\right\rceil$, i.e., a scheme takes $Q$ coherence blocks. Let 
\begin{align}
    \mathbf{h}_{k}^{[m]}(1 + q\mathsf{T}_c), q \in [0:Q-1], m \in [M]
\end{align}
be a random variable that denotes the channel vector at Rx-$k$ in the $(q+1)^{th}$ coherence block when the antenna is operating in mode-$m$. Let 
\begin{align}
    \bm{\mathcal{H}}_k \triangleq \{\mathbf{h}_{k}^{[m]}(1 + q\mathsf{T}_c)\}_{q \in [0:Q-1], m \in [M]}.
\end{align}
Note that Rx-$k$ knows $\bm{\mathcal{H}}_k$ since we assume perfect CSIR.

For ease of proof, let us first define $y_k^{[m]}(t), k \in [K], m \in [M]$ as the signal received by Rx-$k$ at time $t$, with the antenna operating in mode $m$. Specifically,
\begin{align}
	y_k^{[m]}(t)\triangleq\mathbf{h}_k^{[m]}(t) \mathbf{x}(t)+z_k(t).
\end{align}
Also, for compact notation, for $\mathbf{m} = (m(1), m(2), \cdots, m(\mathsf{T}))$, let $\mathbf{y}_{k,\mathsf{T}}^{[\mathbf{m}]}$ denote the collection of the received signals within the $\mathsf{T}$ time slots at Rx-$k$, when the switching pattern is $\mathbf{m}$, i.e.,
\begin{align}
	\mathbf{y}_{k,\mathsf{T}}^{[\mathbf{m}]} \triangleq \bigg(y_k^{[m(1)]}(1), y_k^{[m(2)]}(2), \cdots, y_k^{[m(\mathsf{T})]}(\mathsf{T})\bigg).
\end{align}
Note that, if Rx-$k$ follows the designed switching pattern $\mathbf{m}_k$, the received signal over the $\mathsf{T}$ channel uses, at Rx-$k$, is $\mathbf{y}_{k,\mathsf{T}}^{[\mathbf{m}_k]}$.
Further, let 
\begin{align}
	\mathbf{Y}_{k_M} &\triangleq \bigg(\mathbf{y}_{k,\mathsf{T}}^{[\mathbf{m}_k]}, \mathbf{y}_{k,\mathsf{T}}^{[\Mod{\mathbf{m}_k + \mathbf{1}}{M}]}, \cdots, \mathbf{y}_{k,\mathsf{T}}^{[\Mod{\mathbf{m}_k + (M-1)\mathbf{1}}{M}]}\bigg)\\
 &=\bigg(y_k^{[m_k(1)]}(1), \cdots, y_k^{[\Mod{m_k(1) + (M - 1)}{M}]}(1),\notag\\
 &~~~~y_k^{[m_k(2)]}(2), \cdots, y_k^{[\Mod{m_k(2) + (M-1)}{M}]}(2), \cdots,\notag\\
 &~~~~y_k^{[m_k(\mathsf{T})]}(\mathsf{T}), \cdots, y_k^{[\Mod{m_k(\mathsf{T}) + (M-1)}{M}]}(\mathsf{T})\bigg)\\
 &=\bigg(y_k^{[1]}(1), y_k^{[2]}(1), \cdots, y_k^{[M]}(1),\notag\\
 &~~~~y_k^{[1]}(2), y_k^{[2]}(2), \cdots, y_k^{[M]}(2),\notag\\
 &~~~~\cdots, y_k^{[1]}(\mathsf{T}), y_k^{[2]}(\mathsf{T}), \cdots, y_k^{[M]}(\mathsf{T})\bigg)
\end{align}
so that $\mathbf{Y}_{k_M}$ denotes the collection of the signal received at Rx-$k$ with all the $M$ possible modes, in the $\mathsf{T}$ time slots. Note that, according to the definition, 
\begin{align}
    \mathbf{y}_{k,\mathsf{T}}^{[\mathbf{m}]} \subset \mathbf{Y}_{k_M}, \forall k \in [K], \mathbf{m} \in [M]^{\mathsf{T}}.\label{eq:Mcopy_anypattern}
\end{align}

We also have the following lemma which is important for the converse proof. The lemma essentially says that, in any feasible BCGM or USI scheme, $M$ ``copies'' of Rx-$k$ (each with switching pattern that is obtained by shifting the designed switching pattern $\mathbf{m}_k$ as whole by a distinct value), are able to decode the messages desired by an arbitrary receiver, if provided with the corresponding receiver's side information (which is empty in the BCGM setting). Also, any one of the $M$ ``copies" of Rx-$k$, is able to decode the messages desired by Rx-$k$.
\begin{lemma}\label{lem:decodability}
\begin{align}
&\mbox{\rm Other Rxs: } \forall k, \bar{k} \in [K], k \neq \bar{k}, \notag\\
&\hspace{0.2cm} H(\de_{k}^{*} \mid \si_{k}^{*}, \mathbf{Y}_{\bar{k}_M}, \bm{ \mathcal{H}}_{\bar{k}}) = o(\mathsf{T}),\label{eq:lemma_other}\\
&\mbox{\rm Rx-$k$: } \forall \eta \in [M], \notag\\
&\hspace{0.2cm} H(\de_{k}^{*} \mid \si_{k}^{*}, \mathbf{y}_{k,\mathsf{T}}^{[\Mod{\mathbf{m}_k + \eta\mathbf{1}}{M}]}, \bm{ \mathcal{H}}_k) = o(\mathsf{T}).\label{eq:lemma_rx1}
\end{align}
where $*$ can be substituted with ``BCGM'' when referring to BCGM, and ``USI'' when referring to USI. Note that there is no side information in the BCGM setting, i.e., $\si_{k}^{\gc} = \emptyset$.
\end{lemma}

\proof Let us first consider  BCGM. Note that according to the definition of the problem (all desired messages should be recovered with negligible probability of error), after applying Fano's inequality \cite{Cover_Elements}, we have 
\begin{align}
    H(\de_{k}^{\gc} \mid \mathbf{y}_{k,\mathsf{T}}^{[\mathbf{m}_k]}, \bm{\mathcal{H}}_k) = o(\mathsf{T}).\label{eq:fano_dec}
\end{align}
Recall that $\mathbf{y}_{k,\mathsf{T}}^{\mathbf{m}_k}$ denotes the signals received by Rx-$k$ over $\mathsf{T}$ channel uses, following the designed switching pattern $\mathbf{m}_k$.

For a different Rx-$\bar{k} \neq k$, we note that, if it follows Rx-$k$'s switching pattern $\mathbf{m}_k$ instead of its own swtiching pattern $\mathbf{m}_{\bar{k}}$, it receives $\mathbf{y}_{\bar{k},\mathsf{T}}^{[\mathbf{m}_k]}$. Due to the fact that $\bm{\mathcal{H}}_{\bar{k}}$ is statistically equivalent to $\bm{\mathcal{H}}_{k}$ and the transmitters have no CSIT so $\mathbf{x}(1), \cdots, \mathbf{x}(\mathsf{T})$ are independent of $\bm{\mathcal{H}}_k$ for any $k \in [K]$. Thus, we must have $\mathbf{y}_{\bar{k},\mathsf{T}}^{[\mathbf{m}_k]}, \bm{\mathcal{H}}_{\bar{k}}$ are statistically equivalent to $\mathbf{y}_{k,\mathsf{T}}^{[\mathbf{m}_k]}, \bm{\mathcal{H}}_{k}$, i.e.,
\begin{align}
    (\mathbf{y}_{\bar{k},\mathsf{T}}^{[\mathbf{m}_k]}, \bm{\mathcal{H}}_{\bar{k}}) \sim (\mathbf{y}_{k,\mathsf{T}}^{[\mathbf{m}_k]}, \bm{\mathcal{H}}_{k}), \forall k, \bar{k} \in [K], k \neq \bar{k}.
\end{align}
Combined with \eqref{eq:fano_dec}, we must have 
\begin{align}
    H(\de_{k}^{\gc} \mid \mathbf{y}_{\bar{k},\mathsf{T}}^{[\mathbf{m}_k]}, \bm{\mathcal{H}}_{\bar{k}}) = o(\mathsf{T}).
\end{align}
Note that according to \eqref{eq:Mcopy_anypattern}, $\mathbf{y}_{\bar{k},\mathsf{T}}^{[\mathbf{m}_k]} \subset \mathbf{Y}_{\bar{k}_M}$, thus 
\begin{align}
    H(\de_{k}^{\gc} \mid \mathbf{Y}_{\bar{k}_M}, \bm{\mathcal{H}}_{\bar{k}}) &\leq H(\de_{k}^{\gc} \mid \mathbf{y}_{\bar{k},\mathsf{T}}^{[\mathbf{m}_k]}, \bm{\mathcal{H}}_{\bar{k}})\notag\\
    &= o(\mathsf{T}),
\end{align}
since conditioning reduces entropy. Thus, for the BCGM, \eqref{eq:lemma_other} is proved.

Intuitively, the $M$ ``copies'' (with different offsets) of, say, Rx-$1$  can jointly obtain a signal statistically equivalent to the signal seen by Rx-$2$ regardless of how Rx-$2$ switches its antenna modes. Thus, $M$ ``copies'' of Rx-$1$ are able to decode the messages desired by the Rx-$2$ with negligible probability of error. Similarly, they are able to decode the messages desired by Rx-$3, 4, \cdots, K$ with negligible probability of error.

For \eqref{eq:lemma_rx1}, since the channels for the different modes are i.i.d. and the transmitter has no CSIT, thus 
\begin{align}
    (\mathbf{y}_{k,\mathsf{T}}^{[\mathbf{m}_k]}, \bm{\mathcal{H}}_{k}) \sim (\mathbf{y}_{k,\mathsf{T}}^{[\Mod{\mathbf{m}_k + \eta \mathbf{1}}{M}]}, \bm{\mathcal{H}}_{k}),\notag\\
    \forall k \in [K], \eta \in [M],
\end{align}
combined with \eqref{eq:fano_dec}, \eqref{eq:lemma_rx1} for the BCGM setting is proved. \eqref{eq:lemma_rx1} essentially says that the encoding scheme must enable each ``copy'' of receiver, say Rx-$1$, to decode the desired messages.

The proof for the USI is similar. The only difference is that the $M$ ``copies'' of Rx-$\bar{k}$ need other Rx-$k$'s side information to decode corresponding desired messages. $\hfill \blacksquare$

\subsection{Converse of Theorem \ref{thm:BCGM}}\label{app:converse_BCGM}
To simplify the notation, we eliminate all the subscripts or superscripts that denote we are considering BCGM setting in the proof. Specifically,
\begin{align}
	&M \sum_{n \in \mathcal{V}_1} \mathsf{T} R_n  = MH(\de_{1}) = MH(\de_{1} \mid \bm{\mathcal{H}}_{1})\notag\\
    &\overset{\eqref{eq:lemma_rx1}}{=} \sum_{\eta \in [M]} I(\de_1; \mathbf{y}_{1,\mathsf{T}}^{[\Mod{\mathbf{m}_k + \eta\mathbf{1}}{M}]} \mid \bm{\mathcal{H}}_{1}) + o(\mathsf{T}) \notag\\
	&= \sum_{\eta \in [M]} \bigg(h(\mathbf{y}_{1,\mathsf{T}}^{[\Mod{\mathbf{m}_k + \eta\mathbf{1}}{M}]} \mid \bm{\mathcal{H}}_{1})\notag\\
    &\quad\quad\quad\quad\quad- h(\mathbf{y}_{1,\mathsf{T}}^{[\Mod{\mathbf{m}_k + \eta\mathbf{1}}{M}]} \mid \de_{1}, \bm{\mathcal{H}}_{1})\bigg) + o(\mathsf{T})\notag\\
	&\leq M\mathsf{T}\log(P) - h( \mathbf{Y}_{1_M} \mid \de_{1}, \bm{\mathcal{H}}_{1}) + o(\log P)\label{eq:converse_group_cr}
\end{align}
where \eqref{eq:converse_group_cr} follows from the fact that Gaussian distribution maximizes entropy under power constraint $P$ and conditioning reduces entropy. Meanwhile, we have
\begin{align}
	&\sum_{n \in [N_g]\setminus\mathcal{V}_1}\mathsf{T}R_{n}
    = H(\de_{[2:K]} \setminus \de_{1})\notag\\
    &= H(\de_{[2:K]} \mid \de_{1}) = H(\de_{[2:K]} \mid \de_{1}, \bm{\mathcal{H}}_{1})\notag\\
    &\overset{\eqref{eq:lemma_other}}{=} I(\de_{[2:K]}; \mathbf{Y}_{1_M} \mid \de_1, \bm{\mathcal{H}}_{1}) + o(\mathsf{T})\\
	&= h(\mathbf{Y}_{1_M} \mid \de_{1}, \bm{\mathcal{H}}_{1}) - h(\mathbf{Y}_{1_M} \mid \de_{[K]}, \bm{\mathcal{H}}_{1}) + o(\mathsf{T})\notag\\
	&= h(\mathbf{Y}_{1_M} \mid \de_{1}, \bm{\mathcal{H}}_{1}) - \underbrace{h(z_1(1), \cdots, z_1(\mathsf{T}))}_{o(\log P)} + o(\mathsf{T})\label{eq:converse_group_noise}
\end{align}
where \eqref{eq:converse_group_noise} follows from the fact that given all the messages and channel vectors, the received signal is a function of the noise which is independent of the channel vector and messages. Adding \eqref{eq:converse_group_cr} and \eqref{eq:converse_group_noise}, dividing by $\mathsf{T}$, and considering the symmetry of the receivers, for any $k \in [K]$, we have,
\begin{align}
	M \sum_{n \in \mathcal{V}_k} R_n + \sum_{n\in [N_g]\setminus\mathcal{V}_k}R_{n} \leq M\log(P) + o(\log P).
\end{align}
Thus,
\begin{align}
	M \sum_{n \in \mathcal{V}_k} d_n + \sum_{n \in [N_g]\setminus\mathcal{V}_k}d_{n} \leq M.
\end{align}
Averaging over all the $K$ receivers, we have
\begin{align}
	\sum_{n \in [N_g]}d_{n} &\leq \frac{MK}{GM + K - G} = \frac{N_g M}{(M-1)\nu_g + N_g}.
\end{align}
Thus, $d_{\Sigma} \leq \frac{N_g M}{(M-1)\nu_g + N_g}$.

\begin{remark}\label{rmk:BCGM_Converse_MTx_MRx}
In fact, the statistical equivalence argument is also true even if the number of transmitter antennas is not equal to the number of receiver antennas' modes, i.e., $M_{\mbox{\tiny Tx}} > M_{\mbox{\tiny Rx}}$. Specifically, $M_{\mbox{\tiny Rx}}$ ``copies'' of Rx-$1$ can jointly obtain a signal statistically equivalent to the signal seen by Rx-$2$ and decode the messages desired by Rx-$2$. Similarly, messages desired by Rx-$3, \cdots, K$ can be decoded. Thus $d_{\Sigma} \leq \frac{N_g M_{\mbox{\tiny Rx}}}{(M_{\mbox{\tiny Rx}}-1)\nu_g + N_g}$. 

On the other hand, when $M_{\mbox{\tiny Tx}} < M_{\mbox{\tiny Rx}}$, then $M_{\mbox{\tiny Tx}}$ ``copies'' of Rx-$1$ can already invert the channel matrix and thus obtain whatever is sent out from the transmitter antenna (within bounded noise distortion  that is inconsequential for DoF) and thus decode the messages desired by Rx-$2,3, \cdots, K$. Applying the argument in this section, we obtain $d_{\Sigma} \leq \frac{N_g M_{\mbox{\tiny Tx}}}{(M_{\mbox{\tiny Tx}}-1)\nu_g + N_g}$.
\end{remark}

\subsection{Converse of Theorem \ref{thm:USI}}\label{app:converse_USI}
The converse idea is similar, but a bit more involved in the USI setting. Recall that in USI, all messages are transmitted at the same rate $R$, since we have this restriction in MapReduce which is equivalent to USI.

\begin{align}
	&M \underbrace{\binom{K-1}{G-1}}_{\nu_g} \mathsf{T} R = MH(\de_{1}) = MH(\de_{1} \mid \si_1, \bm{\mathcal{H}}_{1})\notag\\
    &\overset{\eqref{eq:lemma_rx1}}{=} \sum_{\eta \in [M]} I(\de_1; \mathbf{y}_{1,\mathsf{T}}^{[\Mod{\mathbf{m}_k + \eta\mathbf{1}}{M}]} \mid \si_1, \bm{\mathcal{H}}_{1}) + o(\mathsf{T}) \notag\\
	&= \sum_{\eta \in [M]} \bigg(h(\mathbf{y}_{1,\mathsf{T}}^{[\Mod{\mathbf{m}_k + \eta\mathbf{1}}{M}]} \mid \si_1, \bm{\mathcal{H}}_{1})\notag\\
    &\quad\quad\quad\quad\quad - h(\mathbf{y}_{1,\mathsf{T}}^{[\Mod{\mathbf{m}_k + \eta\mathbf{1}}{M}]} \mid \de_{1},\si_1, \bm{\mathcal{H}}_{1})\bigg) + o(\mathsf{T})\notag\\
	&\leq M\mathsf{T}\log(P) - h( \mathbf{Y}_{1_M} \mid \de_1, \si_{[2]}, \bm{\mathcal{H}}_{1}) + o(\mathsf{T}),\label{eq:converse_uni_cr}
\end{align}
where the last step follows from the fact that Gaussian distribution maximizes entropy under power constraint $P$, and conditioning reduces entropy. Meanwhile, we have,
\begin{align}
	&\binom{K-2}{G-1} \mathsf{T} R = H(\de_2\setminus\si_1)\notag\\
    &= H(\de_2 \mid \si_1) = H(\de_2 \mid \de_{1}, \si_{[2]}, \bm{\mathcal{H}}_{1})\notag\\ 
	&\overset{\eqref{eq:lemma_rx1}}{=} I(\de_2; \mathbf{Y}_{1_M} \mid \de_1,\si_{[2]}, \bm{\mathcal{H}}_{1}) + o(\mathsf{T}) \notag\\
	&\leq h(\mathbf{Y}_{1_M} \mid \de_1,\si_{[2]}, \bm{\mathcal{H}}_{1})\notag\\
    &\quad - h(\mathbf{Y}_{1_M} \mid \de_{[2]},\si_{[3]}, \bm{\mathcal{H}}_{1}) + o(\mathsf{T}).\label{eq:uni_induction_2}
\end{align}
The first step is to bound the entropy of the messages that are desired by Rx-$2$ and are not available to Rx-$1$ as side information. Specifically, there are $\binom{K-2}{G-1}$ super-messages that correspond to Rx-$2$ while not to Rx-$1$, each of which contains a message that is desired by Rx-$2$. Meanwhile, there are $\binom{K-2}{G-2}$ super-messages correspond to both Rx-$1$ and Rx-$2$. Though these super-messages also contain messages desired by Rx-$2$, these messages are all known to Rx-$1$ by problem formulation. The last step follows from the fact that conditioning reduces entropy.

Following a similar process, we have the bound,
\begin{align}
	\binom{K-k}{G-1} \mathsf{T} R &\leq h(\mathbf{Y}_{1_M} \mid \de_{[k-1]},\si_{[k]}, \bm{\mathcal{H}}_{1})\notag\\
    &\quad- h(\mathbf{Y}_{1_M} \mid \de_{[k]},\si_{[k+1]}, \bm{\mathcal{H}}_{1}) + o(\mathsf{T}).\label{eq:uni_induction_k}
\end{align}
Here, we bound the entropy of the messages that are desired by Rx-$k$ and are not available to the first $k-1$ receivers as  side information. 

For all $k \in [2:K-G+1]$, adding \eqref{eq:uni_induction_k}  to \eqref{eq:converse_uni_cr} we have 
\begin{align}
	&\left(M\binom{K-1}{G-1} + \binom{K-2}{G-1} + \cdots + \binom{G-1}{G-1}\right) \mathsf{T} R \notag\\
    &= (M-1)\nu_g \mathsf{T} R \notag\\
    &\quad+\underbrace{\left(\binom{K-1}{G-1} + \binom{K-2}{G-1} + \cdots + \binom{G-1}{G-1}\right)}_{=\binom{K}{G} = N_g}\mathsf{T} R \notag\\
	&\leq M\mathsf{T}\log P - h(\mathbf{Y}_{1_M} \mid \de_{[k-G+1]}, \si_{[k-G+2]}) + o(\log P)\notag\\
	&= M\mathsf{T}\log P \underbrace{- h(z_1(1), \cdots, z_1(\mathsf{T})) + o(\log P)}_{o(\log P)},\label{eq:uni_RX1_bound}
\end{align}
where the last step holds as any $K-G+1$ receivers' side information and their desired messages constitute all the $N_g$ super-messages, and the power of the noise is bounded (negligible for DoF). Thus,
\begin{align}
    &\left((M-1)\nu_g + N_g\right) R \leq M\log P + o(\log P)\notag\\
    &\Rightarrow d \leq \frac{M}{(M-1)\nu_g + N_g},
\end{align}
and  $d_{\Sigma} \leq \frac{GN_gM}{(M-1)\nu_g + N_g}$.

\bibliographystyle{IEEEtran}
\bibliography{Thesis}

\end{document}